\documentclass[12pt]{article}

\def\modifymargins#1#2{
\newdimen\addtoh
\newdimen\addtow
\addtoh=#1
\addtow=#2

\advance\topmargin by -\addtoh
\multiply\addtoh by 2
\advance\textheight by \addtoh

\advance\oddsidemargin by -\addtow
\advance\evensidemargin by -\addtow
\multiply\addtow by 2
\advance\textwidth by \addtow
}

\let\citeyear=\cite
\expandafter\ifx\csname ttytty\endcsname\relax
\def\ttytex#1{#1\nop}			% this is for latex stuff
\else

\modifymargins{0pt}{-10pt}
\def\frac#1#2{#1/#2}
\def\_{\char 95}
\expandafter\ifx\csname makeatletter\endcsname\relax\makeatletter\fi
\def\@ttyfig#1#2{\def\specialtext{\special{txt:#2}}\specialtext\egroup}
\def\ttyfig{\relax\bgroup\catcode`\^^M\active\let^^M
\let\-\relax\catcode`\ \active\@ttyfig}
\let\ttytex\ttyfig
\expandafter\ifx\csname makeatother\endcsname\relax\makeatletter\fi
\fi

\expandafter\ifx\csname ttytty\endcsname\relax\modifymargins{60pt}{40pt}%
\else\raggedright\parindent=1cm

\fi

\def\color[#1]#2{}

\def\possnewtheorem#1#2{
\expandafter\ifx\csname #1\endcsname\relax
\newtheorem{#1}{#2}
\fi
}

\long\def\nop#1{}
\def\true{{\sf true}}
\def\false{{\sf false}}
\def\var{V\!ar}

\def\resolve{\mbox{resolve}}
\def\rescn{\mbox{ResCn}}
\def\np{{\rm NP}}
\def\conp{{\rm coNP}}
\def\D#1{\mbox{$\Delta^p_{#1}$}}
\def\Dp{${\rm D}^p$}
\def\Dpk#1{\mbox{$D^p_{#1}$}}
\def\Dptwo{\Dpk{2}}
\def\Dlog#1{\mbox{$\Delta^p_{#1}[\log n]$}}
\def\S#1{\mbox{$\Sigma^p_{#1}$}}
\def\pspace{{\rm PSPACE}}
\possnewtheorem{definition}{Definition}
\possnewtheorem{lemma}{Lemma}
\possnewtheorem{theorem}{Theorem}
\possnewtheorem{algorithm}{Algorithm}
\possnewtheorem{counterexample}{Counterexample}
\def\proof{\noindent {\sl Proof.\ \ }}
\expandafter\ifx\csname shortcite\endcsname\relax
\let\shortcite=\cite
\fi
\def\qed{\hfill{\boxit{}}
  \ifdim\lastskip<\medskipamount \removelastskip\penalty55\medskip\fi}
\def\qedn#1{\hfill{\boxit{}$_#1$}
  \ifdim\lastskip<\medskipamount \removelastskip\penalty55\medskip\fi}
\long\def\boxit#1{\vbox{\hrule\hbox{\vrule\kern3pt
                  \vbox{\kern3pt#1\kern3pt}\kern3pt\vrule}\hrule}}
\let\plural=\relax
\let\cedilla=\c

\title{The ghosts of forgotten things:\\
A study on size after forgetting}
\author{Paolo Liberatore\thanks{%
DIIAG, Sapienza University of Rome.
{\tt liberato@diag.uniroma1.it}
}}
\date{}

\begin{document}

\maketitle

\let\oinput=\input
\def\noinput#1{}

\let\oproof=\proof
\long\def\proof#1\qed{}

\begin{abstract}

Forgetting is removing variables from a logical formula while preserving the
constraints on the other variables. In spite of reducing information, it does
not always decrease the size of the formula and may sometimes increase it. This
article discusses the implications of such an increase and analyzes the
computational properties of the phenomenon. Given a propositional Horn formula,
a set of variables and a maximum allowed size, deciding whether forgetting the
variables from the formula can be expressed in that size is \Dp-hard in \S{2}.
The same problem for unrestricted CNF propositional formulae is \Dptwo-hard in
\S{3}.

\end{abstract}

%input{synopsis.tex}
\section{Introduction}

Several articles mention simplification as an advantage of forgetting, if not
its motivation. Forgetting means deleting pieces of knowledge, and less is
more. Less knowledge is easier to remember, easier to work with, easier to
interpret. To cite a few:

\begin{itemize}

\item ``With an ever growing stream of information, bounded memory and short
response time suggest that not all information can be kept and treated in the
same way. [...] forgetting [...] helps us to deal with information overload and
to put a focus of attention''~\cite{eite-kern-19}.

% \citeyear{eite-kern-19}

\item ``For example, in query answering, if one can determine what is relevant
with respect to a query, then forgetting the irrelevant part of a knowledge
base may yield more efficient query-answering''~\cite{delg-wang-15}.

\item ``Moreover, forgetting may be applicable in summarising a knowledge base
by suppressing lesser details, or for reusing part of a knowledge base by
removing an unneeded part of a larger knowledge base, or in clarifying
relations between predicates''~\cite{delg-17}.

\item ``For performing reasoning tasks (planning, prediction, query answering,
etc.) in an action domain, not all actions of that domain might be necessary.
By instructing the reasoning system to forget about these
unnecessary/irrelevant actions, without changing the causal relations among
fluents, we might obtain solutions using less computational
time/space''~\cite{erde-ferr-07}.

\item ``There are often scenarios of interest where we want to model the fact
that certain information is discarded. In practice, for example, an agent may
simply not have enough memory capacity to remember everything he has
learned''~\cite{fagi-etal-95}.

\item ``The most immediate application of forgetting is to model agents with
limited resources (e.g., robots), or agents that need to deal with vast
knowledge bases (e.g., cloud computing), or more ambitiously, dealing with the
problem of lifelong learning. In all such cases it is no longer reasonable to
assume that all knowledge acquired over the operation of an agent can be
retained indefinitely''~\cite{raja-etal-14}.

\item ``For example, we have a knowledge base $K$ and a query $Q$. It may be
hard to determine if $Q$ is true or false directly from $K$. However, if we
discard or forget some part of $K$ that is independent of $Q$, the querying
task may become much easier''~\cite{wang-etal-05}.

\item ``To some extent, all of these can be reduced to the problem of
extracting relevant segments out of large ontologies for the purpose of
effective management of ontologies so that the tractability for both humans and
computers is enhanced. Such segments are not mere fragments of ontologies, but
stand alone as ontologies in their own right. The intuition here is similar to
views in databases: an existing ontology is tailored to a smaller ontology so
that an optimal ontology is produced for specific
applications''~\cite{eite-06}.

\end{itemize}

These authors are right: if forgetting simplifies the body of knowledge then it
is good
{} for reducing the amount of information to store,
{} for increasing the efficiency of querying it,
{} for clarifying the relationships between facts,
{} for obtaining solutions more easily,
{} for retaining by agents of limited memory,
{} for tailoring knowledge to a specific application.
If forgetting simplifies the body of knowledge, all these motivations are
valid.

If.

What if not? What if forgetting does not simplify the body of knowledge? What
if it complicates it? What if it enlarges it instead of making it smaller?

\begin{figure}
\begin{center}
\setlength{\unitlength}{5000sp}%
\begingroup\makeatletter\ifx\SetFigFont\undefined%
\gdef\SetFigFont#1#2#3#4#5{%
  \reset@font\fontsize{#1}{#2pt}%
  \fontfamily{#3}\fontseries{#4}\fontshape{#5}%
  \selectfont}%
\fi\endgroup%
\begin{picture}(5250,1542)(5206,-4810)
\thinlines
{\color[rgb]{0,0,0}\put(5311,-3436){\line( 1,-1){270}}
\put(5581,-3706){\line(-1,-1){270}}
}%
{\color[rgb]{0,0,0}\put(5581,-3706){\vector( 1, 0){360}}
}%
{\color[rgb]{0,0,0}\put(6121,-3706){\vector( 1, 0){450}}
}%
{\color[rgb]{0,0,0}\put(6751,-3706){\vector(-1, 0){  0}}
\put(6751,-3706){\vector( 1, 0){450}}
}%
{\color[rgb]{0,0,0}\put(6571,-4336){\vector(-4, 1){1260}}
}%
{\color[rgb]{0,0,0}\put(6751,-4336){\vector( 1, 0){450}}
}%
\thicklines
{\color[rgb]{0,0,0}\put(6301,-3301){\line( 0,-1){1260}}
}%
\thinlines
{\color[rgb]{0,0,0}\put(7831,-3931){\line( 0,-1){180}}
\put(7831,-4111){\line( 1, 0){180}}
\put(8011,-4111){\line( 0,-1){ 90}}
\put(8011,-4201){\line( 1, 1){180}}
\put(8191,-4021){\line(-1, 1){180}}
\put(8011,-3841){\line( 0,-1){ 90}}
\put(8011,-3931){\line(-1, 0){180}}
}%
{\color[rgb]{0,0,0}\put(8821,-3436){\line( 1,-1){270}}
\put(9091,-3706){\line(-1,-1){270}}
}%
{\color[rgb]{0,0,0}\put(9091,-3706){\vector( 1, 0){405}}
}%
\thicklines
{\color[rgb]{0,0,0}\put(9811,-3301){\line( 0,-1){1260}}
}%
\put(5221,-3481){\makebox(0,0)[b]{\smash{{\SetFigFont{12}{24.0}
{\rmdefault}{\mddefault}{\updefault}{\color[rgb]{0,0,0}$a$}%
}}}}
\put(5221,-4021){\makebox(0,0)[b]{\smash{{\SetFigFont{12}{24.0}
{\rmdefault}{\mddefault}{\updefault}{\color[rgb]{0,0,0}$b$}%
}}}}
\put(6031,-3751){\makebox(0,0)[b]{\smash{{\SetFigFont{12}{24.0}
{\rmdefault}{\mddefault}{\updefault}{\color[rgb]{0,0,0}$c$}%
}}}}
\put(6661,-3751){\makebox(0,0)[b]{\smash{{\SetFigFont{12}{24.0}
{\rmdefault}{\mddefault}{\updefault}{\color[rgb]{0,0,0}$d$}%
}}}}
\put(7291,-3751){\makebox(0,0)[b]{\smash{{\SetFigFont{12}{24.0}
{\rmdefault}{\mddefault}{\updefault}{\color[rgb]{0,0,0}$e$}%
}}}}
\put(6661,-4381){\makebox(0,0)[b]{\smash{{\SetFigFont{12}{24.0}
{\rmdefault}{\mddefault}{\updefault}{\color[rgb]{0,0,0}$f$}%
}}}}
\put(7291,-4381){\makebox(0,0)[b]{\smash{{\SetFigFont{12}{24.0}
{\rmdefault}{\mddefault}{\updefault}{\color[rgb]{0,0,0}$g$}%
}}}}
\put(6931,-4741){\makebox(0,0)[b]{\smash{{\SetFigFont{12}{24.0}
{\rmdefault}{\mddefault}{\updefault}{\color[rgb]{0,0,0}$forget$}%
}}}}
\put(5671,-4741){\makebox(0,0)[b]{\smash{{\SetFigFont{12}{24.0}
{\rmdefault}{\mddefault}{\updefault}{\color[rgb]{0,0,0}$remember$}%
}}}}
\put(8731,-3481){\makebox(0,0)[b]{\smash{{\SetFigFont{12}{24.0}
{\rmdefault}{\mddefault}{\updefault}{\color[rgb]{0,0,0}$a$}%
}}}}
\put(8731,-4021){\makebox(0,0)[b]{\smash{{\SetFigFont{12}{24.0}
{\rmdefault}{\mddefault}{\updefault}{\color[rgb]{0,0,0}$b$}%
}}}}
\put(9586,-3751){\makebox(0,0)[b]{\smash{{\SetFigFont{12}{24.0}
{\rmdefault}{\mddefault}{\updefault}{\color[rgb]{0,0,0}$c$}%
}}}}
\put(9181,-4741){\makebox(0,0)[b]{\smash{{\SetFigFont{12}{24.0}
{\rmdefault}{\mddefault}{\updefault}{\color[rgb]{0,0,0}$remember$}%
}}}}
\put(10441,-4741){\makebox(0,0)[b]{\smash{{\SetFigFont{12}{24.0}
{\rmdefault}{\mddefault}{\updefault}{\color[rgb]{0,0,0}$forget$}%
}}}}
\end{picture}%
\nop{
            .                                    .             
a --+       .                        a --+       .             
    +--> c ---> d <---> e                +--> c  .
b --+       .                        b --+       .             
^           .                 ==>                .             
|           .                                    .             
+-------------- f ---> g                         .
            .                                    .            
  remember     forget                  remember     forget    
}
\caption{An example of forgetting some variables.
Arrows stands for propositional implications.}
\label{shrink}
\end{center}
\end{figure}
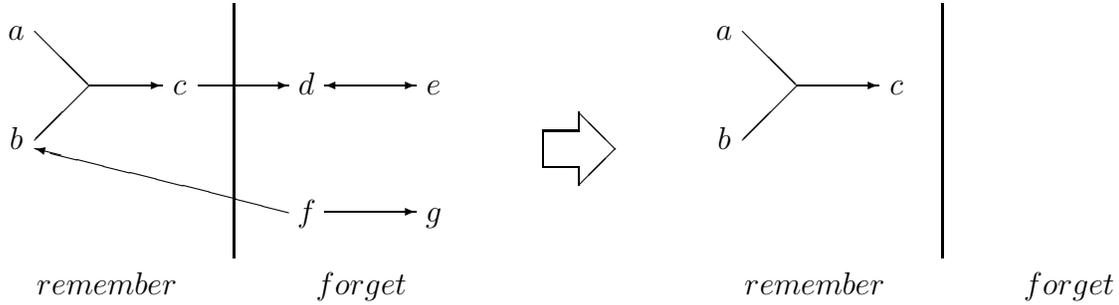

This looks impossible. Forgetting is removing. Removing information, but still
removing. Removing something leaves less, not more. What remains is less than
what before, not more. Forgetting about $d$, $e$, $f$ and $g$ in the formula
depicted on the left of Figure~\ref{shrink} only leaves information about $a$,
$b$ and $c$.

The only information that remains is that $a$ and $b$ imply $c$. All the rest,
like $c$ implying $d$ or $f$ implying $b$ is forgotten. What is left is smaller
than what before because it is only a part of that.

This is the prototypical scenario of forgetting, the first that comes to mind
when thinking about removing information: some information goes away, the rest
remains. The rest is a part of the original. Smaller. Simpler. Easier than that
to store, to query, to interpret. But prototypical does not mean exclusive.

\begin{figure}
\begin{center}
\setlength{\unitlength}{5000sp}%
\begingroup\makeatletter\ifx\SetFigFont\undefined%
\gdef\SetFigFont#1#2#3#4#5{%
  \reset@font\fontsize{#1}{#2pt}%
  \fontfamily{#3}\fontseries{#4}\fontshape{#5}%
  \selectfont}%
\fi\endgroup%
\begin{picture}(2910,2172)(5656,-5080)
\thinlines
{\color[rgb]{0,0,0}\put(5761,-3526){\line( 1,-1){270}}
\put(6031,-3796){\line(-1,-1){270}}
}%
{\color[rgb]{0,0,0}\put(5761,-3796){\vector( 1, 0){720}}
}%
{\color[rgb]{0,0,0}\put(6481,-3706){\vector(-4, 3){720}}
}%
{\color[rgb]{0,0,0}\put(6481,-3886){\vector(-4,-3){720}}
}%
{\color[rgb]{0,0,0}\put(6526,-3931){\vector(-1,-1){765}}
}%
\thicklines
{\color[rgb]{0,0,0}\put(6211,-2941){\line( 0,-1){2070}}
}%
\thinlines
{\color[rgb]{0,0,0}\put(7021,-3751){\line( 0,-1){180}}
\put(7021,-3931){\line( 1, 0){180}}
\put(7201,-3931){\line( 0,-1){ 90}}
\put(7201,-4021){\line( 1, 1){180}}
\put(7381,-3841){\line(-1, 1){180}}
\put(7201,-3661){\line( 0,-1){ 90}}
\put(7201,-3751){\line(-1, 0){180}}
}%
{\color[rgb]{0,0,0}\put(7921,-3526){\line( 1,-1){270}}
\put(8191,-3796){\line(-1,-1){270}}
}%
{\color[rgb]{0,0,0}\put(7921,-3796){\line( 1, 0){270}}
}%
{\color[rgb]{0,0,0}\put(7921,-3481){\line( 1, 0){270}}
\put(8191,-3481){\line(-1,-2){270}}
}%
{\color[rgb]{0,0,0}\put(7921,-3751){\line( 1, 1){270}}
}%
{\color[rgb]{0,0,0}\put(7921,-4111){\line( 1, 0){270}}
\put(8191,-4111){\line(-1, 2){270}}
}%
{\color[rgb]{0,0,0}\put(7921,-3841){\line( 1,-1){270}}
}%
{\color[rgb]{0,0,0}\put(8191,-3481){\line( 0, 1){315}}
\put(8191,-3166){\vector(-1, 0){270}}
}%
{\color[rgb]{0,0,0}\put(8191,-4111){\line( 0,-1){315}}
\put(8191,-4426){\vector(-1, 0){270}}
}%
{\color[rgb]{0,0,0}\put(8191,-3796){\line( 1, 0){ 90}}
\put(8281,-3796){\line( 0,-1){900}}
\put(8281,-4696){\vector(-1, 0){360}}
}%
\thicklines
{\color[rgb]{0,0,0}\put(8371,-2941){\line( 0,-1){2070}}
}%
\put(5671,-3571){\makebox(0,0)[b]{\smash{{\SetFigFont{12}{24.0}
{\rmdefault}{\mddefault}{\updefault}{\color[rgb]{0,0,0}$a$}%
}}}}
\put(5671,-3841){\makebox(0,0)[b]{\smash{{\SetFigFont{12}{24.0}
{\rmdefault}{\mddefault}{\updefault}{\color[rgb]{0,0,0}$b$}%
}}}}
\put(5671,-4111){\makebox(0,0)[b]{\smash{{\SetFigFont{12}{24.0}
{\rmdefault}{\mddefault}{\updefault}{\color[rgb]{0,0,0}$c$}%
}}}}
\put(6571,-3841){\makebox(0,0)[b]{\smash{{\SetFigFont{12}{24.0}
{\rmdefault}{\mddefault}{\updefault}{\color[rgb]{0,0,0}$x$}%
}}}}
\put(5671,-3211){\makebox(0,0)[b]{\smash{{\SetFigFont{12}{24.0}
{\rmdefault}{\mddefault}{\updefault}{\color[rgb]{0,0,0}$n$}%
}}}}
\put(5671,-4741){\makebox(0,0)[b]{\smash{{\SetFigFont{12}{24.0}
{\rmdefault}{\mddefault}{\updefault}{\color[rgb]{0,0,0}$l$}%
}}}}
\put(6031,-5011){\makebox(0,0)[rb]{\smash{{\SetFigFont{12}{24.0}
{\rmdefault}{\mddefault}{\updefault}{\color[rgb]{0,0,0}$remember$}%
}}}}
\put(6391,-5011){\makebox(0,0)[lb]{\smash{{\SetFigFont{12}{24.0}
{\rmdefault}{\mddefault}{\updefault}{\color[rgb]{0,0,0}$forget$}%
}}}}
\put(7831,-3211){\makebox(0,0)[b]{\smash{{\SetFigFont{12}{24.0}
{\rmdefault}{\mddefault}{\updefault}{\color[rgb]{0,0,0}$n$}%
}}}}
\put(7831,-3571){\makebox(0,0)[b]{\smash{{\SetFigFont{12}{24.0}
{\rmdefault}{\mddefault}{\updefault}{\color[rgb]{0,0,0}$a$}%
}}}}
\put(7831,-3841){\makebox(0,0)[b]{\smash{{\SetFigFont{12}{24.0}
{\rmdefault}{\mddefault}{\updefault}{\color[rgb]{0,0,0}$b$}%
}}}}
\put(7831,-4111){\makebox(0,0)[b]{\smash{{\SetFigFont{12}{24.0}
{\rmdefault}{\mddefault}{\updefault}{\color[rgb]{0,0,0}$c$}%
}}}}
\put(7831,-4471){\makebox(0,0)[b]{\smash{{\SetFigFont{12}{24.0}
{\rmdefault}{\mddefault}{\updefault}{\color[rgb]{0,0,0}$m$}%
}}}}
\put(5671,-4471){\makebox(0,0)[b]{\smash{{\SetFigFont{12}{24.0}
{\rmdefault}{\mddefault}{\updefault}{\color[rgb]{0,0,0}$m$}%
}}}}
\put(7831,-4741){\makebox(0,0)[b]{\smash{{\SetFigFont{12}{24.0}
{\rmdefault}{\mddefault}{\updefault}{\color[rgb]{0,0,0}$l$}%
}}}}
\put(8551,-5011){\makebox(0,0)[lb]{\smash{{\SetFigFont{12}{24.0}
{\rmdefault}{\mddefault}{\updefault}{\color[rgb]{0,0,0}$forget$}%
}}}}
\put(8191,-5011){\makebox(0,0)[rb]{\smash{{\SetFigFont{12}{24.0}
{\rmdefault}{\mddefault}{\updefault}{\color[rgb]{0,0,0}$remember$}%
}}}}
\end{picture}%
\nop{
          .                                        .
    n <------+                   n <------+        .
          .  |                            |        .
    a --+ .  |                   a --+----+        .
    b --+--> x ---+              b --+---------+   .
    c --+ .  |    |     ==>      c --+----+    |   .
          .  |    |                       |    |   .
    m <------+    |              m <------+    |   .
          .       |                            |   .
    l <-----------+              l <-----------+   .
          .                                        .
 remember   forget                   remember          forget
}
\caption{A formula that is complicated instead of simplified by forgetting}
\label{enlarge}
\end{center}
\end{figure}
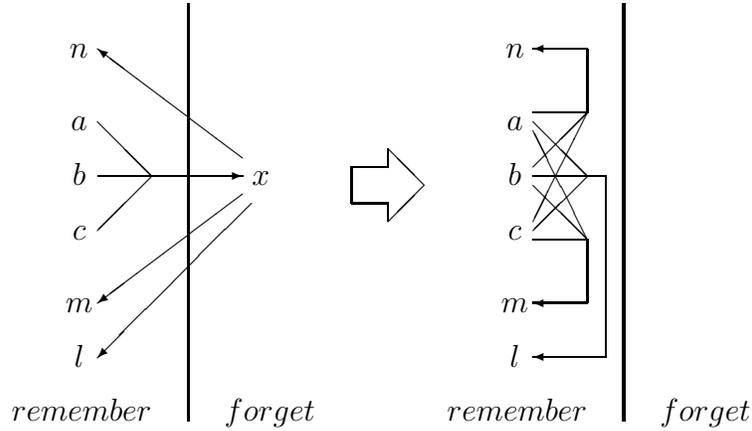

Forgetting $x$ from the formula on the left of Figure~\ref{enlarge} complicates
it instead of simplifying it. Whenever $a$, $b$ and $c$ are the case so is $x$.
And $x$ implies $n$, $m$ and $l$. Like the neck of an hourglass, $x$ funnels
the first three variables in the upper bulb to the last three in the lower.
Without it, these links need to be spelled out one by one:
{} $a$, $b$ and $c$ imply $n$; 
{} $a$, $b$ and $c$ imply $m$; 
{} $a$, $b$ and $c$ imply $l$.
The variable $x$ acts like a shorthand for the first three variables together.
Removing it forces repeating them.

Forgetting $x$ deletes $x$ but not its connections with the other variables.
The lines that go from $a$, $b$ and $c$ to $n$, $l$ and $m$ survive. Like a
ghost, $x$ is no longer there in its body, but in its spirit: its bonds. These
remain, weaved where $x$ was.

% They pass through $x$ like the live pass through the ghost.

% Its haunts the body of knowledge with its these connections.

% What is forgotten is often related to what remains. These bonds weave around
% it, and remain after forgetting. Like ghosts of things no longer there, they
% haunt knowledge by forming complex links in place of multiple single ones.

The formula resulting from forgetting is still quite short, but this is only
because the example is designed to be simple for the sake of clarity. Cases
with larger size increase due to forgetting are easy to find. Forgetting a
single variable never increases size much, but forgetting many may increase
size exponentially.

The size of the formula resulting from forgetting matters for all reasons
cited by the authors above. To summarize, it is important for:

\begin{enumerate}

\item sheer memory needed;

\item the cost of reasoning;
formulae that are difficult for modern solvers are typically large; while
efficiency is not directly related to size, small formulae are usually easy to
solve;

\item interpreting the information;
the size of a formula tells something about how much the forgotten variables
are related to the others.

\end{enumerate}

These points are relevant to different research areas: for example, Delgrande
and Wang~\shortcite{delg-wang-15} mention the second point regarding
disjunctive logic programming; Erdem and Ferraris~\shortcite{erde-ferr-07} do
the same in the context of reasoning about actions. The third point is cited in
the general survey on forgetting by Eiter and
Kern-Isberner~\cite{eite-kern-19} and in the article that generalizes
forgetting across different logics by Delgrande~\cite{delg-17}.

This witnesses that the problem of size after forgetting is relevant to
different logics. Many of them generalize or can express propositional logic or
Horn logic as a subcase. These are the two logics considered here, as greater
common divisors of them.

The figures visualize forgetting as a cut between what is remembered and what
is forgotten. This cut may divide parts that are easy to separate like in the
first figure or parts that are not natural to separate like in the second
figure. The first cut glides following the direction of the fabric of the
knowledge base. The second is resisted by the connections it cuts.

The number of these connections does not tell the difference. How closely they
hold together the parts across the cut does. Even if the links in the first
example were $c \rightarrow d_1, \ldots, c \rightarrow d_{100}$ instead of $c
\rightarrow d$, the result would be the same. What matters is not how many
links connect the parts across the cut, but how they do.

An increase of size gauges the complexity of these connections. The first
example is easy to cut because its implications are easy to ignore: $c$ may
imply $d$, but if $d$ is forgotten this implication is removed and nothing else
changes. The second example is not so easy to cut: forgetting $x$ does not just
remove its implication from $a$, $b$ and $c$; it shifts its burden to the
remaining variables.

A size increase suggests that the forgotten variables are closely connected to
the remaining ones. If forgetting is aimed at subdividing knowledge, it would
be like the chapter on Spain next to that of Samoa and far away from France in
an atlas. The natural division is by continents, not initials of the name. In
general, the natural divisions are by topics, so that things closely connected
stay close to each other. Forgetting about Samoa when describing Spain is
easier than forgetting about France. Neglecting some obscure diplomatic
relations is more natural than neglecting a bordering country.

Forgetting may be abstracting~\cite{mart-etal-15}. Cold weather increases virus
survival, which facilitates virus transmission, which causes flu. Forgetting
about viruses: cold weather causes flu. But forgetting is not always natural as
an abstraction. A low battery level, a bad UPS unit and a black-out cause a
laptop not to start; which causes a report not to be completed, a movie not to
be watched and a game not to be played. Forgetting about the laptop is a
complication more than an abstraction: the three preconditions cause the first
effect, they cause the second effect, and they cause the third effect. If $x$
is the laptop not starting, this is the example in the second figure, where
forgetting increases size. That cold weather causes cold is short, simple, a
basic fact of life for most people. Brevity is the soul of abstraction.

In summary, a short formula is preferred for storage and computational reasons.
The size of the formula after forgetting is important for epistemological
reason, to evaluate how natural a partition or abstraction of knowledge is.
Either way, the question is: how large is a formula after forgetting variables?

The question is not as obvious as it looks. Several formulae represent the same
piece of knowledge. For example, $a \vee (\neg a \wedge b)$ is the same as the
shorter $a \vee b$. The problem of formula size without forgetting eluded
complexity researchers for twenty years: it was the prototypical problem for
which the polynomial hierarchy was created in the seventies~\cite{stoc-76}, but
framing it exactly into one of these classes only succeeded at the end of the
nineties~\cite{uman-01}. This is the problem of whether a formula is equivalent
to another of a given size.

The problem studied in this article is whether forgetting some variables from a
formula is equivalent to a formula of given size.

Forgetting is not complicated. A simple recipe for forgetting $x$ from $F$ is:
replace $x$ with true in $F$, replace $x$ with false in $F$, disjoin the two
resulting formulae~\cite{lang-etal-03}. If $F$ is in conjunctive normal form,
another recipe is: replace all clauses containing $x$ with the result of
resolving them~\cite{wang-15,delg-17,robi-65}. The first solution may not
maintain the syntactic form of the formula. None of them is guaranteed to
produce a minimal one.

Forgetting no variable from a formula results in the formula itself. Insisting
on forgetting something does not change complexity: every formula $F$ is the
result of forgetting $x$ from $F \wedge x$ if $x$ is a variable not in $F$. The
complexity of the size of $F$ is a subcase of the size of forgetting $x$ from
$F$. It is however not an interesting subcase: the question is how much size
decreases or increases due to forgetting. If $F$ has size 100 before forgetting
and 10 after, this looks like a decrease, but is not if $F$ is equivalent to a
formula of size 5 before forgetting and to none of size 9 or less afterwards.
This is a size increase, not a decrease.

% outline of the article

The main results of this article are the complexity characterization of this
problem in the Horn and general propositional case: the problem is \Dp-hard and
belongs to \S{2} when the formula is Horn; it is \Dptwo-hard and in \S{3} for
arbitrary CNF formulae. A detailed plan of the article follows.

After Section~\ref{section-preliminaries} introduces some basic concepts like
resolution, Section~\ref{section-forget} formally defines forgetting and gives
some results about size. Thanks to equivalence, forgetting can be expressed in
several ways: if $F'$ expresses forgetting some variables from a formula, every
formula $F''$ equivalent to $F'$ does. Some equivalent formulations of
forgetting are given, as well as some ways to compute forgetting.

Section~\ref{section-horn} shows the complexity of the problem in the Horn
restriction. It comes before the general case because of its slightly simpler
proofs. The problem is \Dp-hard and belongs to \S{2}.

Section~\ref{section-general} shows that the problem is \Dptwo-hard and belongs
to \S{3} for arbitrary CNF formulae. In both cases, hardness is more difficult
to prove than membership; on the other hand, it extends to logics that include
propositional or Horn logics as subcases. For example, since modal logics
extend propositional logics, the problem of size of forgetting is \Dp{2}-hard.

A number of examples and counterexamples rely on calculating the resolution
closure of a formula, its minimal equivalent formulae, the result of forgetting
a variable from it and the minimal formulae equivalent to that. The program
{\tt minimize.py} does these operations on the formula it reads from another
file, for example
{} {\tt allvariables.py}
or
{} {\tt outresolve.py}.
It is currently available at
{} {\tt https://github.com/paololiberatore/minimize.py}
together with the files that contain the formulae mentioned in this article.

\section{Preliminaries}
\label{section-preliminaries}

\subsection{Formulae}

The formulae in this article are all propositional in conjunctive normal form
(CNF): they are sets of clauses, a clause being the disjunction of some literals
and a literal a propositional variable or its negation. This is not truly a
restriction, as every formula can be turned into CNF without changing its
semantics. A clause is sometimes identified with the set of literals it
contains. For example, a subclause is a subset of a clause.

If $l$ is a negative literal $\neg x$, its negation $\neg l$ is defined as $x$.

The variables a formula $A$ contains are denoted $\var(A)$.

\begin{definition}
\label{size}

The size $||A||$ of a formula $A$ is the number of variable occurrences it
contains.

\end{definition}

This is not the same as the cardinality of $\var(A)$ because a variable may
occur multiple times in a formula. For example, $A = \{a, \neg a \vee b, a \vee
\neg b\}$ has size five because it contains five literal occurrences even if
its variables are only two. The size is obtained by removing from the formula
all propositional operators, commas and parentheses and counting the number of
symbols left.

Other definitions are possible but are not considered in this article. An
alternative measure of size is the total number of symbols a formula contains
(including conjunctions, disjunctions, negations and parenthesis). Another is
the number of clauses (regardless of their length).

The definition of size implies the definition of minimality: a formula is
minimal if it is equivalent to no formula smaller than it. Given a formula, a
minimal equivalent formula is a possibly different but equivalent formula that
is minimal. As an example, $A = \{a, \neg a \vee b, a \vee \neg b\}$ has size
five since it contains five literal occurrences; yet, it is equivalent to $B =
\{a,b\}$, which only contains two literal occurrences. No formula equivalent
to $A$ or $B$ is smaller than that: $B$ is minimal. Minimizing a formula means
obtaining a minimal equivalent formula. This problem has long been
studied~\cite{hema-schn-11,cepe-kuce-08}.

\begin{definition}
\label{clauses-literal}

The clauses of a formula $A$ that contain a literal $l$ are denoted by $A \cap
l = \{c \in A \mid l \in c\}$.

\end{definition}

This notation cannot cause confusion: when is between two sets, the symbol
$\cap$ denotes their intersection; when is between a set and a literal, it
denotes the clauses of the set that contain the literal. This is like seeing $A
\cap l$ as the shortening of $A \cap \mathrm{clauses}(l)$, where
$\mathrm{clauses}(l)$ is the set of all possible clauses that contain the
literal $l$.

\subsection{Resolution}
\label{subsection-resolution}

Resolution is a syntactic derivation mechanism that produces a new clause that
is a consequence of two clauses: $c_1 \vee l, c_2 \vee \neg l \vdash c_1 \vee
c_2$. The result is implicitely removed repetitions. Sometimes $\vdash_R$ is
used in place of $\vdash$ to emphasize the use of resolution as the syntactic
derivation rule. This is unnecessary in this article since no other derivation
rule is ever mentioned.

Unless noted otherwise, tautologic clauses are excluded. Writing
{} $c_1 \vee a, c_2 \vee \neg a \vdash c_1 \vee c_2$
implicitly assumes that none of the three clauses is a tautology unless
explicitly stated. Two clauses that would resolve in a tautology are considered
not to resolve, which is not a limitation~\cite{love-14}.
% page 21: Aid 2. "Ignore tautologies"; applies also when resolution is used
% for deduction, as every clause is entailed by its subclauses
Tautologic clauses are forbidden in formulae, which is not a limitation either
since tautologies are always satisfied. This assumption has normally little
importance, but is crucial to superredundancy, a concept defined in the next
section.

In what follows tautologies are excluded from formulae and from resolution
derivations. As a result, resolving two clauses always generates a clause
different from them.

A resolution proof $F \vdash G$ is a binary forest where the roots are the
clauses of $G$, the leaves the clauses of $F$ and every parent is the result of
resolving its two children.

% The alternative form of direct acyclic graphs can be converted to this by
% duplicating subgraphs. The increased number of nodes does not constitute a
% problem because resolution is not used for automated theorem proving in this
% article but as a theoretical tool.

\begin{definition}

The resolution closure of a formula $F$ is the set $\rescn(F) = \{c \mid F
\vdash c\}$ of all clauses that result from applying resolution zero or more
times from $F$.

\end{definition}

The clauses of $F$ are derivable by zero-step resolutions from $F$. Therefore,
$F \vdash c$ and $c \in \rescn(F)$ hold for every $c \in F$.

The resolution closure is similar to the deductive closure but not identical.
For example, $a \vee b \vee c$ is in the deductive closure of $F = \{a \vee
b\}$ but not in the resolution closure. It is a consequence of $F$ but is not
obtained by resolving clauses of $F$.

All clauses in the resolution closure $\rescn(F)$ are in the deductive closure
but not the other way around. The closures differ because resolution does not
expand clauses: $a \vee b \vee c$ is not a resolution consequence of $a \vee
b$. Adding expansion kills the difference~\cite{lee-67,slag-69}.
% also: inou-91,matu-etal-17

\[
F \models c \mbox{ if and only if }
c' \in \rescn(F) \mbox{ for some } c' \subseteq c
\]

That resolution does not include expansion may suggest that it cannot generate
any non-minimal clause. That would be too good to be true, since a clause would
be minimal just because it is obtained by resolution. In fact, it is not the
case. Expansion is only one of the reasons clauses may not be minimal, as seen
in the formula
{} $\{a \vee b \vee c, a \vee b \vee e, \neg e \vee c \vee d\}$:
the second and third clauses resolve to
{} $a \vee c \vee b \vee d$,
which is however not minimal: it contains the first clause of the formula,
{} $a \vee b \vee c$.
% This example is in the {\tt resolutionnotminimal.py} file of {\tt minimize.py}.

What is the case is that resolution generates all prime
implicates~\cite{lee-67,slag-69}, the minimally entailed clauses. The relation
between $\rescn(F)$ and the deductive closure of $F$ tells that if a clause is
entailed, a subset of it is generated by resolution; since the only entailed
subclause of a prime implicate is itself, it is the only one resolution may
generate. Removing all clauses that contain others from $\rescn(F)$ results in
the set of the prime implicates of $F$.

While $\rescn(F)$ contains all clauses generated by an arbitrary number of
resolutions, some properties used in the following require the clauses obtained
by a single resolution step.

\begin{definition}
\label{resolve-function}

The resolution of two formulae is the set of clauses obtained by resolving each
clause of the first formula with each clause of the second:

\[
\resolve(A,B) = \{c \mid c',c'' \vdash c 
\mbox{ where } c' \in A \mbox{ and } c'' \in B\}
\]

If either of the two formulae comprises a single clause, the abbreviations
{} $\resolve(A,c) = \resolve(A, \{c\})$,
{} $\resolve(c,B) = \resolve(\{c\}, B)$ and
{} $\resolve(c,c') = \resolve(\{c\}, \{c'\})$
are used.

\end{definition}

This set contains only the clauses that results from resolving a single clause
of $A$ with a single clause of $B$. Exactly one resolution of one clause with
one clause. Not zero, not multiple ones. A clause of $A$ is not by itself in
$\resolve(A,B)$ unless it is also the resolvent of another clause of $A$ with a
clause of $B$.

A clause of a formula is superredundant if it is redundant in the resolution
closure of the formula~\cite{libe-22}: $\rescn(F) \backslash \{c\} \models c$.
The following properties of superredundancy and superirredundancy are used in
this article.

\begin{lemma}[\cite{libe-22}]
\label{minimal}

If a formula contains only superirredundant clauses, it is minimal.

\end{lemma}

\begin{lemma}[\cite{libe-22}]
\label{no-resolution}

If no two clauses of $F$ resolve,
then
a clause of $F$ is superredundant
if and only if
$F$ contains a clause that is a strict subset of it.

\end{lemma}

\begin{lemma}[\cite{libe-22}]
\label{superset}

If a clause $c$ of $F$ is superredundant,
it is also superredundant in $F \cup \{c'\}$.

\end{lemma}

\begin{lemma}[\cite{libe-22}]
\label{set-value}

A clause $c$ of $F[\true/x]$ is superredundant
if it is superredundant in $F$, 
{} it contains neither $x$ nor $\neg x$
{} and $F$ does not contain $c \vee \neg x$.
The same holds for $F[\false/x]$ if $F$ does not contain $c \vee x$.

\end{lemma}

\section{Forgetting}
\label{section-forget}

Forgetting is defined semantically: no specific formula is given as the unique
result of forgetting variables from another. Several equivalent formulae may
express forgetting the same variables from the same formula.

\begin{definition}
\label{express-forget}

A formula $B$ expresses forgetting all variables from $A$ except $Y$ if and
only if $\var(B) \subseteq Y$ and $B \models C$ is the same as $A \models C$
for all formulae $C$ such that $\var(C) \subseteq Y$.

\end{definition}

The definition sets a constraint over $B$ rather than uniquely defining a
specific formula. Every formula $B$ fits it as long as it is built over the
right variables and has the right consequences.

Syntax is irrelevant to this definition. As it should: every $B'$ that is
syntactically different but equivalent to $B$ carries the same information.
There is no reason to confer $A[\true/x] \vee A[\false/x]$ a special status
among all formulae holding the same information. Every formula equivalent to
it, every formula that entails the same formulae is an equally valid result of
forgetting.

The definition captures this parity among formulae by not defining forgetting
as a single specific formula and then delegating the definition of its
alternatives to equivalence. If $B$ expresses forgetting some variables from
$A$ and $B'$ is equivalent to $B$ and contains the same variables, then $B'$
also expresses forgetting the same variables from $A$. This is because
equivalence implies equality of consequences.

The following Section~\ref{size-forget} discusses the main focus of the
analysis of this article: the size of a formula when variables are forgotten
from it; it is followed by the equivalent ways of defining forgetting in
Section~\ref{equivalent-forget}; the subsequent Section~\ref{howto-forget}
shows how to actually compute forgetting in general and in two specific cases;
finally, Section~\ref{necessary-forget} proves that in some cases, certain
literals are always in the result of forgetting, which is important when
computing the size after forgetting.

\subsection{Size of forgetting}
\label{size-forget}

Many formulae $B$ express forgetting the same variables $X$ from a formula $A$.
Some may be large and some may be small. Producing an artificially large
formula is straightforward: if
{} $\{a \vee b, b \vee c\}$
expresses forgetting, also
{} $\{a \vee b, b \vee c, \neg a \vee a, a \vee b \vee \neg c\}$ 
does: adding tautologies and consequences does not change the semantics of a
formula. The question is not whether a large expression of forgetting exists.

The question is whether a small expression of forgetting exists. In this
context, ``small'' means ``of polynomial size''. Technically: given a formula
$A$ and a set of variables $X$, does any formula of size polynomial in that of
$A$ express forgetting $X$ from $A$?

Forgetting each variable $x$ from the CNF formula $A$ is expressed by
$A[\true/x] \vee A[\false/x]$, which can be converted back into a CNF of
quadratic size. Forgetting many variables this way produces an exponentially
large formula. Yet, this formula may be equivalent to a short one.

This is not the case for all formulae. That would have unlikely consequences on
the complexity hierarchy~\cite{lang-etal-03}. Yet, it is the case for some
formulae. It depends on the formula. For example, forgetting variables from
negation-free CNF formulae amounts to removing the clauses that contain these
variables. The question is whether expressing forgetting can be done in small
space for a specific formula. This will be proved \Dp-hard and in \S{2} for
Horn formulae, and \Dptwo-hard and in \S{3} for unrestricted CNF formulae.

The existing literature provides mechanisms for forgetting variables from a
formula and results about the minimal size of expressing forgetting for all
formulae. They leave open the question in between: the minimal size of
expressing forgetting for a formula.
An example result of the first kind is: ``Salient features of the solution
provided include linear time complexity, and linear size of the output of
(iterated) forgetting''~\cite{anto-etal-12}: given a formula, the size of
forgetting is linear.
An example result of the second kind is: ``the size of the result of forgetting
may be exponentially large in the size of the input program''~\cite{delg-17}:
forgetting may produce exponentially large formulae when considering all
possible input formulae. It may, not must. For some formulae, forgetting may
not increase size. The only hardness result about this problem has been
published by Zhou~\cite{zhou-14}; it is discussed in a following section. Other
authors reported worst-case results~\cite{erde-ferr-07,eite-wang-06}, and some
the opposite, as certain forgetting mechanisms of certain logics can be
expressed in polynomial size~\cite{gonc-etal-16}.

Forgetting propositional variables is also called variable elimination,
especially in the context of automated reasoning~\cite{e.en-bier-05}: it is a
way to simplify a formula before processing. As such, it has stricter
efficiency requirements than general forgetting. For example, the NiVER
preprocessor "resolves away a variable only if there will be no increase in
space"~\cite{subb-prad-04}. A quadratic increase would be too much, given the
aim of reducing the overall runtime of automated reasoning.

Forgetting is often identified by its dual concept of uniform interpolation,
especially in first-order, modal and description logics~\cite{bilk-07}. While
forgetting is always expressible in exponential space in propositional logics,
uniform interpolants in other logics may be larger, if they exist at all. For
example, their size is at least triple-exponential in certain description
logics, provided that they exist~\cite{niki-rudo-14}. Analogous to the question
of checking their size is checking their existence~\cite{arta-etal-20}.

\subsection{Equivalent conditions}
\label{equivalent-forget}

Forgetting could be based on consistency rather than inference. Or it could be
based on models. It would still be the same, at least in propositional logic
because of the way inference, consistency and models are related. For example,
since mutual consistency is the exact opposite of inference, forget is
equivalently defined in terms of equisatisfiability.

\begin{lemma}
\label{consistent}

A formula $B$ over the variables $Y$ expresses forgetting all variables from
$A$ but $Y$ if and only if $A \wedge D$ is equisatisfiable with $B \wedge D$
for all formulae $D$ over variables $Y$.

\end{lemma}

\proof The definition of $B$ expressing forgetting is that it is built over the
variables $Y$ and that $A \models C$ is the same as $B \models C$ for every
formula $C$ on the alphabet $Y$. The two entailments are respectively the same
as the inconsistency of $A \wedge \neg C$ and $B \wedge \neg C$. They coincide
if and only if $A \wedge D$ and $B \wedge D$ are equisatisfiable, where $D =
\neg C$. In the other way around, $A \wedge D$ and $B \wedge D$ are
equisatisfiable if and only if $A \wedge \neg C$ and $B \wedge \neg C$ are,
where $C = \neg D$.~\qed

The condition mostly used in this article is the restriction of mutual
consistency to sets of literals.

% Every formula is equivalent to a DNF over the same variables, and also to a DNF
% where every term contains all variables of the formula. The consistency of $A
% \wedge D$ is the same as the consistency of $A$ with every term, and the same
% for $B$. The following two theorems follow.

\begin{theorem}
\label{consistent-literals}

A formula $B$ over the variables $Y$ expresses forgetting all variables except
$Y$ from $A$ if and only if $S \cup A$ is equisatisfiable with $S \cup B$ for
all sets of literals $S$ over variables $Y$.

\end{theorem}

% A further restriction is that the sets of literals can be constrained to
% contain all variables not to be forgotten. Such sets have exactly one model
% over these variables, proving that forget does not change if defined in terms
% of models instead of inference.

\begin{theorem}
\label{consistent-literals-complete}

A formula $B$ over the variables $Y$ expresses forgetting all variables except
$Y$ from $A$ if and only if $S \cup A$ is equisatisfiable with $S \cup B$ for
all sets of literals $S$ over variables $Y$ that mention all variables in $Y$.

\end{theorem}

\subsection{How to forget}
\label{howto-forget}

Three properties related to computing forgetting are proved: it can be
performed one variable at time, it can be performed by resolution, and it may
be performed on the independent parts of the formula, if any.

The first property is an almost direct consequence of the definition.

% delg-17, theorem 1, point 6

\begin{lemma}~\cite{delg-17}
\label{transitive}

If $B$ expresses forgetting the variables $Y$ from $A$ and $C$ expresses
forgetting the variables $Z$ from $B$, then $C$ expresses forgetting $Y \cup Z$
from $A$.

\end{lemma}

\proof Since $B$ expresses forgetting $Y$ from $A$, it is build over the
variables $\var(A) \backslash Y$. Since $C$ expresses forgetting $Z$ from $B$
it is build over the variables $\var(B) \backslash Z$, which is a subset of
$\var(A) \backslash Y \backslash Z = \var(A) \backslash (Y \cup Z)$. This is
the first condition for $C$ expressing forgetting $Y \cup Z$ from $A$.

Let $D$ be a formula over the variables $\var(A) \backslash (Y \cup Z)$. This
alphabet is also $\var(A) \backslash Y \backslash Z$, which emphasizes that the
variables of $D$ are a subset of $\var(A) \backslash Y$. The assumption that
$B$ expresses forgetting $Y$ from $A$ implies that $A \models D$ is equivalent
to $B \models D$ since the variables of $D$ are all in $\var(A) \backslash Y$.
The assumption that $C$ expresses forgetting $Z$ from $B$ implies that $B
\models D$ is the same as $C \models D$ since $\var(B) = \var(A) \backslash Y$
and the alphabet of $D$ is $\var(A) \backslash Y \backslash Z = \var(B)
\backslash Z$. The equivalence of $A \models D$ with $B \models D$ and the
equivalence of $B \models D$ with $C \models D$ implies that $A \models D$ is
the same as $C \models D$. This holds for every formula $D$ over the variables
of $\var(A) \backslash (Y \cup Z)$, and defines $C$ expressing forgetting all
variables of $Y \cup Z$ from $A$.~\qed

Forgetting is expressed by $A[\true/x] \vee A[\false/x]$, but this formula does
not maintain the syntactic form of $A$: a CNF like $a \wedge (x \vee b \vee c)$
becomes the non-CNF $((a) \vee (a \wedge (b \vee c))$. While this formula can
be turned into CNF, directly combining clauses is more convenient when working
on CNFs.

Forgetting can be performed by resolution, as proved by Delgrande and
Wassermann~\shortcite{delg-wass-13} in the Horn case and extended to the
general case by Wang~\cite{wang-15} and Delgrande~\shortcite{delg-17}. The
function $\resolve(A,B)$ provided by Definition~\ref{resolve-function} gives
the clauses obtained by resolving each clause of $A$ with each clause of $B$,
if they resolve. The notation $A \cap l$ introduced in
Definition~\ref{clauses-literal} gives the clauses of $A$ that contain the
literal $l$.

\begin{theorem}[{\cite[Theorem~6]{wang-15},\cite[Theorem~6]{delg-17}}]
\label{resolve-out}

The formula
{} $A \backslash (A \cap x) \backslash (A \cap \neg x) \cup
{}  \resolve(A \cap x, A \cap \neg x)$
expresses forgetting $x$ from $A$.

\end{theorem}

Forgetting a single variable is not a limitation because Lemma~\ref{transitive}
tells that forgetting a set of variables can be performed one variable at time:
forgetting $x$ first and $Y \backslash \{x\}$ then is the same as forgetting
$Y$.

The problem is that forgetting this way may produce non-minimal formulae even
from minimal ones. For example,
{} $A = \{a \vee b \vee x, \neg x \vee c, a \vee c\}$
is minimal, but resolving $x$ out to forget it produces
{} $\{a \vee b \vee c, a \vee c\}$,
which is not minimal since the first clause is entailed by the second. The
proof that $A$ is minimal is long and tedious, and is therefore omitted. The
formulae in the {\tt outresolve.py} file of {\tt minimize.py} show similar
examples where the formula obtained by resolving out a variable either contains
a redundant literal or is irredundant although not minimal.

\iffalse

That $A$ is minimal can also be proved with superredundancy, as shown in
section ``minimizing after resolution'' of other.tex. This is not done here
because superredundancy is introduced later.

The other formulae in {\tt outresolve.py} are also described in section
``minimizing after resolution'' of other.tex.

\fi

Since resolving Horn clauses produces Horn clauses, this theorem indirectly
shows that forgetting variables from Horn formulae is expressed by a Horn
formula~\cite{delg-wass-13}.
% Corollary 3 in delg-wass-13
That formula may not be minimal, yet its minimal equivalent formulae cannot be
non-Horn: as mentioned in Section~\ref{subsection-resolution}, resolution
derives all clauses of all minimal equivalent formulae.

% mentioned in section "preliminaries", subsection "resolution":
%
% "Since resolution allows deriving all prime implicates of a
% formula~\cite{lee-67,slag-69}, it derives all clauses of all minimal
% equivalent formulae."

When a formula comprises two independent parts with no shared variable,
forgetting from the formula is the same as forgetting from the two parts
separately. This property is used in the following hardness proofs that merge
two polynomial-time reductions.

\begin{lemma}~\cite{darw-99,lang-etal-03}
\label{independent}

Let $A$ and $B$ be two formulae built over disjoint alphabets:
$\var(A) \cap \var(B) = \emptyset$.
A formula $C$ expresses forgetting the variables $Y$ from $A$ and $D$ expresses
forgetting the variables $Y$ from $B$ if and only if $C \cup D$ expresses
forgetting the variables $Y$ from $A \cup B$.

\end{lemma}

% lang-etal-03 attributes this result to darw-99, which does not mention it;
% probably the result is in the technical report that is the full version of
% this conference article, but is not accessible

\proof The claim is first proved when $Y$ comprises a single variable $x$. By
Theorem~\ref{resolve-out}, forgetting $x$ from $A \cup B$ is expressed by the
following formula.

\[
(A \cup B) \backslash
((A \cup B) \cap x) \backslash
((A \cup B) \cap \neg x) \cup
\resolve((A \cup B) \cap x) \cup ((A \cup B) \cap \neg x))
\]

By definition $(A \cup B) \cap x$ is the set of clauses of $A \cup B$ that
contain $x$. Therefore, it is the union of such clauses of $A$ and of $B$.
In formulae, $(A \cup B) \cap x = (A \cap x) \cup (B \cap x)$. Applying this
equality to the formula expressing forgetting turns into the following.

\[
(A \cup B) \backslash
((A \cap x) \cup (B \cap x)) \backslash
((A \cap \neg x) \cup (B \cap \neg x)) \cup
\resolve((A \cap x) \cup (B \cap x)) \cup
               ((A \cap \neg x) \cup (B \cap \neg x))
\]

Since the variables of $A$ and $B$ are disjoint by assumption, the variable $x$
either is in $A$ or in $B$, but not in both. Only the first case is considered:
$x$ is in $A$ and not in $B$; the other case is analogous. Since $x$ does not
occur in $B$, this formula contains no clause including $x$ and none including
$\neg x$. In formulae, $B \cap x = \emptyset$ and $B \cap \neg x = \emptyset$.
These identities simplify the formula expressing forgetting to the following.

\[
(A \cup B) \backslash
(A \cap x) \backslash
(A \cap \neg x) \cup
\resolve((A \cap x) \cup (A \cap \neg x))
\]

Since $A \cap x$ and $A \cap \neg x$ only contain clauses of $A$ and none of
$B$, the set differences can be applied to $A$ only.

\[
B \cup A \backslash
(A \cap x) \backslash
(A \cap \neg x) \cup
\resolve((A \cap x) \cup (A \cap \neg x))
\]

This is the union of $B$ and
{} $A \backslash (A \cap x) \backslash (A \cap \neg x) \cup
{}  \resolve((A \cap x) \cup (A \cap \neg x))$.
The second part expresses forgetting $x$ from $A$ by Theorem~\ref{resolve-out}.
Since $x$ is not in $B$, this formula $B$ it expresses forgetting $x$ from $B$.
This proves that forgetting a single variable $x$ from $A \cup B$ is expressed
by the union of formula expressing forgetting it from $A$ and from $B$.

\

The claim is proved by iterating over the variables to forget. Let $A$ and $B$
be two formulae built over disjoint alphabets and $Y$ the variables to forget.
If $Y$ comprises a single variable, the claim that forgetting can be done
separately from $A$ and from $B$ is proved above. Otherwise,
Lemma~\ref{transitive} proves that forgetting $Y$ is the same as forgetting an
arbitrary variable $x$ of $Y$ first and then $Y \backslash \{x\}$. What proved
above is that forgetting $x$ can be done by forgetting it separately from $A$
and $B$. This is a proof by induction over the number of the variables to
forget, because it follows on $Y$ from being proven on $Y \backslash
\{x\}$.~\qed

\subsection{Necessary literals}
\label{necessary-forget}

Finding a minimal version of a formula is
difficult~\cite{mccl-56,theo-etal-96,fivs-hlav-12}. Finding a minimal formula
expressing forgetting is further complicated by the addition of forgetting.
Determining the exact complexity of this problem proved difficult; not so much
for membership to classes in the polynomial hierarchy but for hardness.
Fortunately, proving hardness does not require finding the minimal size of
arbitrary formulae, just for the formulae that are targets of the reduction.
\np-hardness is for example proved by translating a formula (to be checked for
satisfiability) to another formula and a set of variables (where the variables
have to be forgotten from the formula). Such a reduction does not generate all
possible formulae. Only for the ones generated by the reduction, the minimal
size after forgetting is necessary.

This is good news, because reductions do not generate all possible formulae.
Rather the opposite: they usually produce formulae of a very specific form.
Still better, a reduction can be altered to simplify computing the minimal size
of the formulae it produces. If the minimal size is difficult to assess for the
formulae produced by a reduction, the reduction itself can be changed to
simplify them.

The reductions used in this article rely on two tricks to allow for simple
proofs. The first is that some clauses of the formulae they generate are in all
minimal-size equivalent formulae; this part of the minimal size is therefore
always the same. The second is that the rest of the minimal size depends on the
presence or absence of certain literals in all formulae expressing forgetting;
where these literals occur if present does not matter, only whether they are
present or not.

The first trick is based on superredundancy~\cite{libe-22}, defined in
Section~\ref{section-preliminaries}.

The second requires proving that a literal is contained in all formulae that
express forgetting. This is preliminarily proved when no forgetting is
involved. When the next lemmas say ``$A$ contains $l$'' they mean: $A$ contains
a clause that contains the literal $l$.

% the lemma also holds on NNF formulae: an NNF formula that does not contain l
% cannot be falsified by making l false: all its literals maintain its value or
% become true (the latter only for -l), and the formula does not contain any
% negation once literals are replaced by their value
%
% the consequences of this lemma on the size of formulae apply to every
% formula: if its premises are met, the formula contains an occurrence of the
% variable of l that becomes l when the formula is turned into NNF; this
% translation does not change the occurrences of the variables, and their
% number is the size of the formula
%
% the lemma does not require S to contain the variable l

\begin{lemma}
\label{necessary-literal}

If $S$ is a set of literals such that $S \cup A$ is consistent, but $S
\backslash \{l\} \cup \{\neg l\} \cup A$ is not, the CNF formula $A$ contains
a clause that contains $l$.

\end{lemma}

\proof Since $S \cup A$ is consistent, it has a model $M$.

The claim is that $A$ contains a clause that contains $l$. This is proved by
contradiction, assuming that no clause of $A$ contains $l$. By construction $S
\backslash \{l\}$ does not contain $l$ either. As a result,
{} $A' = S \backslash \{l\} \cup A$
does not contain $l$. It is still satisfied by $M$ because $M$ satisfies its
superset $S \cup A$. Let $M'$ be the model that sets $l$ to $\false$ and all
other variables the same as $M$. Let $l_1 \vee \cdots \vee l_m$ be an arbitrary
clause of $A'$. Since $M$ satisfies $A$, it satisfies at least one of these
literals $l_i$. Since $A$ does not contain $l$, this literal $l_i$ is either
$\neg l$ or a literal over a different variable. In the first case $M'$
satisfies $l_i = \neg l$ because it sets $l$ to false; in the second because it
sets $l_i$ the same as $M$, which satisfies $l_i$. This happens for all clauses
of $A'$, proving that $M'$ satisfies $A'$.

Since $M'$ also satisfies $\neg l$ because it sets $l$ to false, it satisfies
{} $A' \cup \{\neg l\} = S \backslash \{l\} \cup \{\neg l\} \cup A$,
contrary to its assumed unsatisfiability.~\qed

Since consistency with $S$ and with $S \backslash \{l\} \cup \{\neg l\}$ are
unaffected by syntactic changes, they are the same for all formulae equivalent
to $A$. In other words, if the conditions of the lemma hold for $A$ they also
hold for every formula equivalent to $A$.

This property carries over to formulae expressing forgetting by constraining
$S$ to only contain variables not to be forgotten.

% the lemma requires the variable of l to be in Y, but does not require l to be
% in S

\begin{lemma}
\label{forget-contains}

If $S \cup \{l\}$ is a set of literals over the variables $Y$ such that $S \cup
A$ is consistent, but $S \backslash \{l\} \cup \{\neg l\} \cup A$ is not, every
CNF formula that expresses forgetting all variables except $Y$ from $A$
contains a clause that contains $l$.

\end{lemma}

\proof Let $B$ be a formula expressing forgetting all variables from $A$ but
$Y$. By Theorem~\ref{consistent-literals}, since $S$ is a set of literals
over $Y$, the consistency of $S \cup A$ equates that of $S \cup B$. The same
holds for $S \backslash \{l\} \cup \{\neg l\}$ since its variables are all in
$Y$.

The lemma assumes the consistency of $S \cup A$ and the inconsistency of $S
\backslash \{l\} \cup \{\neg l\} \cup A$. They imply the consistency of $S \cup
B$ and the inconsistency of $S \backslash \{l\} \cup \{\neg l\} \cup B$. These
two conditions imply that $B$ contains $l$ by
Lemma~\ref{necessary-literal}.~\qed

How is this lemma used? To prove that reductions from a problem to the problem
of minimal size of forgetting work. Not all reductions can be proved correct
this way. The ones used in this article are built to allow that. They generate
a formula that contains a certain literal $l$ that may or may not meet the
condition of the lemma. Depending on this, $l$ may or may not be necessary
after forgetting. This is a $+0$ or a $+1$ in the size of the minimal formulae
expressing forgetting. If the other literals occurrences are $k$, the minimal
size is $k+0$ or $k+1$ depending on whether the conditions of
Lemma~\ref{forget-contains} are met.

In order for this to work, the $+0 / +1$ separation is not enough. Equally
important to the $k+0$ vs. $k+1$ size is that the other addend $k$ stays the
same. This is the number of the other literal occurrences. The formulae
produced by the reduction may or may not contain a literal $l$, but this is
useless if the rest of the formula changes. For example, if $k$ changes from
$10$ to $9$ the total size is either $10+0$ or $9+1$, which are the same.
Lemma~\ref{forget-contains} concerns the presence of $l$ in a formula, but this
tells its overall size only when the rest of the formula has a fixed form. This
is ensured by superirredundancy~\cite{libe-22}, defined in
Section~\ref{section-preliminaries}.

%input{superredundancy.tex}
\section{Size after forgetting, Horn case}
\label{section-horn}

How much forgetting variables increases or decreases size? Given a formula $A$
and a set of variables $Y$, how much space forgetting $Y$ from $A$ takes?
Technically, how large is a formula expressing forgetting $Y$ from $A$? A
complexity analysis of a decision problem requires turning it into a yes/no
question. Given $k$, $A$ and $Y$, does a formula $B$ of size bounded by $k$
express forgetting?

This is a decision problem: each of its instances comprises a number $k$, a
formula $A$ and a set of variables $Y$; the solution is yes or no. Yet, it may
not always capture the question of interest. For example, $A$ may be a formula
of size 100 that can be reduced to size 20 by forgetting the variables $Y$.
This looks like a good result: the resulting formula takes much less space to
be stored, checking what can be inferred from it is usually easier, and its
literals are probably related in some simple way. Yet, all of this may be
illusory: formula $A$ has size 100, but only because it is extremely redundant;
it could be reduced to size 10 just by rearrangements, without forgetting
anything. That forgetting can be expressed in size 20 no longer looks good. It
is not even a size decrease, it is a size doubling.

If forgetting was required independent on size, and checking size is a side
question, the problem still makes sense: is forgetting $A$ expressed by a
formula of size $20$? If forgetting is done for size reasons, or for reasons
that depend on size, the problem is not this but rather ``does forgetting
reduce size?'' or ``how much forgetting increases or decreases size?'' These
questions depend on the original size of the formula. The answer is not ``20''.
It is rather ``forgetting increases size from 10 to 20''. It is certainly not
``forgetting decreases size from 100 to 20'', since the formula can be shrunk
more without forgetting.

The solution is to disallow formulae of size 100 that can be reduced to 10
without forgetting. If a formula has size 100, it really has size 100. It is
not the inflated version of a formula of size 10. This way, if size can be
reduced from 100 to 20 when forgetting, this reduction is only due to
forgetting, not to the original formula being larger than necessary.

The following lemmas and theorems include this assumption that the formula is
minimal in size. For example, the problem of checking the size after forgetting
is proved hard for the complexity class \Dptwo\  when the formula is minimal.
It is also proved to be in the class \S{2}. The proof of the latter also holds
when the formula is not minimal: it holds in both cases.

In other words, the problem belongs to \S{2} even if the formula is not
minimal. Without the assumption of minimality, it is already known to be
\S{2}-hard~\cite{zhou-14}. This seems to better characterize complexity than
\Dptwo-hard, but there is a caveat. The \S{2}-hardness proof only holds
allowing the troublesome formulae, the ones that can be reduced size without
forgetting. Actually, it relies on them. It hinges on them. The reduction used
in the \S{2}-hardness proof turns a formula $F$ into a formula such that
forgetting a certain variable from it produces $F \wedge G$, where $G$ is a
formula of size $4$. This specific conjunction $F \wedge G$ can be represented
in space $k$ if and only if $F$ can be represented in space $k-4$. Since the
problem of whether a formula can be represented in a given space is \S{2}-hard,
the proof is technically correct. Yet, it proves the problem hard only in a
specific case where the size after forgetting is the same as the size before
forgetting (plus 4). This is not the question whether forgetting increases or
decreases the size of a formula, it is whether the formula itself can be
reduced size. It is like saying ``this sweater is very warm, I tried it at noon
during my trip to Tahiti''. No surprise it feels warm, it was already warm at
noon in Tahiti. The question is whether it would be warm in Yakutsk as well.

Not assuming minimality is not a detail, is crucial to the \S{2}-hardness
proof. The result of the reduction trivalizes for minimal formulae, as a
minimal formula can always be represented in its size and not less. The
assumption of minimality in the hardness proofs of the present article is a
technical mean to ensure that the size after forgetting is due to forgetting,
not to the original formula being representable in less space than it currently
is.

\

Checking whether forgetting $Y$ from $A$ can be expressed in space $k$ is easy
to be proved in \S{2} if $k$ is unary or polynomially bounded by the size of
$A$: all it takes is checking all formulae of size $k$ for their
equisatisfiability with all sets of literals $S$ over the remaining variables
by Theorem~\ref{consistent-literals}. Hardness is not so easy to prove, and in
fact leaves a gap to membership: it is only proved \Dp-hard in this article.

Not that \Dp-hardness is easy to prove. It requires two long lemmas, one for a
\conp-hardness reduction and one for an \np-hardness reduction. These
reductions have a form that allows them to be merged into a single \Dp-hardness
reduction.

A generic \np-hardness reduction is ``if $F$ is satisfiable then forgetting
takes space less than or equal to $k$ and greater otherwise''. Such reductions
cannot be merged. An additional property is required: forgetting can never be
expressed in size less than $k$. If this is also a property of a \conp-hardness
reduction where the size bound is $l$, the overall size is always $k+l$ or
greater, with $k+l$ being only possible when the first formula is satisfiable
and the second unsatisfiable.

This explains why the lemmas are formulated with ``equal to $k$'' in one case
and ``greater than $k$'' in the other. Their other peculiarity, that the
formula generated by the reduction is required to be minimal, is due to the
reasons explained before.

% the minimality of A is proved by superirredundancy; therefore, no CNF formula
% shorter than A is equivalent to it; minimality is over all CNF formulae, not
% just Horn formulae; but is not among arbitrary formulae

% the size of formulae expressing forgetting is proved by the presence of
% literals: all NNF formulae expressing forgetting contain -xi, -ni, a and -b;
% if F is unsatisfiable they also also contain -a and b; every formula can be
% turned into NNF without changing its occurrences of variables; therefore,
% minimality of size is among all formulae, not just CNFs

% at the same time, existence of a formula of size k in one case is proved in
% the Horn case; therefore, hardness holds regardless of whether the formula
% expressing forgetting is limited to be Horn or CNF or not

\begin{lemma}
\label{horn-conp}

There exists a polynomial algorithm that turns a CNF formula $F$ into a
minimal-size Horn formula $A$, a subset $X_C \subseteq \var(A)$ and a number
$k$ such that forgetting all variables except $X_C$ from $A$ is expressed by a
Horn formula of size $k$ if $F$ is unsatisfiable and only by Horn formulae of
size greater than or equal to $k+2$ if $F$ is satisfiable.

\end{lemma}

\proof Let $F = \{f_1,\ldots,f_m\}$ be a CNF formula built over the alphabet $X
= \{x_1,\ldots,x_n\}$. The reduction employs the fresh variables
{} $E = \{e_1,\ldots,e_n\}$,
{} $T = \{t_1,\ldots,t_n\}$,
{} $C = \{c_1,\ldots,c_m\}$ and
{} $\{a,b\}$.
The formula $A$, the set of variables $X_C$ and the number $k$ are:

\begin{eqnarray*}
A &=&
	\{\neg x_i \vee \neg e_i,
	  \neg x_i \vee t_i,
	  \neg e_i \vee t_i \mid x_i \in X \} \cup			\\
&&
	\{\neg x_i \vee c_j \mid      x_i \in f_j\} \cup
	\{\neg e_i \vee c_j \mid \neg x_i \in f_j\} \cup		\\
&&
	\{\neg t_1 \vee \cdots \vee \neg t_n \vee
	\neg c_1 \vee \cdots \vee \neg c_m \vee
	\neg a \vee b\} \cup						\\
&&
	\{a \vee \neg b\}						\\
X_C &=& X \cup E \cup \{a,b\}						\\
k &=& 2 \times n + 2
\end{eqnarray*}

Before formally proving the claim, how the reduction works is summarized. Some
literals are still necessary after forgetting, and some of them are necessary
only if $F$ is satisfiable. The clauses $\neg x_i \vee \neg e_i$ make $\neg
x_i$ and $\neg e_i$ necessary. The clause $a \vee \neg b$ makes $a$ and $\neg
b$ necessary. If $F$ is always false, then for every value of the variables $X
\cup E$ either some $t_i$ can be set to false (if $x_i = e_i = \false$) or some
$c_j$ can be set to false (because $e_i$ is the negation of $x_i$, and at least
a clause of $F$ is false). This makes the clause
{} $\neg t_1 \vee \cdots \vee \neg t_n \vee
{}  \neg c_1 \vee \cdots \vee \neg c_m \vee
{}  \neg a \vee b$
satisfied regardless of $a$ and $b$. Instead, if the formula is satisfied by an
evaluation of $X$ and $E$ is its opposite, then all $c_j$ and $t_i$ have to be
true, turning
{} $\neg t_1 \vee \cdots \vee \neg t_n \vee
{}  \neg c_1 \vee \cdots \vee \neg c_m \vee
{}  \neg a \vee b$
into $\neg a \vee b$. This makes $\neg a$ and $b$ necessary as well.

\begin{center}
\setlength{\unitlength}{5000sp}%
\begingroup\makeatletter\ifx\SetFigFont\undefined%
\gdef\SetFigFont#1#2#3#4#5{%
  \reset@font\fontsize{#1}{#2pt}%
  \fontfamily{#3}\fontseries{#4}\fontshape{#5}%
  \selectfont}%
\fi\endgroup%
\begin{picture}(3648,2863)(7168,-7859)
\thinlines
{\color[rgb]{0,0,0}\put(7201,-5536){\line( 1, 0){3000}}
}%
\thicklines
{\color[rgb]{0,0,0}\put(7201,-6586){\line( 0,-1){ 75}}
\put(7201,-6661){\line( 1, 0){1200}}
\put(8401,-6661){\line( 0, 1){ 75}}
}%
{\color[rgb]{0,0,0}\put(9601,-6586){\line( 0,-1){ 75}}
\put(9601,-6661){\line( 1, 0){600}}
\put(10201,-6661){\line( 0, 1){ 75}}
}%
\thinlines
{\color[rgb]{0,0,0}\put(9976,-7636){\framebox(150,900){}}
}%
{\color[rgb]{0,0,0}\put(9676,-7636){\framebox(150,300){}}
}%
{\color[rgb]{0,0,0}\put(9751,-5236){\line( 0,-1){225}}
}%
{\color[rgb]{0,0,0}\put(8776,-5836){\line( 1, 0){150}}
}%
{\color[rgb]{0,0,0}\put(9076,-5836){\line( 1, 0){150}}
}%
{\color[rgb]{0,0,0}\put(9376,-5836){\line( 1, 0){150}}
}%
{\color[rgb]{0,0,0}\put(8476,-6136){\line( 1, 0){150}}
}%
{\color[rgb]{0,0,0}\put(8776,-6136){\line( 1, 0){150}}
}%
{\color[rgb]{0,0,0}\put(9376,-6136){\line( 1, 0){150}}
}%
\put(7351,-5461){\makebox(0,0)[b]{\smash{{\SetFigFont{12}{24.0}
{\rmdefault}{\mddefault}{\updefault}{\color[rgb]{0,0,0}$x_1$}%
}}}}
\put(7651,-5461){\makebox(0,0)[b]{\smash{{\SetFigFont{12}{24.0}
{\rmdefault}{\mddefault}{\updefault}{\color[rgb]{0,0,0}$n_1$}%
}}}}
\put(7951,-5461){\makebox(0,0)[b]{\smash{{\SetFigFont{12}{24.0}
{\rmdefault}{\mddefault}{\updefault}{\color[rgb]{0,0,0}$x_2$}%
}}}}
\put(8851,-5461){\makebox(0,0)[b]{\smash{{\SetFigFont{12}{24.0}
{\rmdefault}{\mddefault}{\updefault}{\color[rgb]{0,0,0}$t_2$}%
}}}}
\put(9151,-5461){\makebox(0,0)[b]{\smash{{\SetFigFont{12}{24.0}
{\rmdefault}{\mddefault}{\updefault}{\color[rgb]{0,0,0}$c_1$}%
}}}}
\put(9451,-5461){\makebox(0,0)[b]{\smash{{\SetFigFont{12}{24.0}
{\rmdefault}{\mddefault}{\updefault}{\color[rgb]{0,0,0}$c_2$}%
}}}}
\put(8251,-5461){\makebox(0,0)[b]{\smash{{\SetFigFont{12}{24.0}
{\rmdefault}{\mddefault}{\updefault}{\color[rgb]{0,0,0}$n_2$}%
}}}}
\put(7951,-5911){\makebox(0,0)[b]{\smash{{\SetFigFont{12}{24.0}
{\rmdefault}{\mddefault}{\updefault}{\color[rgb]{0,0,0}$0$}%
}}}}
\put(8251,-5911){\makebox(0,0)[b]{\smash{{\SetFigFont{12}{24.0}
{\rmdefault}{\mddefault}{\updefault}{\color[rgb]{0,0,0}$1$}%
}}}}
\put(7351,-6211){\makebox(0,0)[b]{\smash{{\SetFigFont{12}{24.0}
{\rmdefault}{\mddefault}{\updefault}{\color[rgb]{0,0,0}$1$}%
}}}}
\put(7651,-6211){\makebox(0,0)[b]{\smash{{\SetFigFont{12}{24.0}
{\rmdefault}{\mddefault}{\updefault}{\color[rgb]{0,0,0}$0$}%
}}}}
\put(7951,-6211){\makebox(0,0)[b]{\smash{{\SetFigFont{12}{24.0}
{\rmdefault}{\mddefault}{\updefault}{\color[rgb]{0,0,0}$0$}%
}}}}
\put(8251,-6211){\makebox(0,0)[b]{\smash{{\SetFigFont{12}{24.0}
{\rmdefault}{\mddefault}{\updefault}{\color[rgb]{0,0,0}$1$}%
}}}}
\put(7351,-6511){\makebox(0,0)[b]{\smash{{\SetFigFont{12}{24.0}
{\rmdefault}{\mddefault}{\updefault}{\color[rgb]{0,0,0}$0$}%
}}}}
\put(7651,-6511){\makebox(0,0)[b]{\smash{{\SetFigFont{12}{24.0}
{\rmdefault}{\mddefault}{\updefault}{\color[rgb]{0,0,0}$1$}%
}}}}
\put(7951,-6511){\makebox(0,0)[b]{\smash{{\SetFigFont{12}{24.0}
{\rmdefault}{\mddefault}{\updefault}{\color[rgb]{0,0,0}$0$}%
}}}}
\put(8251,-6511){\makebox(0,0)[b]{\smash{{\SetFigFont{12}{24.0}
{\rmdefault}{\mddefault}{\updefault}{\color[rgb]{0,0,0}$1$}%
}}}}
\put(8551,-5911){\makebox(0,0)[b]{\smash{{\SetFigFont{12}{24.0}
{\rmdefault}{\mddefault}{\updefault}{\color[rgb]{0,0,0}$0/1$}%
}}}}
\put(9751,-5911){\makebox(0,0)[b]{\smash{{\SetFigFont{12}{24.0}
{\rmdefault}{\mddefault}{\updefault}{\color[rgb]{0,0,0}$0/1$}%
}}}}
\put(10051,-5911){\makebox(0,0)[b]{\smash{{\SetFigFont{12}{24.0}
{\rmdefault}{\mddefault}{\updefault}{\color[rgb]{0,0,0}$1$}%
}}}}
\put(9151,-6211){\makebox(0,0)[b]{\smash{{\SetFigFont{12}{24.0}
{\rmdefault}{\mddefault}{\updefault}{\color[rgb]{0,0,0}$0/1$}%
}}}}
\put(8551,-6511){\makebox(0,0)[b]{\smash{{\SetFigFont{12}{24.0}
{\rmdefault}{\mddefault}{\updefault}{\color[rgb]{0,0,0}$1$}%
}}}}
\put(8851,-6511){\makebox(0,0)[b]{\smash{{\SetFigFont{12}{24.0}
{\rmdefault}{\mddefault}{\updefault}{\color[rgb]{0,0,0}$1$}%
}}}}
\put(9151,-6511){\makebox(0,0)[b]{\smash{{\SetFigFont{12}{24.0}
{\rmdefault}{\mddefault}{\updefault}{\color[rgb]{0,0,0}$1$}%
}}}}
\put(9451,-6511){\makebox(0,0)[b]{\smash{{\SetFigFont{12}{24.0}
{\rmdefault}{\mddefault}{\updefault}{\color[rgb]{0,0,0}$1$}%
}}}}
\put(9751,-6211){\makebox(0,0)[b]{\smash{{\SetFigFont{12}{24.0}
{\rmdefault}{\mddefault}{\updefault}{\color[rgb]{0,0,0}$0/1$}%
}}}}
\put(10051,-6211){\makebox(0,0)[b]{\smash{{\SetFigFont{12}{24.0}
{\rmdefault}{\mddefault}{\updefault}{\color[rgb]{0,0,0}$1$}%
}}}}
\put(9751,-6511){\makebox(0,0)[b]{\smash{{\SetFigFont{12}{24.0}
{\rmdefault}{\mddefault}{\updefault}{\color[rgb]{0,0,0}$1$}%
}}}}
\put(10051,-6511){\makebox(0,0)[b]{\smash{{\SetFigFont{12}{24.0}
{\rmdefault}{\mddefault}{\updefault}{\color[rgb]{0,0,0}$1$}%
}}}}
\put(7351,-6961){\makebox(0,0)[b]{\smash{{\SetFigFont{12}{24.0}
{\rmdefault}{\mddefault}{\updefault}{\color[rgb]{0,0,0}$1$}%
}}}}
\put(7651,-6961){\makebox(0,0)[b]{\smash{{\SetFigFont{12}{24.0}
{\rmdefault}{\mddefault}{\updefault}{\color[rgb]{0,0,0}$1$}%
}}}}
\put(7951,-6961){\makebox(0,0)[b]{\smash{{\SetFigFont{12}{24.0}
{\rmdefault}{\mddefault}{\updefault}{\color[rgb]{0,0,0}$0$}%
}}}}
\put(8251,-6961){\makebox(0,0)[b]{\smash{{\SetFigFont{12}{24.0}
{\rmdefault}{\mddefault}{\updefault}{\color[rgb]{0,0,0}$1$}%
}}}}
\put(7351,-7261){\makebox(0,0)[b]{\smash{{\SetFigFont{12}{24.0}
{\rmdefault}{\mddefault}{\updefault}{\color[rgb]{0,0,0}$1$}%
}}}}
\put(7651,-7261){\makebox(0,0)[b]{\smash{{\SetFigFont{12}{24.0}
{\rmdefault}{\mddefault}{\updefault}{\color[rgb]{0,0,0}$0$}%
}}}}
\put(7951,-7261){\makebox(0,0)[b]{\smash{{\SetFigFont{12}{24.0}
{\rmdefault}{\mddefault}{\updefault}{\color[rgb]{0,0,0}$0$}%
}}}}
\put(8251,-7261){\makebox(0,0)[b]{\smash{{\SetFigFont{12}{24.0}
{\rmdefault}{\mddefault}{\updefault}{\color[rgb]{0,0,0}$1$}%
}}}}
\put(7951,-7561){\makebox(0,0)[b]{\smash{{\SetFigFont{12}{24.0}
{\rmdefault}{\mddefault}{\updefault}{\color[rgb]{0,0,0}$0$}%
}}}}
\put(8251,-7561){\makebox(0,0)[b]{\smash{{\SetFigFont{12}{24.0}
{\rmdefault}{\mddefault}{\updefault}{\color[rgb]{0,0,0}$1$}%
}}}}
\put(7651,-7561){\makebox(0,0)[b]{\smash{{\SetFigFont{12}{24.0}
{\rmdefault}{\mddefault}{\updefault}{\color[rgb]{0,0,0}$1$}%
}}}}
\put(7351,-7561){\makebox(0,0)[b]{\smash{{\SetFigFont{12}{24.0}
{\rmdefault}{\mddefault}{\updefault}{\color[rgb]{0,0,0}$0$}%
}}}}
\put(9751,-6961){\makebox(0,0)[b]{\smash{{\SetFigFont{12}{24.0}
{\rmdefault}{\mddefault}{\updefault}{\color[rgb]{0,0,0}$0/1$}%
}}}}
\put(9751,-7261){\makebox(0,0)[b]{\smash{{\SetFigFont{12}{24.0}
{\rmdefault}{\mddefault}{\updefault}{\color[rgb]{0,0,0}$0/1$}%
}}}}
\put(9751,-7561){\makebox(0,0)[b]{\smash{{\SetFigFont{12}{24.0}
{\rmdefault}{\mddefault}{\updefault}{\color[rgb]{0,0,0}$1$}%
}}}}
\put(10051,-6961){\makebox(0,0)[b]{\smash{{\SetFigFont{12}{24.0}
{\rmdefault}{\mddefault}{\updefault}{\color[rgb]{0,0,0}$1$}%
}}}}
\put(10051,-7261){\makebox(0,0)[b]{\smash{{\SetFigFont{12}{24.0}
{\rmdefault}{\mddefault}{\updefault}{\color[rgb]{0,0,0}$1$}%
}}}}
\put(10051,-7561){\makebox(0,0)[b]{\smash{{\SetFigFont{12}{24.0}
{\rmdefault}{\mddefault}{\updefault}{\color[rgb]{0,0,0}$1$}%
}}}}
\put(10051,-5461){\makebox(0,0)[b]{\smash{{\SetFigFont{12}{24.0}
{\rmdefault}{\mddefault}{\updefault}{\color[rgb]{0,0,0}$a \vee \neg b$}%
}}}}
\put(9751,-5161){\makebox(0,0)[b]{\smash{{\SetFigFont{12}{24.0}
{\rmdefault}{\mddefault}{\updefault}{\color[rgb]{0,0,0}$\neg a \vee b$}%
}}}}
\put(8551,-5461){\makebox(0,0)[b]{\smash{{\SetFigFont{12}{24.0}
{\rmdefault}{\mddefault}{\updefault}{\color[rgb]{0,0,0}$t_1$}%
}}}}
\put(9826,-7786){\makebox(0,0)[rb]{\smash{{\SetFigFont{12}{24.0}
{\rmdefault}{\mddefault}{\updefault}{\color[rgb]{0,0,0}$\neg a,b \mbox{ necessary}$}%
}}}}
\put(10201,-7411){\makebox(0,0)[lb]{\smash{{\SetFigFont{12}{24.0}
{\rmdefault}{\mddefault}{\updefault}{\color[rgb]{0,0,0}$a,\neg b \mbox{ necessary}$}%
}}}}
\put(10801,-6511){\makebox(0,0)[lb]{\smash{{\SetFigFont{12}{24.0}
{\rmdefault}{\mddefault}{\updefault}{\color[rgb]{0,0,0}$(\mbox{all } x_i \not= n_i, \mbox{all } f_j \mbox { true})$}%
}}}}
\put(10801,-5911){\makebox(0,0)[lb]{\smash{{\SetFigFont{12}{24.0}
{\rmdefault}{\mddefault}{\updefault}{\color[rgb]{0,0,0}$(x_1=n_1=\false)$}%
}}}}
\put(10801,-6211){\makebox(0,0)[lb]{\smash{{\SetFigFont{12}{24.0}
{\rmdefault}{\mddefault}{\updefault}{\color[rgb]{0,0,0}$(f_1 \mbox{ false})$}%
}}}}
\put(7651,-5911){\makebox(0,0)[b]{\smash{{\SetFigFont{12}{24.0}
{\rmdefault}{\mddefault}{\updefault}{\color[rgb]{0,0,0}$0$}%
}}}}
\put(7351,-5911){\makebox(0,0)[b]{\smash{{\SetFigFont{12}{24.0}
{\rmdefault}{\mddefault}{\updefault}{\color[rgb]{0,0,0}$0$}%
}}}}
\end{picture}%
\nop{
x1 n1 x2 n2  t1 t2 c1 c2  -avb av-b
 0  0  0  1  0/1 -  -  -   0/1  1
 1  0  0  1   -  - 0/1 -   0/1  1
 0  1  0  1   1  1  1  1     1  1
-----------               -------
 0  0  0  1                0/1  1 \                       .
 1  0  0  1                0/1  1   >-- a,-b necessary
 0  1  0  1              +-- 1  1 /
                         |   
                  -a,b necess.
}
\end{center}

The figure shows three models as an example. In the first model, the
assignments $x_1 = e_1 = \false$ allow $t_1$ to take any value (denoted $0/1$);
regardless of the other variables (irrelevant values are marked $-$), $t_1 =
\false$ satisfies the clause
{} $\neg t_1 \vee \cdots \vee \neg t_n \vee
{}  \neg c_1 \vee \cdots \vee \neg c_m \vee
{}  \neg a \vee b$
without the need to also satisfy its subset $\neg a \vee b$; this subclause can
be false and still $A$ is true. In the second model the values of $x_i$ and
$e_i$ are opposite to each other for every $i$, but the clause $f_2 \in F$ is
false; $c_1$ can take any value, including $\false$; this again allows $A$ to
be true even if $\neg a \vee b$ is false. In the third model, the variables
$x_i$ and $e_i$ are all opposite to each other and all clauses of $F$ true; all
$t_i$ and $c_i$ are forced to be true, making $\neg a \vee b$ the only way to
satisfy the clause
{} $\neg t_1 \vee \cdots \vee \neg t_n \vee
{}  \neg c_1 \vee \cdots \vee \neg c_m \vee
{}  \neg a \vee b$.
When removing the intermediate variables $t_i$ and $c_i$, all that matters is
whether $\neg a \vee b$ was allowed to be false for some values of the removed
variables or not. This is the case for the first two models but not the third,
where $\neg a$ and $b$ are necessary.

\

\noindent {\bf Minimality.} The minimality of $A$ is proved applying
Lemma~\ref{set-value} to remove some clauses so that the remaining ones do not
resolve and Lemma~\ref{no-resolution} applies. Lemma~\ref{minimal} proves $A$
minimal since it only contains superirredundant clauses.

Substituting the variables $a$, $b$ with false removes
{} $\neg t_1 \vee \cdots \vee \neg t_n \vee
{}  \neg c_1 \vee \cdots \vee \neg c_m \vee
{}  \neg a \vee b$ and
{} $a \vee \neg b$
from $A$. The remaining clauses contain $x_i,e_i$ only negative and $t_i,c_j$
only positive. Therefore, these clauses do not resolve. Since they do not
contain each other, Lemma~\ref{no-resolution} proves them superirredundant.
They are also superirredundant in $A$ by Lemma~\ref{set-value} since $A$ does
not contain any of their supersets.

The superirredundancy of the remaining two clauses is proved by substituting
all $x_i,e_i$ with false. This substitution removes the clauses
{} $\neg x_i \vee \neg e_i$,
{} $\neg x_i \vee t_i$,
{} $\neg e_i \vee t_i$,
{} $\neg x_i \vee c_j$ and
{} $\neg e_i \vee c_j$
because they all contain either $\neg x_i$ or $\neg e_i$. The two remaining
clauses are
{} $\neg t_1 \vee \cdots \vee \neg t_n \vee
{}  \neg c_1 \vee \cdots \vee \neg c_m \vee
{}  \neg a \vee b$ and
{} $a \vee \neg b$.
They have opposite literals, but resolving them results in tautologies. As a
result, $F = \rescn(F)$. Since none of the two entails the other, they are
irredundant in $\rescn(F)$ and are therefore superirredundant. By
Lemma~\ref{set-value}, they are superirredundant in $A$ as well since $A$ does
not contain a superset of them.

\

\noindent {\bf Formula $F$ is unsatisfiable.} Forgetting all variables except
$X_C$ from $A$ is expressed by
{} $B = \{\neg x_i \vee \neg e_i\} \cup \{a \vee \neg b\}$,
a Horn formula of the required variables $X_C = X \cup E \cup \{a, b\}$ and
size $||B|| = 2 \times n + 2 = k$.

Theorem~\ref{consistent-literals-complete} proves that $B$ expresses forgetting
if every set of literals $S$ that contains exactly all variables $X_C = X \cup
E \cup \{a,b\}$ is satisfiable with $A$ if and only if it is satisfiable with
$B$. Two cases are possible.

\begin{description}

\item[$\{x_i,e_i\} \subseteq S$ for some $i$]; the clause $\neg x_i \vee \neg
e_i$ in both $A$ and $B$ is falsified by $S$; both $A \cup S$ and $B \cup S$
are unsatisfiable;

\item[$\{x_i,e_i\} \subseteq S$ for no $i$]; since $S$ contains either $x_i$
or $\neg x_i$ for each $i$ and either $e_i$ or $\neg e_i$ for each $i$,
either $\neg x_i \in S$ or $\neg e_i \in S$; as a result, all clauses $\neg
x_i \vee \neg e_i$ are satisfied in $A \cup S$ and $B \cup S$, and can
therefore be disregarded from this point on; the only remaining clause of $B$
is $a \vee \neg b$;

if $S$ contains $\neg a$ and $b$, then $B$ is not satisfied; but $A$ contains
the same clause $a \vee \neg b$, so it is not satisfied either; if $S$ contains
both $a$ and $b$ or both $\neg a$ and $\neg b$, then $B$ is satisfied, and $A$
is also satisfied by setting all variables $t_i$ and $c_j$ to true; therefore,
the only sets $S$ that may differ when added to $A$ and $B$ are those
containing $a$ and $\neg b$; these sets are consistent with $B$; they make the
clause $a \vee \neg b$ of $A$ redundant, and resolve with the clause
{} $\neg t_1 \vee \cdots \vee \neg t_n \vee
{}  \neg c_1 \vee \cdots \vee \neg c_m \vee
{}  \neg a \vee b$
making it subsumed by
{} $\neg t_1 \vee \cdots \vee \neg t_n \vee
{}  \neg c_1 \vee \cdots \vee \neg c_m$.

Two subcases are considered:

\begin{description}

\item[$\{\neg x_i,\neg e_i\} \subseteq S$ for some $i$] the remaining clauses
of $A$ are satisfied by setting $t_i$ to $\false$, all $t_z$ with $z \not= i$
to true and all $c_j$ to true; in particular, the clause
{} $\neg t_1 \vee \cdots \vee \neg t_n \vee
{}  \neg c_1 \vee \cdots \vee \neg c_m$
is satisfied because of $t_i = \false$;

\item[$\{\neg x_i,\neg e_i\} \subseteq S$ for no $i$]; at this point, also
$\{x_i,e_i\} \subseteq S$ for no $i$; as a result, $S$ contains either
$\{x_i,\neg e_i\}$ or $\{\neg x_i,e_i\}$, which means that it implies $x_i
\not\equiv e_i$; the clauses $\neg e_i \vee c_j$ are therefore equivalent to
$x_i \vee c_j$; by assumption, at least a clause of $F$ is false for every
possible value of the variables $X$; let $f_j$ be such a clause for the only
truth evaluation on $X$ that satisfies $S$; by setting all variables $t_i$ and
all $c_z$ with $z \not= j$ to true and $c_j$ to false, this assignment
satisfies all clauses; in particular, the clauses
{} $\neg x_i \vee c_j$ and $x_i \vee c_j$
are satisfied even if $c_j$ is false because $f_j$ is false in $S$, which
implies that all literals $x_i$ and $\neg x_i$ it contains are false; the
clause
{} $\neg t_1 \vee \cdots \vee \neg t_n \vee
{}  \neg c_1 \vee \cdots \vee \neg c_m$
is satisfied because of $c_j = \false$.

\end{description}

\end{description}

All of this proves that $B$ is the result of forgetting all variables except
$X_C$ from $A$.

% The minimality of $B$ follows from the following point: its size is $k$, and
% the following point proves that every formula expressing forgetting has size
% $k$ or more. It also follows from Lemma~\ref{no-resolution} since its clauses
% do not share variables.

\

\noindent {\bf Minimal number of literals.} Every CNF formula $B$ that
expresses forgetting all variables except $X_C = X \cup E \cup \{a,b\}$ from
$A$ contains at least $k = 2 \times n + 2$ literal occurrences.

This is proved by showing that $B$ contains the literals $\neg x_i$, $\neg
e_i$, $a$ and $\neg b$. This is in turn proved by Lemma~\ref{forget-contains}:
for each of them $l$, a set $S$ is shown consistent with $A$ while
{} $S \backslash \{l\} \cup \{\neg l\}$
is not.

For the literals $\neg x_i$ and $a$ the set $S$ contains all $\neg x_i$, all
$e_i$, $a$ and $b$. It is consistent with $A$ because both are satisfied by the
model that sets all $x_i$ to false and all $e_i$, $t_i$, $c_i$, $a$ and $b$ to
true. Replacing $\neg x_i$ with $x_i$ makes $S$ inconsistent with $\neg x_i
\vee \neg e_i$. Replacing $a$ with $\neg a$ makes $S$ inconsistent with $a \vee
\neg b$.

For the literals $\neg e_i$ and $\neg b$, the set $S$ contains all $x_i$, all
$\neg e_i$, $\neg a$ and $\neg b$. It is consistent with $A$ because both are
satisfied by the model that assigns all $x_i$, $t_i$ and $c_i$ to true and all
$e_i$, $a$ and $b$ to false. Replacing $\neg e_i$ with $e_i$ makes $S$
inconsistent with $\neg x_i \vee \neg e_i$. Replacing $\neg b$ with $b$ makes
it inconsistent with $a \vee \neg b$.

This proves that every formula obtained by forgetting all variables except
$X_C$ from $A$ contains all the $k = 2 \times n + 2$ literals $\neg x_i$, $\neg
e_i$, $a$ and $\neg b$.

\

\noindent {\bf Formula $F$ is satisfiable.} If this is the case, every CNF
formula $B$ that expresses forgetting all variables except $X_C$ from $A$
contains the literals $\neg a$ and $b$. These two literals are in addition to
the $k$ literals of the previous point, raising the minimal number of literals
to $k + 2$.

That $B$ contains $\neg a$ is proved by exhibiting a set of literals $S$ that
is consistent with $A$ while $S \backslash \{l\} \cup \{\neg l\}$ is not where
$l = \neg a$. This implies that $B$ contains $\neg a$ by
Lemma~\ref{forget-contains}. A similar set with $l = b$ shows that $B$ also
contains $b$.

Let $M$ be a model of $F$. The set of literals $S$ contains $x_i$ or $\neg x_i$
depending on whether $M$ satisfies $x_i$; it contains $e_i$ or $\neg e_i$
depending on whether $M$ falsifies $x_i$; it also contains $\neg a$ and $\neg
b$. This set is consistent with $A$ because they are both satisfied by the
model that extends $M$ by setting each $e_i$ opposite to $x_i$, all $t_i$
and $c_i$ to true and $a$ and $b$ to false. In particular, the clause
{} $\neg t_1 \vee \cdots \vee \neg t_n \vee
{}  \neg c_1 \vee \cdots \vee \neg c_m \vee
{}  \neg a \vee b$
is satisfied because it contains $\neg a$.

Replacing $\neg a$ with $a$ makes $S$ no longer consistent with $A$. Let $S' =
S \backslash \{\neg a\} \cup \{\neg \neg a\}$. This set has the same literals
over $x_i$ and $e_i$ of $S$. Since $M$ satisfies $F$, for each of its clauses
$f_j$ at least a literal in $f_j$ is true in $M$. If this literal is $x_i$,
then $S'$ contains $x_i$; since $x_i$ is in $f_j$, formula $A$ contains the
clause $\neg x_i \vee c_j$; therefore, $S' \cup A \models c_j$. If the literal
of $f_j$ that is true in $M$ is $\neg x_i$, then $S'$ contains $e_i$; since
$\neg x_i$ is in $f_j$, formula $A$ contains $\neg e_i \vee c_j$; therefore,
$S' \cup A \models c_j$. This proves that regardless of whether the literal of
$f_j$ that is true in $M$ is positive or negative, if $F$ is consistent then
$S' \cup A$ implies $c_j$. This is the case for every $j$ since $M$ satisfies
all clauses of $F$. Since $S'$ contains either $x_i$ or $e_i$ for every $i$ and
$A$ contains both $\neg x_i \vee t_i$ and $\neg e_i \vee t_i$ for every $i$,
$S' \cup A$ also implies all variables $t_j$. Since $S' = S \backslash \{\neg
a\} \cup \{a\}$ also contains $a$ and $\neg b$, it is inconsistent with
{} $\neg t_1 \vee \cdots \vee \neg t_n \vee
{}  \neg c_1 \vee \cdots \vee \neg c_m \vee
{}  \neg a \vee b$.
This proves that $\neg a$ is in $B$.

The similar set $S$ that contains $a$ and $b$ leads to the same point where all
variables $c_j$ and $t_j$ are implied, making
{} $\neg t_1 \vee \cdots \vee \neg t_n \vee
{}  \neg c_1 \vee \cdots \vee \neg c_m \vee
{}  \neg a \vee b$
consistent with $\{a,b\}$ but not with $\{a,\neg b\}$. This proves that $b$ is
also in $B$.~\qed

This lemma shows a polynomial reduction from propositional unsatisfiability to
the problem of forget size in the Horn case. As for all polynomial reductions,
it translates a formula $F$ without knowing its satisfiability, which however
affects the minimal size of expressing forgetting.

Being a polynomial reduction from propositional unsatisfiability, it proves the
forgetting size problem \conp-hard. Yet, the lemma is not formulated this way.
It instead predicates about the reduction itself. Only this way it could
include the additional property that the minimal size is either $k$ or at least
$k+2$. This allows merging it with another reduction to form a proof of
\Dp-hardness.

\

The following lemma also shows the problem \np-hard: a formula $F$ is
satisfiable if and only if forgetting some variables from $A$ can be expressed
in a certain space. However, its statement refers to the reduction itself for
the same reason of the previous lemma: it is necessary to the following proof
of \Dp-hardness.

% the minimality of A is proved by superirredundancy; as a result, no CNF
% formula shorter than A is equivalent to it; minimality is over CNF all
% formulae, not just Horn formulae

% the proof of minimal size of formulae expressing forgetting uses
% superirredundancy: they all contain AT, AC and AB; they also contain other
% clauses, but part of their size is due to the superirredundancy of these; as
% a result, minimality is among CNFs and not arbitrary formulae like in the
% previous reduction; it is however not limited to Horn formulae

% the formula expressing forgetting in size k if any is Horn; therefore,
% hardness holds both by restricting it to be Horn or not

\begin{lemma}
\label{horn-np}

There exists a polynomial algorithm that turns a CNF formula $F$ into a
minimal-size Horn formula $A$, a subset $X_C \subseteq \var(A)$ and a number
$k$ such that forgetting all variables except $X_C$ from $A$ is expressed by a
Horn formula of size $k$ if $F$ is satisfiable and only by Horn formulae of
size greater than $k$ otherwise.

\end{lemma}

\proof Let the formula be $F = \{f_1,\ldots,f_m\}$ and $X=\{x_1,\ldots,x_n\}$
its variables. The formula $A$ is built over an extended alphabet comprising
the variables
{} $X = \{x_1,\ldots,x_n\}$
and the additional variables
{} $O = \{o_1,\ldots,o_n\}$,
{} $E = \{e_1,\ldots,e_n\}$,
{} $P = \{p_1,\ldots,p_n\}$,
{} $T = \{t_1,\ldots,t_n\}$,
{} $C = \{c_1,\ldots,c_m\}$,
{} $R = \{r_1,\ldots,r_n\}$,
{} $S = \{s_1,\ldots,s_n\}$ and
{} $q$.

The formula $A$, the set of variables $X_C$ and the integer $k$ are as follows.

\begin{eqnarray*}
A &=& A_F \cup A_T \cup A_C \cup A_B					\\
A_F &=&
\{x_i \vee \neg o_i, o_i \vee \neg q \mid x_i \in X\} \cup
\{e_i \vee \neg p_i, p_i \vee \neg q \mid x_i \in X\}			\\
A_T &=&
\{\neg x_i \vee t_i, \neg e_i \vee t_i \mid x_i \in X\}			\\
A_C &=&
\{\neg x_i \vee c_j \mid      x_i \in f_j \} \cup
\{\neg e_i \vee c_j \mid \neg x_i \in f_j \}				\\
A_B &=&
\{\neg t_1 \vee \cdots \vee \neg t_n \vee
  \neg c_1 \vee \cdots \vee \neg c_m \vee
  x_i \vee \neg r_i, r_i \vee \neg q \mid x_i \in X\} \cup		\\
&&
\{\neg t_1 \vee \cdots \vee \neg t_n \vee
  \neg c_1 \vee \cdots \vee \neg c_m \vee
  e_i \vee \neg s_i, s_i \vee \neg q \mid x_i \in X\}			\\
X_C &=& X \cup E \cup T \cup C \cup R \cup S \cup \{q\}			\\
k &=& 2 \times n + ||A_T|| + ||A_C|| + ||A_B||
\end{eqnarray*}

Before formally proving that the reduction works, a short summary of why it
works is given. The variables to forget are $O \cup P$. A way to forget them is
to turn $A_F$ into $A_R = \{x_i \vee \neg q, e_i \vee \neg q \mid x_i \in X\}$.
The other clauses of $A$ are superirredundant: all minimal equivalent formulae
contain them. The bound $k$ allows only one clause of $A_R$ for each $i$.
Combined with the clauses of $A_T$ they entail $t_i \vee \neg q$. If $F$ is
satisfiable, they also combine with the clauses $A_C$ to imply all clauses $c_j
\vee \neg q$. Resolving these clauses with $A_B$ produces all clauses $x_i \vee
\neg q$ and $e_i \vee \neg q$, including the ones not in the formula. This way,
a formula that contains one clause of $A_R$ for each index $i$ implies all of
$A_R$, but only if $F$ is satisfiable.

The following figure shows how $e_1 \vee q$ is derived from $x_1 \vee q$ and
$e_2 \vee q$, when the formula is $F = \{f_1,f_2\}$ where $f_1 = x_1 \vee x_2$
and $f_2 = \neg x_1 \vee \neg x_2$. These clauses translate into
{} $A_C = \{
{}	\neg x_1 \vee c_1,
{}	\neg x_2 \vee c_1,
{}	\neg e_1 \vee c_2,
{}	\neg e_2 \vee c_2
{} \}$.

\begin{center}
\setlength{\unitlength}{5000sp}%
\begingroup\makeatletter\ifx\SetFigFont\undefined%
\gdef\SetFigFont#1#2#3#4#5{%
  \reset@font\fontsize{#1}{#2pt}%
  \fontfamily{#3}\fontseries{#4}\fontshape{#5}%
  \selectfont}%
\fi\endgroup%
\begin{picture}(5277,3675)(6589,-7726)
\thinlines
{\color[rgb]{0,0,0}\put(7501,-4486){\line( 0,-1){225}}
}%
{\color[rgb]{0,0,0}\put(7501,-5011){\line( 0,-1){300}}
}%
{\color[rgb]{0,0,0}\put(7501,-5986){\line( 0,-1){300}}
}%
{\color[rgb]{0,0,0}\put(6601,-4936){\line( 1,-1){300}}
}%
{\color[rgb]{0,0,0}\put(6601,-5236){\line( 1, 1){300}}
}%
{\color[rgb]{0,0,0}\put(7501,-6586){\line( 0,-1){225}}
}%
{\color[rgb]{0,0,0}\put(8551,-5311){\line( 1, 0){1500}}
}%
{\color[rgb]{0,0,0}\put(8551,-6061){\line( 1, 0){1500}}
}%
{\color[rgb]{0,0,0}\put(8551,-6586){\line( 1, 0){1500}}
}%
{\color[rgb]{0,0,0}\put(8551,-4711){\line( 1, 0){1500}}
}%
{\color[rgb]{0,0,0}\put(6601,-5986){\line( 1,-1){300}}
}%
{\color[rgb]{0,0,0}\put(6601,-6286){\line( 1, 1){300}}
}%
{\color[rgb]{0,0,0}\put(10501,-7111){\vector( 1, 0){1050}}
}%
{\color[rgb]{0,0,0}\put(10051,-4411){\vector( 0,-1){2475}}
}%
{\color[rgb]{0,0,0}\put(7051,-4711){\vector( 1, 0){900}}
}%
{\color[rgb]{0,0,0}\put(7051,-4786){\vector( 2,-1){900}}
}%
{\color[rgb]{0,0,0}\put(7051,-6511){\vector( 2, 1){900}}
}%
{\color[rgb]{0,0,0}\put(7051,-6586){\vector( 1, 0){900}}
}%
{\color[rgb]{0,0,0}\put(11026,-7411){\line( 0, 1){300}}
}%
\put(7501,-4411){\makebox(0,0)[b]{\smash{{\SetFigFont{12}{24.0}
{\rmdefault}{\mddefault}{\updefault}{\color[rgb]{0,0,0}$\neg x_1 \vee t_1$}%
}}}}
\put(7501,-6961){\makebox(0,0)[b]{\smash{{\SetFigFont{12}{24.0}
{\rmdefault}{\mddefault}{\updefault}{\color[rgb]{0,0,0}$\neg e_2 \vee t_2$}%
}}}}
\put(6751,-4711){\makebox(0,0)[b]{\smash{{\SetFigFont{12}{24.0}
{\rmdefault}{\mddefault}{\updefault}{\color[rgb]{0,0,0}$x_1 \vee \neg q$}%
}}}}
\put(6751,-5116){\makebox(0,0)[b]{\smash{{\SetFigFont{12}{24.0}
{\rmdefault}{\mddefault}{\updefault}{\color[rgb]{0,0,0}$e_1 \vee \neg q$}%
}}}}
\put(6751,-6211){\makebox(0,0)[b]{\smash{{\SetFigFont{12}{24.0}
{\rmdefault}{\mddefault}{\updefault}{\color[rgb]{0,0,0}$x_2 \vee \neg q$}%
}}}}
\put(6751,-6586){\makebox(0,0)[b]{\smash{{\SetFigFont{12}{24.0}
{\rmdefault}{\mddefault}{\updefault}{\color[rgb]{0,0,0}$e_2 \vee \neg q$}%
}}}}
\put(8251,-4711){\makebox(0,0)[b]{\smash{{\SetFigFont{12}{24.0}
{\rmdefault}{\mddefault}{\updefault}{\color[rgb]{0,0,0}$\neg q \vee t_1$}%
}}}}
\put(8251,-6586){\makebox(0,0)[b]{\smash{{\SetFigFont{12}{24.0}
{\rmdefault}{\mddefault}{\updefault}{\color[rgb]{0,0,0}$\neg q \vee t_2$}%
}}}}
\put(7501,-5461){\makebox(0,0)[b]{\smash{{\SetFigFont{12}{24.0}
{\rmdefault}{\mddefault}{\updefault}{\color[rgb]{0,0,0}$\neg x_1 \vee c_1$}%
}}}}
\put(7501,-7711){\makebox(0,0)[b]{\smash{{\SetFigFont{12}{24.0}
{\rmdefault}{\mddefault}{\updefault}{\color[rgb]{0,0,0}$\neg e_1 \vee c_2$}%
}}}}
\put(7501,-7486){\makebox(0,0)[b]{\smash{{\SetFigFont{12}{24.0}
{\rmdefault}{\mddefault}{\updefault}{\color[rgb]{0,0,0}$\neg x_2 \vee c_1$}%
}}}}
\put(8251,-5311){\makebox(0,0)[b]{\smash{{\SetFigFont{12}{24.0}
{\rmdefault}{\mddefault}{\updefault}{\color[rgb]{0,0,0}$\neg q \vee c_1$}%
}}}}
\put(7501,-5911){\makebox(0,0)[b]{\smash{{\SetFigFont{12}{24.0}
{\rmdefault}{\mddefault}{\updefault}{\color[rgb]{0,0,0}$\neg e_2 \vee c_2$}%
}}}}
\put(8251,-6061){\makebox(0,0)[b]{\smash{{\SetFigFont{12}{24.0}
{\rmdefault}{\mddefault}{\updefault}{\color[rgb]{0,0,0}$\neg q \vee c_2$}%
}}}}
\put(9976,-4261){\makebox(0,0)[b]{\smash{{\SetFigFont{12}{24.0}
{\rmdefault}{\mddefault}{\updefault}{\color[rgb]{0,0,0}$\neg t_1 \vee \neg t_2 \vee \neg c_1 \vee \neg c_2 \vee e_1 \vee \neg s_1$}%
}}}}
\put(10051,-7111){\makebox(0,0)[b]{\smash{{\SetFigFont{12}{24.0}
{\rmdefault}{\mddefault}{\updefault}{\color[rgb]{0,0,0}$\neg q \vee e_1 \vee \neg s_1$}%
}}}}
\put(11026,-7561){\makebox(0,0)[b]{\smash{{\SetFigFont{12}{24.0}
{\rmdefault}{\mddefault}{\updefault}{\color[rgb]{0,0,0}$s_1 \vee \neg q$}%
}}}}
\put(11851,-7111){\makebox(0,0)[b]{\smash{{\SetFigFont{12}{24.0}
{\rmdefault}{\mddefault}{\updefault}{\color[rgb]{0,0,0}$e_1 \vee \neg q$}%
}}}}
\end{picture}%
\nop{
      -x1 v t1           -t1 v -t2 v -c1 v -c2 v n1 v -r1
          |                    |
       ---+---> -q v t1 -------+
x1 v -q                        |
       ---+---> -q v c2 -------+
          |                    |
      -x1 v c2                 |
                               |
(n1 v -q)                      |
                               |
                               |
(x2 v -q)                      |
                               |
      -n2 v c1                 |
          |                    |
      ----+---> -q v c1 -------+
n2 v -q                        |
      ----+---> -q v t2 -------+
          |                    |
      -n2 v t2                 V
                         -q v n1 v -r1 ----+----> n1 v -q
                                           |
      -x2 v c1                          r1 v -q
      -n1 v c2
}
\end{center}

For each index $i$, at least one among $x_i \vee q$ and $e_i \vee q$ is
necessary for deriving $q \vee t_i$, which is required for these derivations to
work. Alternatively, $q \vee t_i$ may be selected. Either way, for each index
$i$ at least a two-literal clause is necessary.

The claim is formally proved in four steps: first, a non-minimal way to forget
all variables except $X_C$ is shown; second, its superirredundant clauses are
identified; third, an equivalent formula of size $k$ is built if $F$ is
satisfiable; fourth, the necessary clauses in every equivalent formula are
identified; fifth, if $F$ is unsatisfiable every equivalent formula is proved
to have size greater than $k$.

\

{\bf Effect of forgetting.}

Theorem~\ref{resolve-out} proves that forgetting all variables not in $X_C$,
which are $O \cup P$, is expressed by resolving out these variables. Since
$o_i$ occurs only in $x_i \vee \neg o_i$ and $o_i \vee \neg q$, the result is
$x_i \vee \neg q$. The same holds for $p_i$. The resulting clauses are denoted
$A_R$:

\[
A_R = \{x_i \vee \neg q, e_i \vee \neg q \mid x_i \in X\}
\]

\

{\bf Superirredundancy.}

The claim requires $A$ to be minimal, which follows from all its clauses being
superirredundant by Lemma~\ref{minimal}. Most of them survive forgetting; the
reduction is based on these being superirredundant. Instead of proving
superirredundancy in two different but similar formulae, it is proved in their
union.

In particular, the clauses
{} $A_F \cup A_T \cup A_C \cup A_B$
are shown superirredundant in
{} $A_F \cup A_R \cup A_T \cup A_C \cup A_B$.
Lemma~\ref{superset} implies that they are also superirredundant in its subsets
{} $A_F \cup A_T \cup A_C \cup A_B$
and
{} $A_R \cup A_T \cup A_C \cup A_B$,
the formula before and after forgetting.

To be precise, the latter is just one among the formulae expressing forgetting.
Yet, its superirredundant clauses are in all minimal CNF formulae equivalent to
it. Therefore, all minimal CNF formulae expressing forgetting contain them.

Superirredundancy is proved via Lemma~\ref{set-value}: a substitution simplify
{} $A_F \cup A_R \cup A_T \cup A_C \cup A_B$
enough to prove superirredundancy easily, for example because its clauses do
not resolve and Lemma~\ref{no-resolution} applies.

\begin{itemize}

\item

Replacing all variables $x_i$, $e_i$, $t_i$ and $c_j$ with true removes from
{} $A_F \cup A_R \cup A_T \cup A_C \cup A_B$
all clauses in $A_R \cup A_T \cup A_C$, all clauses of $A_F$ but $o_i \vee \neg
q$ and $p_i \vee \neg q$ and all clauses of $A_B$ but $r_i \vee \neg q$ and
$s_i \vee \neg q$. The remaining clauses contain only the literals $o_i$,
$p_i$, $r_i$, $s_i$ and $\neg q$. Therefore, they do not resolve. Since none is
contained in another, they are all superirredundant by
Lemma~\ref{no-resolution}. This proves the superirredundancy of all clauses
$o_i \vee \neg q$, $p_i \vee \neg q$, $r_i \vee \neg q$ and $s_i \vee \neg q$.

\item

Replacing all variables $q$, $o_i$, $p_i$, $r_i$ and $s_i$ with false removes
from
{} $A_F \cup A_R \cup A_T \cup A_C \cup A_B$
all clauses but $A_T \cup A_C$. These clauses contain only the literals $\neg
x_i$, $\neg e_i$, $t_i$ and $c_j$. Therefore, they do not resolve. Since they
are not contained in each other, Lemma~\ref{no-resolution} proves them
superirredundant.

\item

Replacing all variables $q$, $r_i$ and $s_i$ with false and all variables $t_i$
and $c_i$ with true removes from
{} $A_F \cup A_R \cup A_T \cup A_C \cup A_B$
all clauses but $x_i \vee \neg o_i$ and $e_i \vee \neg p_i$. They do not
resolve because they do not share variables. Lemma~\ref{no-resolution} proves
them superirredundant because they do not contain each other.

\item

Replacing all variables with false except for all variables $t_i$ and $c_j$ and
the two variables $x_h$ and $r_h$ removes all clauses from
{} $A_F \cup A_R \cup A_T \cup A_C \cup A_B$
but
{} $\neg x_h \vee t_h$,
{} $\neg t_1 \vee \cdots \vee \neg t_n \vee
{}  \neg c_1 \vee \cdots \vee \neg c_m \vee
{}  x_h \vee \neg r_h$
and all clauses
{} $\neg x_h \vee c_j$ with $x_h \in f_j$.
They only resolve in tautologies. Therefore, their resolution closure only
contains them. Removing 
{} $\neg t_1 \vee \cdots \vee \neg t_n \vee
{}  \neg c_1 \vee \cdots \vee \neg c_m \vee
{}  x_h \vee \neg r_h$
from the resolution closure leaves only
{} $\neg x_h \vee t_h$
and all clauses
{} $\neg x_h \vee c_j$ with $x_h \in f_j$.
They do not resolve since they do not contain opposite literals. Since
{} $\neg t_1 \vee \cdots \vee \neg t_n \vee
{}  \neg c_1 \vee \cdots \vee \neg c_m \vee
{}  x_h \vee \neg r_h$
is not contained in them, it is not entailed by them. This proves it
superirredundant. A similar replacement proves the superirredundancy of each
{} $\neg t_1 \vee \cdots \vee \neg t_n \vee
{}  \neg c_1 \vee \cdots \vee \neg c_m \vee
{}  e_h \vee \neg s_h$.

\end{itemize}

These points prove that the clauses
{} $A_F \cup A_T \cup A_C \cup A_B$
are superirredundant in the formula before forgetting and the clauses
{} $A_T \cup A_C \cup A_B$
are superirredundant in the formula after forgetting. The only clauses that may
be superredundant are $A_R$ in the formula after forgetting.

\

{\bf Formula $F$ is satisfiable.}

Let $M$ be a model satisfying $F$. Forgetting all variables except $X_C$ is
expressed by $A_R' \cup A_T \cup A_C \cup A_B$, where $A_R'$ comprises the
clauses $x_i \vee \neg q$ such that $M \models x_i$ and the clauses $e_i \vee
\neg q$ such that $M \models \neg x_i$. This Horn formula has size $k$. It
expresses forgetting because it is equivalent to $A_R \cup A_T \cup A_C \cup
A_B$. This is proved by showing that it entails every clause in $A_R$.

Since $M$ satisfies every clause $f_j \in F$, it satisfies at least a literal
of $f_j$: for some $x_i$, either $x_i \in f_j$ and $M \models x_i$ or $\neg x_i
\in f_j$ and $M \models \neg x_i$. By construction, $x_i \in f_j$ implies $\neg
x_i \vee c_j \in A_C$ and $\neg x_i \in f_j$ implies $\neg e_i \vee c_j \in
A_C$. Again by construction, $M \models x_i$ implies $x_i \vee \neg q \in A_R'$
and $M \models \neg x_i$ implies $e_i \vee \neg q \in A_R'$. As a result,
either
{} $x_i \vee \neg q \in A_R'$ and $\neg x_i \vee c_j \in A_C$
or 
{} $e_i \vee \neg q \in A_R'$ and $\neg e_i \vee c_j \in A_C$.
In both cases, the two clause resolve in $c_j \vee q$.

Since $M$ satisfies either $x_i$ or $\neg x_i$, either $x_i \vee \neg q \in
A_R'$ or $e_i \vee \neg q \in A_R'$. The first clause resolve with $\neg x_i
\vee t_i$ and the second with $\neg e_i \vee t_i$. The result is $t_i \vee \neg
q$ in both cases.

Resolving all these clauses $t_i \vee \neg q$ and $c_j \vee q$ with
{} $\neg t_1 \vee \cdots \vee \neg t_n \vee
{}  \neg c_1 \vee \cdots \vee \neg c_m \vee
{}  x_i \vee \neg r_i$
and then with $r_i \vee \neg q$, the result is $x_i \vee \neg q$. In the same
way, resolving these clauses with
{} $\neg t_1 \vee \cdots \vee \neg t_n \vee
{}  \neg c_1 \vee \cdots \vee \neg c_m \vee
{}  e_i \vee \neg s_i$
and $s_i \vee \neg q$ produces $e_i \vee \neg q$. This proves that all clauses
of $A_R$ are entailed.

\

{\bf Necessary clauses}

All CNF formulae that are equivalent to $A_R \cup A_T \cup A_C \cup A_B$ and
have minimal size contain
{} $A_T \cup A_C \cup A_B$
because these clauses are superirredundant. Therefore, these formulae are
{} $A_N \cup A_T \cup A_C \cup A_B$
for some set of clauses $A_N$. This set $A_N$ is now proved to contain either
{} $x_h \vee \neg q$,
{} $x_h \vee \neg r_h$,
{} $e_h \vee \neg q$,
{} $e_h \vee \neg s_h$
or
{} $t_h \vee \neg q$
for each index $h$. Let $M$ and $M'$ be the following models.

\begin{eqnarray*}
M &=&
\{x_i = e_i = t_i = \true \mid i \not= h\} \cup
\{x_h = e_h = t_h = \false\} \cup \\
&&
\{c_j = \true\} \cup
\{q = \true\} \cup
\{r_i = \true, s_i = \true\}
\\
M' &=&
\{x_i = e_i = t_i = \true \mid i \not= h\} \cup
\{x_h = e_h = t_h = \true \} \cup \\
&&
\{c_j = \true\} \cup
\{q = \true\} \cup
\{r_i = \true, s_i = \true\}
\end{eqnarray*}

The five clauses are falsified by $M$. Since the two of them $xh \vee \neg q$
and $e_h \vee \neg q$ are in $A_R$, this set is also falsified by $M$. As a
result, $M$ is not a model of
{} $A_R \cup A_T \cup A_C \cup A_B$.
This formula is equivalent to
{} $A_N \cup A_T \cup A_C \cup A_B$,
which is therefore falsified by $M$. In formulae,
{} $M \not\models A_N \cup A_T \cup A_C \cup A_B$.

The formula $A_N \cup A_T \cup A_C \cup A_B$ contains a clause falsified by
$M$. Since $M \models A_T \cup A_C \cup A_B$, this clause is in $A_N$ but not
in $A_T \cup A_C \cup A_B$. In formulae, $M \not\models c$ for some $c \in A_N$
and $c \not\in A_T \cup A_C \cup A_B$. This clause is entailed by
{} $A_R \cup A_T \cup A_C \cup A_B$
because this formula entails all of
{} $A_N \cup A_T \cup A_C \cup A_B$,
and $c$ is in $A_N$.
In formulae,
{} $A_R \cup A_T \cup A_C \cup A_B \models c$.

This clause $c$ contains either
{} $x_h$, $e_h$ or $t_h$.
This is proved by deriving a contradiction from the assumption that $c$ does
not contain any of these three literals. Since $M \not\models c$, the clause
$c$ contains only literals that are falsified by $M$. Not all of them: it does
not contain
{} $x_h$, $e_h$ and $t_h$
by assumption. It does not contain $\neg x_h$, $\neg e_h$ and $\neg t_h$ either
because it would otherwise be satisfied by $M$. As a result, $c$ is also
falsified by $M'$, which is the same as $M$ but for the values of
{} $x_h$, $e_h$ and $t_h$.
At the same time, $M'$ satisfies
{} $A_R \cup A_T \cup A_C \cup A_B$,
contradicting
{} $A_R \cup A_T \cup A_C \cup A_B \models c$.
This contradiction proves that $c$ contains either
{} $x_h$, $e_h$ or $t_h$.

From the fact that $c$ contains either $x_h$, $e_h$ or $t_h$, that is a
consequence of $A_R \cup A_T \cup A_C \cup A_B$, and that is in a minimal-size
formula, it is now possible to prove that $c$ contains either
{} $x_h \vee \neg q$,
{} $x_h \vee \neg r_h$,
{} $e_h \vee \neg q$,
{} $e_h \vee \neg s_h$
or
{} $t_h \vee \neg q$.

Since $c$ is entailed by
{} $A_R \cup A_T \cup A_C \cup A_B$,
a subset of $c$ follows from resolution from it:
{} $A_R \cup A_T \cup A_C \cup A_B \vdash c'$ with $c' \subseteq c$.
This implies
{} $A_N \cup A_T \cup A_C \cup A_B \models c'$
by equivalence. If $c' \subset c$, then
{} $A_N \cup A_T \cup A_C \cup A_B$
would not be minimal because it contained a non-minimal clause $c \in A_N$.
Therefore,
{} $A_R \cup A_T \cup A_C \cup A_B \vdash c$.

The only two clauses of
{} $A_R \cup A_T \cup A_C \cup A_B$
that contain $x_h$ are $x_h \vee \neg q$ and
{} $\neg t_1 \vee \cdots \vee \neg t_n \vee
{}   \neg c_1 \vee \cdots \vee \neg c_m \vee
{}   x_h \vee \neg r_h$.
They contain either $\neg q$ or $\neg r_h$. These literals are only resolved
out by clauses containing their negations $q$ and $r_h$. No clause contains $q$
and the only clause that contains $r_h$ is $r_h \vee \neg q$, which contains
$\neg q$. If a result of resolution contains $x_h$, it also contains either
$\neg q$ or $\neg r_h$. This applies to $c$ because it is a result of
resolution.

The same applies if $c$ contains $e_h$: it also contains either $\neg q$ or
$\neg s_i$.

The case of $t_h \in c$ is a bit different. The only two clauses of
{} $A_R \cup A_T \cup A_C \cup A_B$
that contain $t_h$ are $\neg x_h \vee t_h$ and $\neg e_h \vee t_h$. Since both
are in $A_T$ and $c \not\in A_T$, they are not $c$. The first clause $\neg x_h
\vee t_h$ only resolves with $x_i \vee \neg q$ or
{} $\neg t_1 \vee \cdots \vee \neg t_n \vee
{}   \neg c_1 \vee \cdots \vee \neg c_m \vee
{}   x_h \vee \neg r_h$,
but resolving with the latter generates a tautology. The result of resolving
$\neg x_h \vee t_h$ with $x_i \vee \neg q$ is $t_h \vee \neg q$; no clause
contains $q$. Therefore, $c$ can only be $t_h \vee \neg q$. The second clause
$\neg e_h \vee t_h$ leads to the same conclusion.

In summary, $c$ contains either
{} $x_h \vee \neg q$,
{} $x_h \vee \neg r_i$,
{} $e_h \vee \neg q$,
{} $e_h \vee \neg s_i$ or
{} $t_h \vee \neg q$.
In all these cases it contains at least two literals. This is the case for
every index $h$; therefore, $A_N$ contains at least $n$ clauses of two
literals. Every minimal CNF formula equivalent to
{} $A_R \cup A_T \cup A_C \cup A_B$
has size at least $2 \times n$ plus the size of $A_T \cup A_C \cup A_B$. This
sum is exactly $k$. This proves that every minimal CNF formula expressing
forgetting contains at least $k$ literal occurrences. Worded differently, every
CNF formula expressing forgetting has size at least $k$.

\

{\bf Formula $F$ is unsatisfiable}

The claim is that no CNF formula of size $k$ expresses forgetting if $F$ is
unsatisfiable. This is proved by deriving a contradiction from the assumption
that such a formula exists.

It has been proved that every CNF formula expressing forgetting is equivalent
to
{} $A_R \cup A_T \cup A_C \cup A_B$
and that the minimal equivalent CNF formulae are
{} $A_N \cup A_T \cup A_C \cup A_B$
for some set $A_N$ that contains clauses that include either
{} $x_h \vee \neg q$,
{} $x_h \vee \neg r_i$,
{} $e_h \vee \neg q$,
{} $e_h \vee \neg s_i$ or
{} $t_h \vee \neg q$
for each index $h$.

If $A_N$ contains other clauses, or more than one clause for each $h$, or these
clauses contain other literals, the size of
{} $A_N \cup A_T \cup A_C \cup A_B$
is larger than
{} $k = 2 \times n + ||A_T|| + ||A_C|| + ||A_B||$,
contradicting the assumption. This proves that every formula of size $k$ that
is equivalent to
{} $A_R \cup A_T \cup A_C \cup A_B$
is equal to
{} $A_N \cup A_T \cup A_C \cup A_B$
where $A_N$ contains exactly one clause among
{} $x_h \vee \neg q$,
{} $x_h \vee \neg r_i$,
{} $e_h \vee \neg q$,
{} $e_h \vee \neg s_i$ or
{} $t_h \vee \neg q$
for each index $h$.

The case
{} $x_h \vee \neg r_h \in A_N$
is excluded. It would imply
{} $A_R \cup A_T \cup A_C \cup A_B \models x_h \vee \neg r_h$,
which implies the redundancy of
{} $\neg t_1 \vee \cdots \vee \neg t_n \vee
{}   \neg c_1 \vee \cdots \vee \neg c_m \vee
{}   x_h \vee \neg r_h \in A_B$
contrary to its previously proved superirredundancy.
A similar argument proves
{} $e_h \vee \neg s_h \not\in A_N$.

The conclusion is that every formula of size $k$ that is equivalent to
{} $A_R \cup A_T \cup A_C \cup A_B$
is equal to
{} $A_N \cup A_T \cup A_C \cup A_B$
where $A_N$ contains exactly one clause among
{} $x_h \vee \neg q$,
{} $e_h \vee \neg q$,
{} $t_h \vee \neg q$
for each index $h$.

If $F$ is unsatisfiable, all such formulae are proved to be satisfied by a
model that falsifies
{} $A_R \cup A_T \cup A_C \cup A_B$,
contrary to the assumed equivalence.

Let $M$ be the model that assigns $q = \true$ and $t_i=\true$, and assigns
$x_i=\true$ and $e_i=\false$ if $x_i \vee \neg q \in A_N$ and $x_i=\false$ and
$e_i=\true$ if $e_i \vee \neg q \in A_N$ or $t_i \vee \neg q \in A_N$. All
clauses of $A_N$ and $A_T$ are satisfied by $M$.

This model $M$ can be extended to satisfy all clauses of $A_C \cup A_B$. Since
$F$ is unsatisfiable, $M$ falsifies at least a clause $f_j \in F$. Let $M'$ be
the model obtained by extending $M$ with the assignments of $c_j$ to false, all
other variables in $C$ to true and all variables $r_i$ and $s_i$ to true. This
extension satisfies all clauses of $A_B$ either because it sets $c_j$ to false
or because it sets $r_i$ and $s_i$ to true. It also satisfies all clauses of
$A_C$ that do not contain $c_j$ because it sets all variables of $C$ but $c_j$
to true.

The only clauses that remain to be proved satisfied are the clauses of $A_C$
that contain $c_j$. They are
{} $\neg x_i \vee c_j$ for all $x_i \in f_j$
and
{} $\neg e_i \vee c_j$ for all $\neg x_i \in f_j$.
Since $M'$ falsifies $f_j$, it falsifies every $x_i \in f_j$; therefore, it
satisfies $\neg x_i \vee c_j$. Since $M'$ falsifies $f_j$, it falsifies every
$\neg x_i \in f_j$; since by construction it assigns $e_i$ opposite to $x_i$,
it falsifies $e_i$ and therefore satisfies $\neg e_i \vee c_j$.

This proves that $M'$ satisfies
{} $A_N \cup A_T \cup A_C \cup A_B$.
It does not satisfy
{} $A_R \cup A_T \cup A_C \cup A_B$.
If $x_1 \vee \neg q \in A_N$, then $M'$ sets $x_1$ to true and $e_1$ to false;
therefore, it does not satisfy $e_1 \vee \neg q \in A_R$. Otherwise, $M'$ sets
$x_1$ to false and $e_1$ to true; therefore, it does not satisfy $x_1 \vee \neg
q$.

This contradicts the assumption that
{} $A_N \cup A_T \cup A_C \cup A_B$
is equivalent to
{} $A_R \cup A_T \cup A_C \cup A_B$.
The assumption that it has size $k$ is therefore false.~\qed

Forgetting variables from the formulae $A$ produced by the reduction can be
done in polynomial time, just not minimally. In other words, if the reduction
produces a formula $A$ and a set of variables $X_C$, forgetting $X_C$ from $A$
can always be expressed by a possibly non-minimal formula $B$ in polynomial
time. The complete translation from $F$ to $B$ and $k$ provides an alternative
proof of \np-hardness of the minimality problem without
forgetting~\cite{hamm-koga-93}: given a Horn formula $B$, is there any formula
of size $k$ equivalent to it?

In the other way around, the existing proof of \np-hardness of the minimality
problem without forgetting~\cite{hamm-koga-93} could be used to show that the
problem with forgetting is \np-hard. Yet, proving \np-hardness is not the final
aim of this section. It is the \Dp-hardness when the formula is minimal.
Minimality invalidates the existing proof. Lifting the \np-hardness reduction
to \Dp-hardness requires forgetting never be expressible in size less than $k$,
which the existing proof does not guarantee.

\

The problem of size after forgetting is the target of both a reduction from
propositional satisfiability and from propositional unsatisfiability. This
alone proves it both \np-hard and \conp-hard. These reductions have the
additional property that forgetting variables from the formulae they generate
cannot be expressed in size less than $k$. This allows merging them into a
single \Dp-hardness proof.

% in both reductions, the minimality of the Horn formula is proved by
% superirredundancy and is therefore ensured among all CNFs formulae, not just
% Horn formulae; but is not among arbitrary formulae

% in the second reduction the minimal size of formulae that express forgetting
% is by superirredundancy; therefore, it holds among both CNF and Horn
% formulae, but not among all formulae

% the formula of size k expressing forgetting, if any, is Horn in both cases;
% therefore, the proof holds both restricting it to be Horn or not

\begin{lemma}
\label{horn-hard}

Checking whether forgetting some variables from a minimal-size Horn formula is
expressed by a CNF or Horn formula bounded by a certain size is \Dp-hard.

\end{lemma}

\proof For every CNF formula $F$, Lemma~\ref{horn-conp} ensures the existence
of a minimal-size Horn formula $A$, a set of variables $X_A$ and an integer $k$
such that forgetting all variables except $X_A$ from $A$ is expressed by a Horn
formula of size $k$ if $F$ is unsatisfiable and is only expressed by larger CNF
formulae otherwise.

For every CNF formula $G$, Lemma~\ref{horn-np} ensures the existence of a
minimal-size Horn formula $B$, a set of variables $X_B$ and an integer $l$ such
that forgetting all variables except $X_B$ from $B$ is expressed by a Horn
formula of size $l$ if $G$ is satisfiable and is only expressed by larger CNF
formulae otherwise.

The prototypical \Dp-hard problem is that of establishing whether a formula $F$
is satisfiable and another $G$ is unsatisfiable. If the alphabets of the two
formulae $G$ and $F$ are not disjoint, they can be made so by renaming one of
them to fresh variables because renaming does not affect satisfiability. The
same applies to the formulae $B$ and $A$ respectively build from them according
to Lemma~\ref{horn-conp} and Lemma~\ref{horn-np} because renaming does not
change the minimal size of forgetting either. Lemma~\ref{independent} proves
that $A \cup B$ can be minimally expressed by $C \cup D$ where $C$ minimally
expresses forgetting from $A$ and $D$ from $B$. The size of these two formulae
are $l$ and $k$ if $G$ is unsatisfiable and $F$ satisfiable. If $G$ is
satisfiable, then $D$ is larger than $k$ while $C$ is still large at least $l$;
the minimal expression of forgetting $A \cup B$ is therefore strictly larger
than $k+l$. The same happens if $F$ is unsatisfiable.

This proves that the problem of checking the satisfiability of a formula and
the unsatisfiability of another reduces to the problem of checking the size of
the minimal expression of forgetting from Horn formulae.~\qed

Proving hardness takes most of this section, but still leaves a gap between the
complexity lower bound it shows and the upper bound in the next theorem. The
problem is \Dp-hard, which is just a bit above \np-hardness and \conp-hardness,
but belongs to a class of the next level of the polynomial hierarchy: \S{2}.

\begin{theorem}
\label{horn-complexity}

Checking whether forgetting some variables from a Horn formula is expressed by
a CNF or Horn formula bounded by a certain size expressed in unary is \Dp-hard
and in \S{2}, and remains hard even if the formula is restricted to be of
minimal size.

\end{theorem}

\proof The problem belongs to \S{2} because it can be expressed as the
existence of a formula of the given size or less that expresses forgetting the
given variables from the formula. In turn, expressing forgetting is by
Theorem~\ref{consistent-literals-complete} the same as the equiconsistency with
a set of literals containing all variables not to be forgotten. This condition
can be expressed by the following metaformula where $A$ is the formula,
$Y${\plural} are the variables not to be forgotten and $k$ the size bound.

\[
\exists B ~.~
	||B|| \leq k \mbox{ and }
	\forall S ~.~ \var(S) \subseteq Y \Rightarrow
		S \cup A \not\models \bot
		\Leftrightarrow
		S \cup B \not\models \bot
\]

Both $B$ and $S$ are bounded in size: the first by $k$, the second by the
number of variables in $Y$. Since consistency is polynomial for Horn formulae,
this is a $\exists\forall$QBF, which proves membership to \S{2}.

Hardness for \Dp\  is proved by Lemma~\ref{horn-hard}.~\qed

The assumption that the size bound is represented in unary is technical. When
formulated as a decision problem, the size of forgetting is the question
whether forgetting certain variables $X$ from a formula $A$ is expressed by a
formula of size $k$, but the actual problem is to find such a formula. If $k$
is exponential in the size of $A$, a formula of size $k$ may very well exist,
but is unpractical to represent. Unless $A$ is very small. The requirement that
$k$ is in unary forces the input of the problem to be as large as the expected
output. If the available space is enough for storing a resulting formula of
size $k$, it is also enough for storing an input string of length $k$, which
$k$ in unary is. In the other way around, representing $k$ in unary witnesses
the ability of storing a resulting formula of size $k$. The similar assumption
``$k$ is polynomial in the size of $A$'' fails to include the case where $A$ is
very small but the space available for expressing forgetting is large.

The problem is \Dp-hard, and belongs to \S{2}. Theorem~\ref{horn-complexity}
leaves a gap between the lower bound of \Dp\  and the upper bound of \S{2}.
According to what proved so far, the problem could be as easy as \Dp-complete
or as hard as \S{2}-complete. Nothing in the results obtained so far favors
either possibility. Actually, nothing indicates for certain that the problem is
complete for either class; it could be complete for any class in between, like
\Dlog{2}\  or \D{2}.

Anecdotal evidence hints that the problem is \S{2}-complete. The analogous
problems without forgetting for unrestricted formulae kept a gap between \np\ 
and \S{2} for twenty years before being closed as
\S{2}-complete~\cite{stoc-76,uman-01}. Proving membership was easy; proving
hardness was not.

This is a common pattern, not limited to formula minimization: in many cases,
hardness is more difficult to prove than membership. Not always, but hardness
proofs are often more complicated than membership proofs. The above lemmas are
an example: several pages of proof for hardness, ten lines for membership. A
proof of \S{2}-hardness may very well exist but is just difficult to find. As
it was for the problem without forgetting.

All of this is anecdotal. Technically, the complexity of the problem could be
anything in between \Dp\  and \S{2}.

As a personal opinion, not based on the technical results, the author of this
article would bet on the problem being \S{2}-complete. The missing proof of
\S{2}-hardness could be an extension of that of Lemma~\ref{horn-np}, since both
\S{2} and \np\  are based on an existential quantification. The extension of an
already difficult proof would be further complicated by the addition of an
inner universal quantification.

A way to partly close the issue is to further restrict the Horn case to make
the problem to be \D{2}-complete. The gap would close to its lower end for such
a class of formulae.

\section{Size after forgetting, general case}
\label{section-general}

The complexity analysis for general CNF formulae mimics that of the Horn case.
Two reductions prove the problem hard for the two basic classes of a level of
the polynomial hierarchy. They are merged into a single proof that slightly
increases the lower bound. A membership proof for a class of the next level
ends the analysis.

The difference is that the level of the polynomial hierarchy is the second
instead of the first. The two reductions prove the problem hard for \P{2} and
\S{2}. They are merged into a \Dptwo-hardness proof. Finally, the problem is
located within \S{3}.

As for the Horn case, the first lemma proves the problem \P{2}-hard, but is
formulated in terms of the reduction because the reduction is needed to raise
the lower bound to \Dptwo-hard.

% being proved by superirredundancy, the minimality of A is among CNF formulae

% the minimal size of formulae expressing forgetting when the formula is
% invalid is proved by the presence of literals; it therefore applies to every
% formula, not just CNFs

\begin{lemma}
\label{general-p2}

There exists a polynomial algorithm that turns a CNF formula $F$ into a
minimal-size CNF formula $A$, a subset $X_C \subseteq \var(A)$ and a number $k$
such that forgetting all variables from $A$ except $X_C$ is expressed by a CNF
formula of size $k$ if $\forall X \exists Y . F$ is valid and only by CNF
formulae of size $k+2$ or greater otherwise.

\end{lemma}

\proof Let $F = \{f_1, \ldots, f_m\}$ and $X = \{x_1,\ldots,x_n\}$. Checking
the validity of $\forall X \exists Y . F$ remains \P{2}-hard even if $F$ is
satisfiable: if $F$ is not satisfiable, $\forall X \exists Y . F$ can be turned
into the equivalent formula $\forall X \cup \{s\} \exists Y . s \vee F$, and $s
\vee F$ is satisfiable.

The reduction is based on an extended alphabet with the additional fresh
variables
{} $E = \{e_1,\ldots,e_n\}$,
{} $D = \{d_1,\ldots,d_m\}$ and
{} $\{a,b,q,r\}$.
The formula $A$, the set of variables $X_C$ and the number $k$ are:

\begin{eqnarray*}
A &=&
	\{ f_j \vee c_j \vee q \mid f_j \in F\} \cup			\\
&&
	\{\neg c_j \vee r \mid f_j \in F\} \cup				\\
&&
	\{\neg r \vee \neg a \vee b \vee q\} \cup			\\
&&
	\{a \vee \neg b \vee q\} \cup					\\
&&
	\{x_i \vee e_i \mid x_i \in X\}					\\
X_C &=& X \cup E \cup \{a,b,q\}						\\
k &=& 2 \times n + 3
\end{eqnarray*}

A short explanation of how the reduction works precedes its formal proof. The
key is how a model over $X \cup \{q\}$ extends to a model of $A$, in particular
its possible values of $a$ and $b$. All models over $X \cup \{q\}$ that satisfy
$q$ can be extended to satisfy $A$: all clauses not containing $q$ are
satisfied by setting $r=\true$ and $e_i$ opposite to $x_i$; satisfaction is not
affected by the values $a$ and $b$. The remaining models set $q = \false$. For
these models, the truth of a clause $f_j$ makes $f_j \vee c_j \vee q$ satisfied
even if $c_j = \false$. In turn, $c_j = \false$ satisfies
{} $\neg c_j \vee r$
even if $r = \false$, which satisfies
{} $\neg r \vee \neg a \vee b \vee q$
regardless of the values of $a$ and $b$; the values of $a$ and $b$ only need to
satisfy $a \vee \neg b \vee q$. Otherwise, the falsity of $f_j$ imposes $c_j =
\true$ to satisfy $f_j \vee c_j \vee q$, which makes
{} $\neg c_j \vee r$ 
require $r=\true$, which turns
{} $\neg r \vee \neg a \vee b \vee q$
into
{} $\neg a \vee b \vee q$,
making the literals $\neg a$ and $b$ necessary in addition to $a$ and $\neg b$.

% The only function of the clauses $x_i \vee e_i$ is to make all literals $x_i$
% necessary in $A$.

The proof comprises four steps: first, $A$ is proved minimal as required by the
claim of the lemma; second, $k$ literals that are in every formula that
expresses forgetting regardless of the validity of the QBF are identified;
third, a formula of size $k$ expressing forgetting when the QBF is valid is
determined; fourth, every formula expressing forgetting contains at least two
further literals if the QBF is invalid.

\

\noindent {\bf Minimality of $A$.}

Follows from Lemma~\ref{minimal} since all clauses of $A$ are superirredundant.
This is in turn proved by showing substitutions that disallow all resolutions,
which proves the superredundancy of the remaining clauses by
Lemma~\ref{set-value} and Lemma~\ref{no-resolution}.

The substitution that replaces with $\true$ the variables $a$, $b$, $r$, all
$e_i$ and all $c_j$ with $j \not= h$ removes all clauses but $f_h \vee c_h \vee
q$, which is therefore superirredundant.

The clauses $\neg c_j \vee r$ are proved superirredundant by substituting $q$
and $e_i$ with true, which removes all other clauses. The clauses $\neg c_j
\vee r$ do not resolve because they do not contain opposite literals.

Two other clauses are proved superirredundant by the substitution that replaces
all $e_i$ with $\true$, all $c_j$ with false, and $X \cup Y$ with some values
that satisfy $F$; such values exist because $F$ is by assumption satisfiable.
This substitution removes all clauses but
{} $\neg r \vee \neg a \vee b \vee q$
and
{} $a \vee \neg b \vee q$,
which only resolve in tautologies.

Finally, the clauses $x_h \vee e_h$ are proved superirredundant by replacing
$q$ and $r$ with $\true$, which removes all other clauses. Since the clauses
$x_h \vee e_h$ only contain positive literals, they do not resolve.

\

\noindent {\bf Necessary literals.}

Regardless of the validity of $\forall X \exists Y . F$, the literals $X \cup E
\cup \{a,\neg b,q\}$ are necessary in every CNF formula that expresses
forgetting all variables except $X_C$ from $A$. This is proved by
Lemma~\ref{forget-contains}, exhibiting a set of literals $S$ such that $S \cup
A$ is consistent, but $S \backslash \{l\} \cup \{\neg l\} \cup A$ is not for
every $l \in X \cup E \cup \{a,\neg b,q\}$.

The first set is
{} $S = \{x_i, \neg e_i, a, b, \neg q\}$,
which is consistent with $A$ because of the model that satisfies $S$ and
assigns $r$ and except variables $c_j$ to $\true$. Changing $x_i$ to $\neg x_i$
violates the clause $x_i \vee e_i$. Changing $a$ to false violates $a \vee \neg
b \vee q$. This proves that $a$ and all variables $x_i$ are necessary by
Lemma~\ref{forget-contains}.

The second set is
{} $S = \{\neg x_i, e_i, \neg a, \neg b, \neg q\}$,
which is consistent with $A$ because of the model that satisfies $S$ and
assigns $r$ and all variables $c_j$ to $\true$. Changing $e_i$ to $\neg e_i$
violates $x_i \vee e_i$, changing $b$ to true violates $a \vee \neg b \vee q$.
This proves that $e_i$ and $\neg b$ are necessary by
Lemma~\ref{forget-contains}.

The third set is
{} $\{S = x_i, e_i, \neg a, b, q\}$,
which is consistent with $A$ because of the model that satisfies $S$ and
assigns $r$ and all variables $c_j$ to $\true$. Changing $q$ to false violates
the clause $a \vee \neg b \vee q$, proving that $q$ is necessary.

In summary, all literals in $X \cup E \cup \{a, \neg b, q\}$ are in every CNF
formula expressing forgetting all variables except $X_C$ from $A$. These
literals are $2 \times n + 3$. This is a part of the claim: no CNF formula
expressing forgetting is smaller than $2 \times n + 3$.

\

\noindent {\bf Forgetting when $\forall X \exists Y . F$ is valid}

If $\forall X \exists Y . F$ is valid, forgetting is expressed by
{} $B = \{a \vee \neg b \vee q\} \cup \{x_i \vee e_i \mid x_i \in X\}$,
which has the required size $k = 2 \times n + 3$ and variables $XC = X \cup E
\cup \{a,b,q\}$. Theorem~\ref{consistent-literals-complete} proves that this
formula expresses forgetting: every set $S$ of literals of $X_C$ that contains
all variables of $X_C$ is consistent with $B$ if and only if it is consistent
with $A$.

Since $B$ only contains clauses of $A$, every set of literals $S$ that is
consistent with $A$ is also consistent with $B$. The claim follows from proving
the converse for every set of literals $S$ over $X_C$ that contains all
variables of $X_C$.

The assumption is that $S \cup B$ is consistent; the claim is that $S \cup A$
is consistent. Since $S \cup B$ is consistent, it has a model $M$. Let $M_X$ be
its restriction to the variables $X$ and $M_Y'$ to $Y$. By assumption, $\forall
X \exists Y . F$ is valid. Therefore, $M_X \cup M_Y$ satisfies $F$ for some
truth evaluation $M_Y$ over $Y$. Since $S$ is satisfied by $M$ and does not
contain any variable $Y$, it is also satisfied by $M \backslash M_Y' \cup M_Y$.
The truth evaluation
{} $M_C = \{c_j = \false \mid f_j \in F\} \cup \{r = \false\}$
satisfies all clauses $\neg c_j \vee r$ and $\neg r \vee \neg a \vee b \vee q$.
Since $M_X \cup M_Y$ satisfies all clauses $f_j \in F$, the union $M \backslash
M_Y' \cup M_Y \cup M_C$ satisfies all clauses $f_j \vee c_j \vee q$ of $A$.
This proves that $M \backslash M_Y' \cup M_Y \cup M_C \cup M_O$ satisfies all
clauses of $A$ that $B$ does not contain.

\

\noindent {\bf Forgetting when $\forall X \exists Y . F$ is invalid}

All CNF formulae that express forgetting have been proved to contain $X \cup E
\cup \{a, \neg b, q\}$. If $\forall X \exists Y . F$ is invalid, they all
contain $\neg a$ and $b$ as well.

This is proved by Lemma~\ref{forget-contains}: a set of literals $S$ over $X_C$
is shown to be consistent with $A$ while $S \backslash \{\neg a\} \cup \{a\}$
is not. A similar set is shown for $b$.

Since $\forall X \exists Y . F$ is invalid, for some model $M_X$ over $X$ the
model $M_X \cup M_Y$ falsifies $F$ for every model $M_Y$ over $Y$. The required
set $S$ is built from $M_X$: it contains the literals over $x_i$ that are
satisfied by $M_X$ and $\neg a$, $\neg b$ and $\neg q$.

\[
S =
\{x_i \mid M_X \models x_i\} \cup \{\neg x_i \mid M_X \models \neg x_i\}
\cup
\{\neg a, \neg b, \neg q\}
\]

By construction, $M_X$ satisfies the first part of $S$.
The model
{} $M_O = \{a = \false, b = \false, q = \false\}$
satisfies the second. Therefore, $M_X \cup M_O$ satisfies $S$.

The consistency of $S \cup A$ is shown by proving that $M_X \cup M_O$ can be
extended to the other variables to satisfy $A$. This extension is $M_X \cup M_Y
\cup M_O \cup M_N \cup M_C$, where $M_Y$ is an arbitrary model over $Y$, $M_N$
assigns every $e_i$ opposite to $x_i$ in $M_X$ and
{} $M_C$ is $\{c_j = \true \mid f_j \in F\} \cup \{r = \true\}$.
The clauses $f_j \vee c_j \vee q$ are satisfied because $c_j$ is true, the
clauses $\neg c_j \vee r$ because $r$ is $\true$, the clause $\neg r \vee \neg
a \vee b \vee q$ because $a$ is $\false$, $a \vee \neg b \vee q$ because $b$ is
$\false$, the clauses $x_i \vee e_i$ because $M_N \models e_i$ if $M_X
\not\models x_i$.

This proves that $M_X \cup M_O \cup M_N \cup M_C$ satisfies $S \cup A$, which
is therefore satisfiable.

The claim is a consequence of $S' = S \backslash \{\neg a\} \cup \{a\}$ being
inconsistent with $A$.

\[
S' =
\{x_i \mid M_X \models x_i\} \cup \{\neg x_i \mid M_X \models \neg x_i\}
\cup
\{a, \neg b, \neg q\}
\]

This is proved by contradiction: a model $M'$ is assumed to satisfy $S' \cup
A$. Since $M'$ satisfies $S'$, it assigns the variables $x_i$ the same as
$M_X$. Let $M_Y$ be the restriction of $M'$ to the variables $Y$. By
assumption, $M_X$ is a model over $X$ that cannot be extended to $Y$ to satisfy
$F$. As a result, $M_X \cup M_Y \not\models F$. Therefore, $M'$ falsifies at
least a clause $f_j \in F$. Since $M'$ satisfies $f_j \vee c_j \vee q$ but
falsifies both $f_j$ and $q$, it satisfies $c_j$. It also satisfies $r$ because
it satisfies $\neg c_j \vee r$ and falsifies $c_j$. Since $M'$ satisfies $S'$
it satisfies $a$ and falsifies $b$ and $q$. The conclusion is that all literals
of $\neg r \vee \neg a \vee b \vee q \in A$ are false, contrary to the
assumption that $M'$ satisfies $A$.

A similar set $S$ with $a$ and $b$ in place of $\neg a$ and $\neg b$ proves
that expressing forgetting also requires $b$.~\qed

The second lemma is again about a reduction. Its statement implies that the
problem is \S{2}-hard, but it predicates about the reduction rather than the
hardness. This allows it to be merged with the first lemma into a proof of
\Dptwo-hardness. The existing proof of \S{2}-hardness of the problem without
forgetting~\cite{uman-01} also proves the problem with forgetting \S{2}-hard,
but does not allow such a merging and does not hold in the restriction of
minimal formulae.

% the minimality of A is proved by superirredundancy; the calculation of the
% minimal size of formulae expressing forgetting involves superirredundancy;
% therefore, minimality is among CNF formulae in both cases, not among
% arbitrary formulae

\begin{lemma}
\label{general-s2}

There exists a polynomial algorithm that turns a DNF formula $F = f_1 \vee
\cdots \vee f_m$ over variables $X \cup Y$ into a minimal-size CNF formula $A$,
a subset $X_C \subseteq \var(A)$ and a number $k$ such that forgetting all
variables except $X_C$ from $A$ is expressed by a CNF formula of size $k$ if
$\exists X \forall Y . F$ is valid, and only by larger CNF formulae otherwise.

\end{lemma}

\proof Let $F = f_1 \vee \cdots \vee f_m$ be the DNF formula over variables $X
\cup Y$. The reduction employs additional variables:
{} $O = \{o_i \mid x_i \in X\}$,
{} $E = \{e_i \mid x_i \in X\}$,
{} $P = \{p_i \mid x_i \in X\}$,
{} $T = \{t_i \mid x_i \in X\}$,
{} $D = \{d_j \mid f_j \in F\}$,
{} $R = \{r_i \mid x_i \in X\}$,
{} $S = \{s_i \mid x_i \in X\}$
and $q$. The formula $A$, the alphabet $X_C$ and the number $k$ are as follows.

\begin{eqnarray*}
A &=& A_F \cup A_T \cup A_D \cup A_B					\\
A_F &=&
\{ x_i \vee \neg o_i, o_i \vee q \mid x_i \in X \} \cup
\{ e_i \vee \neg p_i, p_i \vee q \mid x_i \in X \}			\\
A_T &=& \{ \neg x_i \vee t_i ,~ \neg e_i \vee t_i \mid x_i \in X \}	\\
A_D &=& \{ \neg (f_j[e_i/\neg x_i]) \vee d_j \mid f_j \in F \}		\\
A_B &=&
\{ \neg t_1 \vee \cdots \vee \neg t_n \vee \neg d_j \vee x_i \vee \neg r_i,
   r_i \vee q \mid x_i \in X ~, f_j \in F\} \cup			\\
&&
\{ \neg t_1 \vee \cdots \vee \neg t_n \vee \neg d_j \vee e_i \vee \neg s_i,
   s_i \vee q \mid x_i \in X ,~ f_j \in F\}				\\
X_C &=& X \cup E \cup Y \cup T \cup D \cup R \cup S \cup \{q\}		\\
k &=& 2 \times n + ||A_T \cup A_D \cup A_B||
\end{eqnarray*}

% $\neg f_j[e_i/\neg x_i]$ is $f_l$ with every $\neg x_i$ replaced by $e_i$,
% the result negated and turned into a disjunction; therefore, it contains
% $x_i$ and $e_i$ only negated; it is not the converse: first negation and
% conversion into a disjunction, then substitution; that would still allow a
% model over $X$ to be extended to $E$ by setting each $e_i$ opposite to $x_i$,
% but would not allow the resulting disjunction to resolve with $x_i \vee q$ or
% $e_i \vee q$

% forgetting is proved to be expressed by the following formula, which may not
% be minimal
% \[ A_R = \{x_i \vee q, e_i \vee q \mid x_i \in X\} \]

The reduction works because every minimal CNF formula that expresses forgetting
contains at least one among $x_h \vee q$, $e_h \vee q$ and $t_h \vee q$ for
each $h$, and all of $A_T \cup A_D \cup A_B$. This proves the lower bound $k$.
If the QBF is valid, for some evaluation over $X$ the formula $F$ is true
regardless of $Y$. Choosing the clauses $x_h \vee q$, $e_h \vee q$ or $t_h \vee
q$ that correspond to this model, some clause $A_D$ imply $q \vee d_j$, which
allows $A_B$ to entail all remaining clauses. If the QBF is not valid, no
clause $q \vee d_j$ is entailed.

The formal proof requires five steps: first, every formula expressing
forgetting is equivalent to a certain formula $A_R \cup A_T \cup A_D \cup A_B$;
second, $A$ is a minimal CNF formula and the clauses of $A_T \cup A_D \cup A_B$
are in all minimal CNF formulae equivalent to $A_R \cup A_T \cup A_D \cup A_B$;
third, forgetting is expressed by a formula of size $k$ if the QBF is valid;
fourth, every minimal CNF formula expressing forgetting contains either $x_h
\vee q$, $e_h \vee q$ or $t_h \vee q$ for each $h$; fifth, if the QBF is
invalid then forgetting is only expressed by formulae larger than $k$.

\

{\bf Effect of forgetting.}

The variables to forget are $O \cup P$. Each is contained only in two clauses
of $A$, with opposite signs. Resolving them produces the clauses in the
following set $A_R$.

\[
A_R = \{x_i \vee q, e_i \vee q \mid x_i \in X\}
\]

By Theorem~\ref{resolve-out}, forgetting is expressed by $A_R \cup A_T \cup A_D
\cup A_B$. Therefore, all formulae that express forgetting are equivalent to
this formula.

\

{\bf Superirredundancy.}

All clauses of
{} $A_F \cup A_T \cup A_D \cup A_B$
are proved superirredundant in
{} $A_F \cup A_R \cup A_T \cup A_D \cup A_B$.
Both $A$ and
{} $A_R \cup A_T \cup A_D \cup A_B$
are subsets of this formula; therefore, the superirredundant clauses are
superirredundant in both formulae by Lemma~\ref{superset}. Since $A$ comprises
exactly them, it is minimal thanks to Lemma~\ref{minimal}. Since all formulae
expressing forgetting are equivalent to
{} $A_R \cup A_T \cup A_D \cup A_B$,
where
{} $A_T \cup A_D \cup A_B$
are superirredundant, these clauses are in all formulae expressing forgetting.

Superirredundancy is proved applying a substitution to the formula so that the
resulting clauses do not resolve and are not contained in one another. This
condition proves them superirredundant by Lemma~\ref{no-resolution}.
Lemma~\ref{set-value} implies their superirredundancy in the original formula.

Replacing all variables $X$, $E$, $T$ and $D$ with $\true$ removes from the
formula
{} $A_F \cup A_R \cup A_T \cup A_D \cup A_B$
all clauses but
{} $o_i \vee q$, $p_i \vee q$, $r_i \vee q$ and $s_i \vee q$.
These clauses do not resolve because they only contain positive literals. None
is contained in another.

Replacing all variables $R$ and $S$ with $\false$ and all variables $T$, $D$
and $q$ with $\true$ removes from the formula
{} $A_F \cup A_R \cup A_T \cup A_D \cup A_B$
all clauses but the clauses $x_i \vee \neg o_i$ and $e_i \vee \neg p_i$. They
are not contained in one another; they do not resolve because they do not
contain opposite literals.

Replacing all variables $O$, $P$, $R$ and $S$ with false and $D$ and $q$ with
$\true$ removes all clauses but $\neg x_i \vee t_i$ and $\neg e_i \vee t_i$.
These clauses do not resolve because they do not contain opposite literals;
they are not contained in one another.

Replacing all variables $O$, $P$, $R$ and $S$ with $\false$ and $T$, $D
\backslash \{d_h\}$ and $q$ with $\true$ removes all clauses but
{} $(\neg f_h[e_i/\neg x_i]) \vee d_h$,
which is therefore superirredundant.

The last substitution replaces all variables $X \backslash \{x_h\}$, $E$, $O$,
$P$, $R \backslash \{r_h\}$ and $S$ with $\false$, all variables $D \backslash
\{d_l\}$ and $q$ with $\true$, all variables $y_i$ such that $y_i \in \neg
f_l[e_i/\neg x_i]$ to $\true$ and all such that $\neg y_i \in \neg f_l[e_i/\neg
x_i]$ to $\false$. This substitution removes all clauses but
{} $\neg x_h \vee t_h$,
{} $\neg t_1 \vee \cdots \vee \neg t_n \vee \neg d_l \vee x_h \vee \neg r_i$
and possibly
{} $\neg (f_l[e_i/\neg x_i]) \vee d_l$.
The latter clause is removed if it contains some variable $y_i$. It is removed
if it contains some literal $\neg x_i$ with $i \not= h$. It is removed if it
contains some literal $\neg e_i$. The only other literals it may contain are
$\neg x_h$ and $d_l$; it contains both: $d_l$ by construction, $\neg x_h$
because otherwise $f_l$ would be empty. The remaining clauses are therefore
{} $\neg x_h \vee t_h$,
{} $\neg t_1 \vee \cdots \vee \neg t_n \vee \neg d_l \vee x_i \vee \neg r_i$
and possibly
{} $\neg x_h \vee d_h$.
These clauses only resolve in tautologies, which proves the second
superirredundant. A similar argument holds for
{} $\neg t_1 \vee \cdots \vee \neg t_n \vee \neg d_l \vee e_i \vee \neg s_i$.

\

{\bf Validity of $\exists X \forall Y . F$.}

Let $M$ be a model over variables $X$ that makes $F$ true regardless of the
values of $Y$. Let $A_R' \subseteq A_R$ be the set of clauses $x_i \vee q$ such
that $M \models x_i$ and $e_i \vee q$ such that $M \models \neg x_i$. This set
has size $2 \times n$. Therefore, $A_R' \cup A_T \cup A_D \cup A_B$ has size $k
= 2 \times n + ||A_T \cup A_D \cup A_B||$. This formula expresses forgetting if
it is equivalent to $A_R \cup A_T \cup A_D \cup A_B$, which is the case if
$A_R' \cup A_T \cup A_D \cup A_B \models A_R$. The claim is proved by showing
that $A_R' \cup A_T \cup A_D \cup A_B$ entails $A_R$.

Either $x_i \vee q$ or $e_i \vee q$ is in $A_R'$ for every $i$ and these
clauses respectively resolve with $\neg x_i \vee t_i$ and $\neg e_i \vee t_i$,
producing $t_i \vee q$ in both cases. Each clause
{} $\neg t_1 \vee \cdots \vee \neg t_n \vee \neg d_j \vee x_i \vee \neg r_i$
resolve with them and with $r_i \vee q$ to $\neg d_j \vee x_i \vee q$. This
clause further resolves with
{} $\neg (f_j[e_i/\neg x_i]) \vee d_j$
to produce
{} $\neg (f_j[e_i/\neg x_i]) \vee x_i \vee q$.
This proves that 
{} $A_R' \cup A_T \cup A_D \cup A_B$ implies every clause
{} $\neg (f_j[e_i/\neg x_i]) \vee x_i \vee q$ with $f_j \in F$.
The following equivalence holds.

\ttytex{
\begin{eqnarray*}
\{ \neg (f_j[e_i/\neg x_i]) \vee x_i \vee q | f_j \in F \}
& \equiv &
	\left(\bigwedge \{\neg (f_j[e_i/\neg x_i]) \mid f_j \in F\} \right)
	\vee x_i \vee q							\\
& \equiv &
	\neg \left(\bigvee \{f_j[e_i/\neg x_i] \mid f_j \in F\} \right)
	\vee x_i \vee q							\\
& \equiv & \neg F[e_i/\neg x_i] \vee x_i \vee q
\end{eqnarray*}
}{
{ -fj[ni/-xi] v xi v q | fj in F } =
                           = &{ -fj[ni/-xi] | fj in F } v xi v q
                           = -(v { fj[ni/-xi] | fj in F }) v xi v q
                           = -F[ni/-xi] v x_i v q
}

Since $A_R' \cup A_T \cup A_D \cup A_B$ implies the first set, it implies the
last formula: $A_R' \cup A_T \cup A_D \cup A_B \models \neg F[e_i/\neg x_i]
\vee x_i \vee q$.

Since $M$ satisfies $F$ regardless of $Y$, it follows that
{} $\{x_i \mid M \models x_i\} \cup \{\neg x_i \mid M \models \neg x_i\}
{}  \models F$.
Replacing each $\neg x_i$ with $e_i$ in both sides of this entailment turns
into
{} $\{x_i \mid M \models x_i\} \cup \{e_i \mid M \models \neg x_i\}
{}  \models F[e_i/\neg x_i]$.
Disjoining both terms with $q$ results into
{} $A_R' \models F[e_i/\neg x_i] \vee q$.

This entailment and the previously proved
{} $A_R' \cup A_T \cup A_D \cup A_B \models
{}  \neg F[e_i/\neg x_i] \vee x_i \vee q$
imply $A_R' \cup A_T \cup A_D \cup A_B \models x_i \vee q$.

The same holds for $e_i \vee q$ by symmetry. Therefore, $A_R' \cup A_T \cup A_D
\cup A_B$ implies every clause of $A_R$.

\

{\bf Necessary clauses.}

All formulae that express forgetting are equivalent to $A_R \cup A_T \cup A_D
\cup A_B$ and therefore contain all its superirredundant clauses $A_T \cup A_D
\cup A_B$. As a result, they have the form $A_N \cup A_T \cup A_D \cup A_B$ for
some set of clauses $A_N$. It is now shown that all equivalent CNF formulae of
minimal size contain a clause that include either
{} $x_h \vee q$,
{} $x_h \vee \neg r_h$,
{} $e_h \vee q$,
{} $e_h \vee \neg s_h$,
or
{} $t_h \vee q$
for each $h$.

Since
{} $A_N \cup A_T \cup A_D \cup A_B$
is equivalent to
{} $A_R \cup A_T \cup A_D \cup A_B$,
it entails
{} $x_h \vee q \in A_R$.
This clause is not satisfied by the following model.

\begin{eqnarray*}
M &=&
\{x_i = e_i = t_i = \true \mid i \not= h\}		\cup
\{x_h = e_h = t_h = \false\}				\cup		\\
&&
\{d_j = \true \mid f_j \in F\}				\cup
\{r_i = s_i = \true\}					\cup
\{q = \false\}
\end{eqnarray*}

This model satisfies all clauses of $A_T \cup A_D \cup A_B$. If $A_N$ also
satisfied it,
{} $A_N \cup A_T \cup A_D \cup A_B$
would have a model that falsifies $x_h \vee q$, which it instead entails. As a
result, $A_N$ contains a clause $c$ that $M$ falsifies. Since
{} $A_N \cup A_T \cup A_D \cup A_B$
is a formula of minimal size, it entails no proper subset of $c$. By
equivalence, the same applies to
{} $A_R \cup A_T \cup A_D \cup A_B$.

\begin{eqnarray*}
&& M \not\models c				\\
&& A_R \cup A_T \cup A_D \cup A_B \models c	\\
&& A_R \cup A_T \cup A_D \cup A_B \models c'
   \mbox{ implies } c' \not\subset c
\end{eqnarray*}

If $c$ contains neither $x_h$, $e_h$ nor $t_h$, it would still be falsified by
the model that is the same as $M$ except that it assigns $x_h$, $e_h$ and $t_h$
to $\true$. This model satisfies
{} $A_R \cup A_T \cup A_D \cup A_B$.
As a result,
{} $A_R \cup A_T \cup A_D \cup A_B \cup \neg(c)$
is consistent, contradicting $A_R \cup A_T \cup A_D \cup A_B \models c$. This
proves that $c$ contains either $x_h$, $e_h$ or $t_h$.

Since these three variables are negative in $M$ and $M \not\models c$, they are
positive in $c$. In other words, $c$ contains either $x_h$, $e_h$ or $t_h$
unnegated.

Since $c$ is entailed by $A_R \cup A_T \cup A_D \cup A_B$, but none of its
proper subsets does, it follows from resolution:
{} $A_R \cup A_T \cup A_D \cup A_B \vdash c$.

If $c$ contains $x_h$, it also contains either $q$ or $\neg r_h$. This is
proved as follows. Since $c$ is the tree of a resolution tree and contains
$x_h$, this literal is also in one of the leaves of resolution. The only
clauses of
{} $A_R \cup A_T \cup A_D \cup A_B$
containing $x_h$ are $x_h \vee q$ and all clauses
{} $\neg t_1 \vee \cdots \vee \neg t_n \vee \neg d_j \vee x_h \vee \neg r_h$.
The first does not resolve over $q$ because the formula does not contain $\neg
q$. The other clauses only resolve over $r_h$ with $r_h \vee q$, which
introduces $q$, which again cannot be removed by resolution. Since $c$ is
obtained by resolution, if it contains $x_h$ it also contains either $\neg r_h$
or $q$.

By symmetry, if $c$ contains $e_h$ it also contains either $\neg s_h$ or $q$.

The other case is that $c$ contains $t_h$. The only clauses of
{} $A_R \cup A_T \cup A_D \cup A_B$
that contain $t_h$ are $\neg x_h \vee t_h$ and $\neg e_h \vee t_h$. These
clauses are satisfied by $M$ while $c$ is not, therefore $c$ is not one of
them. The first clause $\neg x_h \vee t_h$ only resolves over $x_h$ with $x_h
\vee q$ and all clauses
{} $\neg t_1 \vee \cdots \vee \neg t_n \vee \neg d_j \vee x_h \vee \neg r_h$,
but resolving with the latter only generates tautologies. Therefore, the first
step of resolution is necessarily $\neg x_h \vee t_h, x_h \vee q \vdash t_h
\vee q$. Since the formula does not contain $q$, every clause obtained from
resolution that contains $t_h$ also contains $q$. This also includes $c$. The
same holds for symmetry for $\neg e_h \vee t_h$.

This proves that every minimal-size CNF formula expressing forgetting contains
a clause that includes either
{} $x_h \vee q$,
{} $x_h \vee \neg r_h$,
{} $e_h \vee q$,
{} $e_h \vee \neg s_h$,
or
{} $t_h \vee q$
for each $h$.

\

{\bf Falsity of $\exists X \forall Y . F$.}

The falsity of $\exists X \forall Y . F$ contradicts the existence of a
minimal-size CNF formula of size $k$ expressing forgetting. The relevant
results proved so far are: every CNF formula expressing forgetting has size $k$
or more and is equivalent to $A_R \cup A_T \cup A_D \cup A_B$; the minimal-size
such formulae are $A_N \cup A_T \cup A_D \cup A_B$ where $A_N$ contains a
clause that includes either
{} $x_h \vee q$,
{} $x_h \vee \neg r_h$,
{} $e_h \vee q$,
{} $e_h \vee \neg s_h$,
or
{} $t_h \vee q$
for each $h$.

A formula $A_N \cup A_T \cup A_D \cup A_B$ of size $k$ expressing forgetting,
if any, is minimal since no smaller formula expresses forgetting. Therefore,
$A_N$ includes a clause containing one of the five disjunctions for each $h$.
Since these are not in $A_T \cup A_D \cup A_B$, the size of such formulae is $k
= 2 \times n + ||A_T \cup A_D \cup A_B||$ if every clause of $A_N$ is exactly
one of the above disjunction for each $h$. If $A_N$ contains more than one
clause for some $h$ or the clause for some $h$ contains more than two literals
or $A_N$ contains other clauses, the formula is not minimal.

The case
{} $x_h \vee \neg r_h \in A_N$
can be excluded: it makes
{} $\neg t_1 \vee \cdots \vee \neg t_n \vee \neg d_j \vee x_h \vee \neg r_h \in
{}   A_N$
redundant in
{} $A_N \cup A_T \cup A_D \cup A_B$,
contradicting the minimality of this formula. The case $e_h \vee \neg s_h \in
A_N$ is excluded in the same way.

These exclusion leaves $A_N$ to contain exactly one among
{} $x_h \vee q$,
{} $e_h \vee q$,
and
{} $t_h \vee q$
for each $h$ and nothing else.

The final step of the proof is that no such $A_N$ makes
{} $A_N \cup A_T \cup A_D \cup A_B$
equivalent to
{} $A_R \cup A_T \cup A_D \cup A_B$
if $\exists X \forall Y . F$ is invalid. Nonequivalence is proved by exhibiting
a model of the first formula that does not satisfy the second.

Let $M_X$ be the model over $X$ that contains $x_i = \true$ if $x_i \vee q \in
A_N$ and $x_i = \false$ otherwise. Let $M_N$ be the model that assigns every
$e_i$ opposite to $x_i$ and $M_T=\{t_i = \true\}$. By construction, $M_X \cup
M_N \cup M_T \cup \{q=\false\}$ satisfies all clauses of $A_N$. It also
falsifies either $x_i \vee q$ or $e_i \vee q$ for each $i$ because it assigns
$\false$ to $q$ and to either $x_i$ or $e_i$. It therefore falsifies $A_R$.

Since $\exists X \forall Y . F$ is invalid, every model over $X$ falsifies $F$
with a model over $Y$. Let $M_Y$ be the model over $Y$ such that $M_X \cup M_Y
\models \neg F$. Since $F = f_1 \vee \cdots \vee f_m$, it holds $M_X \cup M_Y
\models \neg f_j$ for every $f_j \in F$. It follows $M_X \cup M_Y \cup M_N
\models \neg f_j[e_i/\neg x_i]$ since $M_N$ assigns every $e_i$ opposite to
$x_i$.

Merging the results proved in the latter two paragraphs, $M_X \cup M_T \cup
\{q=\false\} \cup M_N \cup M_Y$ satisfies both $A_N$ and $\neg f_j[e_i/\neg
x_i]$ for every $f_j \in F$.

This model can be extended to a model of $A_N \cup A_T \cup A_D \cup A_B$ by
adding $M_O = \{d_j = \false\} \cup \{r_i = \true\} \cup \{s_i = \true\}$. The
clauses of $A_N$ are already proved satisfied. The clauses $\neg x_i \vee t_i
\in A_T$ are satisfied because $M_T$ contains $t_i = \true$. The clauses $(\neg
f_j[e_i/\neg x_i]) \vee d_j$ are satisfied because $\neg f_j[e_i/\neg x_i]$ is.
The clauses of $A_B$ are satisfied because each contains either $\neg d_j$,
$r_i$ or $s_i$, and these literals are true in $M_O$.

This proves that $M_X \cup M_T \cup \{q=\false\} \cup M_N \cup M_Y \cup M_O$
satisfies $A_N \cup A_T \cup A_D \cup A_B$. It does not satisfy $A_R$, which
means that it falsifies $A_R \cup A_T \cup A_D \cup A_B$. This proves that
{} $A_N \cup A_T \cup A_D \cup A_B$
is not equivalent to
{} $A_R \cup A_T \cup A_D \cup A_B$.

In summary, assuming that the QBF is not valid and that a CNF formula of size
$k$ expresses forgetting, it is proved that the formula does not express
forgetting. This contradiction shows that no formula of size $k$ expresses
forgetting if the QBF is not valid.~\qed

As anticipated, the two reductions merge into one that proves the problem of
forgetting size \Dptwo-hard.

\begin{lemma}
\label{general-hard}

Checking whether forgetting a given set of variables from a minimal-size CNF
formula is expressed by a CNF formula bounded by a certain size is \Dptwo-hard.

\end{lemma}

\proof For every $\forall$QBF Lemma~\ref{general-p2} ensures the existence of a
minimal-size CNF formula $A$, a set of variables $X_A$ and an integer $k$ such
that forgetting all variables except $X_A$ from $A$ is expressed by a CNF
formula of size $k$ if the QBF is valid and is only expressed by larger CNF
formulae otherwise.

For every $\exists$QBF Lemma~\ref{general-s2} ensures the existence of a
minimal-size CNF formula $B$, a set of variables $X_B$ and an integer $l$ such
that forgetting all variables except $X_B$ from $B$ is expressed by a CNF
formula of size $l$ if the QBF is valid and is only expressed by larger CNF
formulae otherwise.

% the class \Dptwo\  is defined as the set of problems that are the
% intersection of a \S{2} problem and a \P{2} problem; since $\exists$QBF is
% hard for \S{2} and $\forall$QBF is hard for \P{2}, their combination is hard
% for \Dptwo

A \Dptwo-hard problem is that of establishing whether an $\exists$QBF and a
$\forall$QBF are both valid. If their alphabets are not disjoint they can be
made so by renaming one of them to fresh variables since renaming does not
affect validity. The same applies to the formulae $B$ and $A$ respectively
build from them according to Lemma~\ref{general-p2} and Lemma~\ref{general-s2}
because renaming does not change the minimal size of forgetting either.
Lemma~\ref{independent} proves that forgetting from $A \cup B$ is expressed by
$C \cup D$ where $C$ expresses forgetting from $A$ and $D$ from $B$. The
minimal size of two such CNF formulae are respectively $k$ and $l$. If the QBFs
are both valid, they are exactly $k$ and $l$ large. Otherwise, they are
strictly larger than either $k$ or $l$. The sum is $k+l$ if both QBFs are valid
and is larger than $k+l$ otherwise.~\qed

The next theorem adds a complexity class membership to the hardness of the
problem of size of forgetting proved in the previous lemma.

\begin{theorem}
\label{general-complexity}

Checking whether forgetting some variables from a CNF formula is expressed by a
CNF formula of a certain size expressed in unary is \Dptwo-hard and in \S{3},
and remains hard even if the CNF formula is restricted to be of minimal size.

\end{theorem}

\proof Membership to \S{3} is proved first. The problem is the existence of a
CNF formula of the given size or less that expresses forgetting the given
variables from the formula. Theorem~\ref{consistent-literals-complete}
reformulates forgetting in terms of equiconsistency with a set of literals
containing all variables not to be forgotten. Forgetting withing a certain size
is formalized by the following metaformula where $A$ is the formula, $Y$ the
variables not to be forgotten and $k$ the size bound.

\[
\exists B ~.~
	\var(B) \subseteq Y ,~
	||B|| \leq k \mbox{ and }
	\forall S ~.~ \var(S) \subseteq Y \Rightarrow
		\exists M ~.~ M \models S \cup A
		\Leftrightarrow
		\exists M' ~.~ M' \models S \cup B
\]

All four quantified entities are bounded in size: $B$ by $k$, $S$ and $M'$ by
the number of variables in $Y$ and $M$ by the number of variables in $A$. This
is therefore a $\exists\forall\exists$QBF, which proves membership to \S{3}.

Hardness to \Dptwo\  is proved by Lemma~\ref{general-hard} in the restriction
where $A$ is minimal.~\qed

The technical assumption that the size bound is expressed in unary is the same
as in Theorem~\ref{horn-complexity}; it is motivated after that theorem.

%input{split.tex}
%input{python.tex}
%input{other.tex}
\section{Conclusions}

Forgetting variables from formulae may increase size, instead of decreasing it.
This phenomenon is already recognized as a
problem~\cite{delg-wang-15,bert-etal-19}. Deciding whether it takes place or
not for a specific formula and variables to forget is difficult. While checking
inference is polynomial in the Horn case, checking whether forgetting is
expressed by a formula of a certain maximal size is at least \Dp-hard, which
implies it both \np-hard and \conp-hard; the same for the general case, where
inference is \conp-hard but checking size after forgetting is at least
\Dptwo-hard.

The precise characterization of complexity is an open problem. For Horn
formulae the problem is proved \Dp-hard and is in \S{2}, which leaves a large
gap between the lower and the upper bound. The problem could be complete for
\Dp. Or for \S{2}. Or for a class in between, like \Dlog{2} or \D{2}. The same
for the general case, where the problem is in \S{3} but only hard for \Dptwo.
The definition of the problem suggests that the actual complexity is the same
as its upper bounds, \S{2}-complete and \S{3}-complete, but this is yet to be
proved. Finding a subcase where the upper bound lowers to the hardness level
would also be of interest.

While the problem is investigated in the general propositional case and in its
Horn restriction, many other subcases of propositional logic are relevant.
Forgetting is very easy on formulae in DNNF~\cite{darw-01}, as it amounts to
simply removing literals. It is also easy for the Krom
restriction~\cite{wang-15} as resolving binary clauses always produces binary
clauses, which are at most quadratically many. It may not on other tractable
cases in Post's lattice~\cite{nord-zanu-08}.

Forgetting has variants and is defined for many logics other than propositional
logic. The problem of size applies to all of them. What is its complexity? This
article characterizes it for one version of forgetting in propositional logic.
The other versions and the other logics are still open. Some results may apply
to them as well. Logic programs embed Horn clauses; the hardness results for
the Horn case may hold for them as well. More generally, how hard it is to
check whether forgetting in logic programs is expressed within a certain size?
How hard it is in first-order logic? In description logics? How hard it is when
forgetting literals rather than variables?

The size after forgetting matters not only when forgetting variables, but also
literals~\cite{lang-etal-03}, possibly with varying variables~\cite{moin-07}.
All variants inhibit the values of some variables to matter: forgetting
variables makes their values irrelevant to the satisfaction of the formula;
forgetting literals makes only the true or false value not to matter; varying
variables allows some other variables to change. These variants generalize
forgetting variables, inheriting the problem of size with the same complexity
at least.

Forgetting applies to frameworks other than propositional logic. The problem of
size applies to them as well.

Forgetting from logic programs~\cite{wang-etal-05,wang-etal-14,gonc-etal-16} is
usually backed by the need of solving conflicts rather than an explicit need of
reducing size. Yet, an increase in size is recognized as a problem:
``Whereas relying on such methods to produce a concrete program is important in
the sense of being a proof that such a program exists, it suffers from several
severe drawbacks when used in practice: In general, it produces programs with a
very large number of rules''~\cite{bert-etal-19};
``It can also be observed that forgetting an atom results in at worst a
quadratic blowup in the size of the program. [...] While this may seem
comparatively modest, it implies that forgetting a set of atoms may result in
an exponential blowup''~\cite{delg-wang-15}.

Another common area of application of forgetting is first-order
logic~\cite{lin-reit-94}. Size after forgetting is related to bounded
forgetting~\cite{zhou-zhan-11}, which is forgetting with a constraint on the
number of nested quantifiers. The difference is that the bound is an additional
constraint rather than a limit to check. Bounded forgetting still involves a
measure (the number of quantifiers), but forcing the result by that measure
makes it close to bounding {\pspace} problems~\cite{libe-05}. Enforcing size
rather than checking it is another possible direction of expansion of the
present article.

As are the other logics where forgetting is applied like
{} description logics~\cite{eite-06,zhan-16}
and
{} modal logics~\cite{zhan-zhou-09,vand-etal-09},
where forgetting is often referred to as its dual concept of uniform
interpolant, and also
{} temporal logics~\cite{feng-etal-20},
{} logics for reasoning about actions~\cite{erde-ferr-07,raja-etal-14}, and
{} defeasible logics~\cite{anto-etal-12}.

\appendix
\let\proof=\oproof
\section{Proofs}

\def\relemma#1{\hfill\break\noindent {\bf Lemma~\ref{#1}}}
\def\retheorem#1{\hfill\break\noindent {\bf Theorem~\ref{#1}}}
\def\recorollary#1{\hfill\break\noindent {\bf Corollary~\ref{#1}}}
\def\recounterexample#1{\hfill\break\noindent {\bf Counterexample~\ref{#1}}}

\relemma{consistent}
{\em 
A formula $B$ over the variables $Y$ expresses forgetting all variables from
$A$ but $Y$ if and only if $A \wedge D$ is equisatisfiable with $B \wedge D$
for all formulae $D$ over variables $Y$.
}

\

\proof The definition of $B$ expressing forgetting is that it is built over the
variables $Y$ and that $A \models C$ is the same as $B \models C$ for every
formula $C$ on the alphabet $Y$. The two entailments are respectively the same
as the inconsistency of $A \wedge \neg C$ and $B \wedge \neg C$. They coincide
if and only if $A \wedge D$ and $B \wedge D$ are equisatisfiable, where $D =
\neg C$. In the other way around, $A \wedge D$ and $B \wedge D$ are
equisatisfiable if and only if $A \wedge \neg C$ and $B \wedge \neg C$ are,
where $C = \neg D$.~\qed

\retheorem{consistent-literals}
{\em
A formula $B$ over the variables $Y$ expresses forgetting all variables except
$Y$ from $A$ if and only if $S \cup A$ is equisatisfiable with $S \cup B$ for
all sets of literals $S$ over variables $Y$.

}

\proof The claim is proved in the two directions: first, $B$ expressing
forgetting implies the equisatisfiability; second, equisatisfiability implies
that $B$ expresses forgetting.

The first direction is that if $B$ expresses forgetting then $S \cup A$ and $S
\cup B$ are equisatisfiable. The two satisfiability conditions are respectively
equivalent to the entailments $A \not\models \neg S$ and $B \not\models \neg
S$. They coincide if $A \models \neg S$ and $B \models \neg S$ coincide. Since
$S$ is a set of literals over $Y$, the entailed formula $\neg S$ contains only
variables of $Y$. By definition of forgetting, $A \models \neg S$ is the same
as $B \models \neg S$.

The converse direction is that if $S \cup A$ and $S \cup B$ are equisatisfiable
for every set of literals over $Y$ then $B$ expresses forgetting all variables
except $Y$ from $A$. Forgetting is defined as $A \models C$ being the same as
$B \models C$ for every formula $C$ over $Y$. Like every other formula, $C$ can
be turned into an equivalent CNF $D = \{d_1,\ldots,d_m\}$ over the same
alphabet. Because of equivalence, the two entailments to prove equal are
respectively the same as $A \models D$ and $B \models D$. Since $D$ is a
conjunction, it is entailed if and only if all its clauses are entailed. The
two entailments are therefore respectively equivalent to $A \models d_i$ for
every $d_i \in D$ and $B \models d_i$ for every $d_i \in D$. They coincide if
$A \models d_i$ is the same as $B \models d_i$ for every clause $d_i$ over $Y$.
The two latter entailments are respectively the same as the unsatisfiability of
$A \wedge \neg d_i$ and $B \wedge \neg d_i$. Since $d_i$ is a clause, $\neg
d_i$ is a set of literals. The unsatisfiability of $A \cup \neg d_i$ coincides
with that of $B \cup \neg d_i$ because $\neg d_i$ is a set of literals over
$Y$, and the assumption is exactly the equisatisfiability of $A \cup S$ and $B
\cup S$ for every set of literals over $Y$.~\qed

\retheorem{consistent-literals-complete}
{\em
A formula $B$ over the variables $Y$ expresses forgetting all variables except
$Y$ from $A$ if and only if $S \cup A$ is equisatisfiable with $S \cup B$ for
all sets of literals $S$ over variables $Y$ that contain all variables in $Y$.
}

\proof If $B$ expresses forgetting, then $S \cup A$ and $S \cup B$ are
equisatisfiable for every set of literals $S$ over $Y$, including the ones that
contain all variables of $Y$.

In the other direction, if $S$ is a set of literals over $Y$ that does not
contain all variables of $Y$, it can be expressed as $S \equiv S_1 \vee \cdots
\vee S_m$ where each $S_i$ is a set of literals that contains all variables of
$Y$. The satisfiability of $S \cup A$ coincides with the satisfiability of some
$S_i \cap A$, and the satisfiability of $S \cup B$ with that of $S_i \cup B$.
By assumption $S_i \cup A$ is equisatisfiable with $S_i \cup B$ for every
$S_i$.~\qed

\relemma{necessary-literal}
{\em
If $S$ is a set of literals such that $S \cup A$ is consistent, but $S
\backslash \{l\} \cup \{\neg l\} \cup A$ is not, the CNF formula $A$ contains
$l$.
}

\

\proof Since $S \cup A$ is consistent, it has a model $M$.

The claim is that $A$ contains a clause that contains $l$. This is proved by
contradiction, assuming that no clause of $A$ contains $l$. By construction $S
\backslash \{l\}$ does not contain $l$ either. As a result,
{} $A' = S \backslash \{l\} \cup A$
does not contain $l$. It is still satisfied by $M$ because $M$ satisfies its
superset $S \cup A$. Let $M'$ be the model that sets $l$ to $\false$ and all
other variables the same as $M$. Let $l_1 \vee \cdots \vee l_m$ be an arbitrary
clause of $A'$. Since $M$ satisfies $A$, it satisfies at least one of these
literals $l_i$. Since $A$ does not contain $l$, this literal $l_i$ is either
$\neg l$ or a literal over a different variable. In the first case $M'$
satisfies $l_i = \neg l$ because it sets $l$ to false; in the second because it
sets $l_i$ the same as $M$, which satisfies $l_i$. This happens for all clauses
of $A'$, proving that $M'$ satisfies $A'$.

Since $M'$ also satisfies $\neg l$ because it sets $l$ to false, it satisfies
{} $A' \cup \{\neg l\} = S \backslash \{l\} \cup \{\neg l\} \cup A$,
contrary to its assumed unsatisfiability.~\qed

\relemma{forget-contains}
{\em

If $S \cup \{l\}$ is a set of literals over the variables $Y$ such that $S \cup
A$ is consistent, but $S \backslash \{l\} \cup \{\neg l\} \cup A$ is not, every
CNF formula that expresses forgetting all variables except $Y$ from $A$ contains
$l$.

}

\

\proof Let $B$ be a formula expressing forgetting all variables from $A$ but
$Y$. By Theorem~\ref{consistent-literals}, since $S$ is a set of literals
over $Y$, the consistency of $S \cup A$ equates that of $S \cup B$. The same
holds for $S \backslash \{l\} \cup \{\neg l\}$ since its variables are all in
$Y$.

The lemma assumes the consistency of $S \cup A$ and the inconsistency of $S
\backslash \{l\} \cup \{\neg l\} \cup A$. They imply the consistency of $S \cup
B$ and the inconsistency of $S \backslash \{l\} \cup \{\neg l\} \cup B$. These
two conditions imply that $B$ contains $l$ by
Lemma~\ref{necessary-literal}.~\qed

\relemma{horn-conp}
{\em
There exists a polynomial algorithm that turns a CNF formula $F$ into a
minimal-size Horn formula $A$, a subset $X_C \subseteq \var(A)$ and a number
$k$ such that forgetting all variables except $X_C$ from $A$ is expressed by a
Horn formula of size $k$ if $F$ is unsatisfiable and only by Horn formulae of
size greater than or equal to $k+2$ if $F$ is satisfiable.
}

\

\proof Let $F = \{f_1,\ldots,f_m\}$ be a CNF formula built over the alphabet $X
= \{x_1,\ldots,x_n\}$. The reduction employs the fresh variables
{} $E = \{e_1,\ldots,e_n\}$,
{} $T = \{t_1,\ldots,t_n\}$,
{} $C = \{c_1,\ldots,c_m\}$ and
{} $\{a,b\}$.
The formula $A$, the set of variables $X_C$ and the number $k$ are:

\begin{eqnarray*}
A &=&
	\{\neg x_i \vee \neg e_i,
	  \neg x_i \vee t_i,
	  \neg e_i \vee t_i \mid x_i \in X \} \cup			\\
&&
	\{\neg x_i \vee c_j \mid      x_i \in f_j ,~ f_j \in F\} \cup
	\{\neg e_i \vee c_j \mid \neg x_i \in f_j ,~ f_j \in F\} \cup	\\
&&
	\{\neg t_1 \vee \cdots \vee \neg t_n \vee
	\neg c_1 \vee \cdots \vee \neg c_m \vee
	\neg a \vee b\} \cup						\\
&&
	\{a \vee \neg b\}						\\
X_C &=& X \cup E \cup \{a,b\}						\\
k &=& 2 \times n + 2
\end{eqnarray*}

Before formally proving the claim, how the reduction works is summarized. Some
literals are still necessary after forgetting, and some of them are necessary
only if $F$ is satisfiable. The clauses $\neg x_i \vee \neg e_i$ make $\neg
x_i$ and $\neg e_i$ necessary. The clause $a \vee \neg b$ makes $a$ and $\neg
b$ necessary. If $F$ is always false, then for every value of the variables $X
\cup E$ either some $t_i$ can be set to false (if $x_i = e_i = \false$) or some
$c_j$ can be set to false (because $e_i$ is the negation of $x_i$, and at least
a clause of $F$ is false). This makes the clause
{} $\neg t_1 \vee \cdots \vee \neg t_n \vee
{}  \neg c_1 \vee \cdots \vee \neg c_m \vee
{}  \neg a \vee b$
satisfied regardless of $a$ and $b$. Instead, if the formula is satisfied by an
evaluation of $X$ and $E$ is its opposite, then all $c_j$ and $t_i$ have to be
true, turning
{} $\neg t_1 \vee \cdots \vee \neg t_n \vee
{}  \neg c_1 \vee \cdots \vee \neg c_m \vee
{}  \neg a \vee b$
into $\neg a \vee b$. This makes $\neg a$ and $b$ necessary as well.

\begin{center}
\setlength{\unitlength}{5000sp}%
\begingroup\makeatletter\ifx\SetFigFont\undefined%
\gdef\SetFigFont#1#2#3#4#5{%
  \reset@font\fontsize{#1}{#2pt}%
  \fontfamily{#3}\fontseries{#4}\fontshape{#5}%
  \selectfont}%
\fi\endgroup%
\begin{picture}(3648,2863)(7168,-7859)
\thinlines
{\color[rgb]{0,0,0}\put(7201,-5536){\line( 1, 0){3000}}
}%
\thicklines
{\color[rgb]{0,0,0}\put(7201,-6586){\line( 0,-1){ 75}}
\put(7201,-6661){\line( 1, 0){1200}}
\put(8401,-6661){\line( 0, 1){ 75}}
}%
{\color[rgb]{0,0,0}\put(9601,-6586){\line( 0,-1){ 75}}
\put(9601,-6661){\line( 1, 0){600}}
\put(10201,-6661){\line( 0, 1){ 75}}
}%
\thinlines
{\color[rgb]{0,0,0}\put(9976,-7636){\framebox(150,900){}}
}%
{\color[rgb]{0,0,0}\put(9676,-7636){\framebox(150,300){}}
}%
{\color[rgb]{0,0,0}\put(9751,-5236){\line( 0,-1){225}}
}%
{\color[rgb]{0,0,0}\put(8776,-5836){\line( 1, 0){150}}
}%
{\color[rgb]{0,0,0}\put(9076,-5836){\line( 1, 0){150}}
}%
{\color[rgb]{0,0,0}\put(9376,-5836){\line( 1, 0){150}}
}%
{\color[rgb]{0,0,0}\put(8476,-6136){\line( 1, 0){150}}
}%
{\color[rgb]{0,0,0}\put(8776,-6136){\line( 1, 0){150}}
}%
{\color[rgb]{0,0,0}\put(9376,-6136){\line( 1, 0){150}}
}%
\put(7351,-5461){\makebox(0,0)[b]{\smash{{\SetFigFont{12}{24.0}
{\rmdefault}{\mddefault}{\updefault}{\color[rgb]{0,0,0}$x_1$}%
}}}}
\put(7651,-5461){\makebox(0,0)[b]{\smash{{\SetFigFont{12}{24.0}
{\rmdefault}{\mddefault}{\updefault}{\color[rgb]{0,0,0}$n_1$}%
}}}}
\put(7951,-5461){\makebox(0,0)[b]{\smash{{\SetFigFont{12}{24.0}
{\rmdefault}{\mddefault}{\updefault}{\color[rgb]{0,0,0}$x_2$}%
}}}}
\put(8851,-5461){\makebox(0,0)[b]{\smash{{\SetFigFont{12}{24.0}
{\rmdefault}{\mddefault}{\updefault}{\color[rgb]{0,0,0}$t_2$}%
}}}}
\put(9151,-5461){\makebox(0,0)[b]{\smash{{\SetFigFont{12}{24.0}
{\rmdefault}{\mddefault}{\updefault}{\color[rgb]{0,0,0}$c_1$}%
}}}}
\put(9451,-5461){\makebox(0,0)[b]{\smash{{\SetFigFont{12}{24.0}
{\rmdefault}{\mddefault}{\updefault}{\color[rgb]{0,0,0}$c_2$}%
}}}}
\put(8251,-5461){\makebox(0,0)[b]{\smash{{\SetFigFont{12}{24.0}
{\rmdefault}{\mddefault}{\updefault}{\color[rgb]{0,0,0}$n_2$}%
}}}}
\put(7951,-5911){\makebox(0,0)[b]{\smash{{\SetFigFont{12}{24.0}
{\rmdefault}{\mddefault}{\updefault}{\color[rgb]{0,0,0}$0$}%
}}}}
\put(8251,-5911){\makebox(0,0)[b]{\smash{{\SetFigFont{12}{24.0}
{\rmdefault}{\mddefault}{\updefault}{\color[rgb]{0,0,0}$1$}%
}}}}
\put(7351,-6211){\makebox(0,0)[b]{\smash{{\SetFigFont{12}{24.0}
{\rmdefault}{\mddefault}{\updefault}{\color[rgb]{0,0,0}$1$}%
}}}}
\put(7651,-6211){\makebox(0,0)[b]{\smash{{\SetFigFont{12}{24.0}
{\rmdefault}{\mddefault}{\updefault}{\color[rgb]{0,0,0}$0$}%
}}}}
\put(7951,-6211){\makebox(0,0)[b]{\smash{{\SetFigFont{12}{24.0}
{\rmdefault}{\mddefault}{\updefault}{\color[rgb]{0,0,0}$0$}%
}}}}
\put(8251,-6211){\makebox(0,0)[b]{\smash{{\SetFigFont{12}{24.0}
{\rmdefault}{\mddefault}{\updefault}{\color[rgb]{0,0,0}$1$}%
}}}}
\put(7351,-6511){\makebox(0,0)[b]{\smash{{\SetFigFont{12}{24.0}
{\rmdefault}{\mddefault}{\updefault}{\color[rgb]{0,0,0}$0$}%
}}}}
\put(7651,-6511){\makebox(0,0)[b]{\smash{{\SetFigFont{12}{24.0}
{\rmdefault}{\mddefault}{\updefault}{\color[rgb]{0,0,0}$1$}%
}}}}
\put(7951,-6511){\makebox(0,0)[b]{\smash{{\SetFigFont{12}{24.0}
{\rmdefault}{\mddefault}{\updefault}{\color[rgb]{0,0,0}$0$}%
}}}}
\put(8251,-6511){\makebox(0,0)[b]{\smash{{\SetFigFont{12}{24.0}
{\rmdefault}{\mddefault}{\updefault}{\color[rgb]{0,0,0}$1$}%
}}}}
\put(8551,-5911){\makebox(0,0)[b]{\smash{{\SetFigFont{12}{24.0}
{\rmdefault}{\mddefault}{\updefault}{\color[rgb]{0,0,0}$0/1$}%
}}}}
\put(9751,-5911){\makebox(0,0)[b]{\smash{{\SetFigFont{12}{24.0}
{\rmdefault}{\mddefault}{\updefault}{\color[rgb]{0,0,0}$0/1$}%
}}}}
\put(10051,-5911){\makebox(0,0)[b]{\smash{{\SetFigFont{12}{24.0}
{\rmdefault}{\mddefault}{\updefault}{\color[rgb]{0,0,0}$1$}%
}}}}
\put(9151,-6211){\makebox(0,0)[b]{\smash{{\SetFigFont{12}{24.0}
{\rmdefault}{\mddefault}{\updefault}{\color[rgb]{0,0,0}$0/1$}%
}}}}
\put(8551,-6511){\makebox(0,0)[b]{\smash{{\SetFigFont{12}{24.0}
{\rmdefault}{\mddefault}{\updefault}{\color[rgb]{0,0,0}$1$}%
}}}}
\put(8851,-6511){\makebox(0,0)[b]{\smash{{\SetFigFont{12}{24.0}
{\rmdefault}{\mddefault}{\updefault}{\color[rgb]{0,0,0}$1$}%
}}}}
\put(9151,-6511){\makebox(0,0)[b]{\smash{{\SetFigFont{12}{24.0}
{\rmdefault}{\mddefault}{\updefault}{\color[rgb]{0,0,0}$1$}%
}}}}
\put(9451,-6511){\makebox(0,0)[b]{\smash{{\SetFigFont{12}{24.0}
{\rmdefault}{\mddefault}{\updefault}{\color[rgb]{0,0,0}$1$}%
}}}}
\put(9751,-6211){\makebox(0,0)[b]{\smash{{\SetFigFont{12}{24.0}
{\rmdefault}{\mddefault}{\updefault}{\color[rgb]{0,0,0}$0/1$}%
}}}}
\put(10051,-6211){\makebox(0,0)[b]{\smash{{\SetFigFont{12}{24.0}
{\rmdefault}{\mddefault}{\updefault}{\color[rgb]{0,0,0}$1$}%
}}}}
\put(9751,-6511){\makebox(0,0)[b]{\smash{{\SetFigFont{12}{24.0}
{\rmdefault}{\mddefault}{\updefault}{\color[rgb]{0,0,0}$1$}%
}}}}
\put(10051,-6511){\makebox(0,0)[b]{\smash{{\SetFigFont{12}{24.0}
{\rmdefault}{\mddefault}{\updefault}{\color[rgb]{0,0,0}$1$}%
}}}}
\put(7351,-6961){\makebox(0,0)[b]{\smash{{\SetFigFont{12}{24.0}
{\rmdefault}{\mddefault}{\updefault}{\color[rgb]{0,0,0}$1$}%
}}}}
\put(7651,-6961){\makebox(0,0)[b]{\smash{{\SetFigFont{12}{24.0}
{\rmdefault}{\mddefault}{\updefault}{\color[rgb]{0,0,0}$1$}%
}}}}
\put(7951,-6961){\makebox(0,0)[b]{\smash{{\SetFigFont{12}{24.0}
{\rmdefault}{\mddefault}{\updefault}{\color[rgb]{0,0,0}$0$}%
}}}}
\put(8251,-6961){\makebox(0,0)[b]{\smash{{\SetFigFont{12}{24.0}
{\rmdefault}{\mddefault}{\updefault}{\color[rgb]{0,0,0}$1$}%
}}}}
\put(7351,-7261){\makebox(0,0)[b]{\smash{{\SetFigFont{12}{24.0}
{\rmdefault}{\mddefault}{\updefault}{\color[rgb]{0,0,0}$1$}%
}}}}
\put(7651,-7261){\makebox(0,0)[b]{\smash{{\SetFigFont{12}{24.0}
{\rmdefault}{\mddefault}{\updefault}{\color[rgb]{0,0,0}$0$}%
}}}}
\put(7951,-7261){\makebox(0,0)[b]{\smash{{\SetFigFont{12}{24.0}
{\rmdefault}{\mddefault}{\updefault}{\color[rgb]{0,0,0}$0$}%
}}}}
\put(8251,-7261){\makebox(0,0)[b]{\smash{{\SetFigFont{12}{24.0}
{\rmdefault}{\mddefault}{\updefault}{\color[rgb]{0,0,0}$1$}%
}}}}
\put(7951,-7561){\makebox(0,0)[b]{\smash{{\SetFigFont{12}{24.0}
{\rmdefault}{\mddefault}{\updefault}{\color[rgb]{0,0,0}$0$}%
}}}}
\put(8251,-7561){\makebox(0,0)[b]{\smash{{\SetFigFont{12}{24.0}
{\rmdefault}{\mddefault}{\updefault}{\color[rgb]{0,0,0}$1$}%
}}}}
\put(7651,-7561){\makebox(0,0)[b]{\smash{{\SetFigFont{12}{24.0}
{\rmdefault}{\mddefault}{\updefault}{\color[rgb]{0,0,0}$1$}%
}}}}
\put(7351,-7561){\makebox(0,0)[b]{\smash{{\SetFigFont{12}{24.0}
{\rmdefault}{\mddefault}{\updefault}{\color[rgb]{0,0,0}$0$}%
}}}}
\put(9751,-6961){\makebox(0,0)[b]{\smash{{\SetFigFont{12}{24.0}
{\rmdefault}{\mddefault}{\updefault}{\color[rgb]{0,0,0}$0/1$}%
}}}}
\put(9751,-7261){\makebox(0,0)[b]{\smash{{\SetFigFont{12}{24.0}
{\rmdefault}{\mddefault}{\updefault}{\color[rgb]{0,0,0}$0/1$}%
}}}}
\put(9751,-7561){\makebox(0,0)[b]{\smash{{\SetFigFont{12}{24.0}
{\rmdefault}{\mddefault}{\updefault}{\color[rgb]{0,0,0}$1$}%
}}}}
\put(10051,-6961){\makebox(0,0)[b]{\smash{{\SetFigFont{12}{24.0}
{\rmdefault}{\mddefault}{\updefault}{\color[rgb]{0,0,0}$1$}%
}}}}
\put(10051,-7261){\makebox(0,0)[b]{\smash{{\SetFigFont{12}{24.0}
{\rmdefault}{\mddefault}{\updefault}{\color[rgb]{0,0,0}$1$}%
}}}}
\put(10051,-7561){\makebox(0,0)[b]{\smash{{\SetFigFont{12}{24.0}
{\rmdefault}{\mddefault}{\updefault}{\color[rgb]{0,0,0}$1$}%
}}}}
\put(10051,-5461){\makebox(0,0)[b]{\smash{{\SetFigFont{12}{24.0}
{\rmdefault}{\mddefault}{\updefault}{\color[rgb]{0,0,0}$a \vee \neg b$}%
}}}}
\put(9751,-5161){\makebox(0,0)[b]{\smash{{\SetFigFont{12}{24.0}
{\rmdefault}{\mddefault}{\updefault}{\color[rgb]{0,0,0}$\neg a \vee b$}%
}}}}
\put(8551,-5461){\makebox(0,0)[b]{\smash{{\SetFigFont{12}{24.0}
{\rmdefault}{\mddefault}{\updefault}{\color[rgb]{0,0,0}$t_1$}%
}}}}
\put(9826,-7786){\makebox(0,0)[rb]{\smash{{\SetFigFont{12}{24.0}
{\rmdefault}{\mddefault}{\updefault}{\color[rgb]{0,0,0}$\neg a,b \mbox{ necessary}$}%
}}}}
\put(10201,-7411){\makebox(0,0)[lb]{\smash{{\SetFigFont{12}{24.0}
{\rmdefault}{\mddefault}{\updefault}{\color[rgb]{0,0,0}$a,\neg b \mbox{ necessary}$}%
}}}}
\put(10801,-6511){\makebox(0,0)[lb]{\smash{{\SetFigFont{12}{24.0}
{\rmdefault}{\mddefault}{\updefault}{\color[rgb]{0,0,0}$(\mbox{all } x_i \not= n_i, \mbox{all } f_j \mbox { true})$}%
}}}}
\put(10801,-5911){\makebox(0,0)[lb]{\smash{{\SetFigFont{12}{24.0}
{\rmdefault}{\mddefault}{\updefault}{\color[rgb]{0,0,0}$(x_1=n_1=\false)$}%
}}}}
\put(10801,-6211){\makebox(0,0)[lb]{\smash{{\SetFigFont{12}{24.0}
{\rmdefault}{\mddefault}{\updefault}{\color[rgb]{0,0,0}$(f_1 \mbox{ false})$}%
}}}}
\put(7651,-5911){\makebox(0,0)[b]{\smash{{\SetFigFont{12}{24.0}
{\rmdefault}{\mddefault}{\updefault}{\color[rgb]{0,0,0}$0$}%
}}}}
\put(7351,-5911){\makebox(0,0)[b]{\smash{{\SetFigFont{12}{24.0}
{\rmdefault}{\mddefault}{\updefault}{\color[rgb]{0,0,0}$0$}%
}}}}
\end{picture}%
\nop{
x1 n1 x2 n2  t1 t2 c1 c2  -avb av-b
 0  0  0  1  0/1 -  -  -   0/1  1
 1  0  0  1   -  - 0/1 -   0/1  1
 0  1  0  1   1  1  1  1     1  1
-----------               -------
 0  0  0  1                0/1  1 \                       .
 1  0  0  1                0/1  1   >-- a,-b necessary
 0  1  0  1              +-- 1  1 /
                         |   
                  -a,b necess.
}
\end{center}

The figure shows three models as an example. In the first model, the
assignments $x_1 = e_1 = \false$ allow $t_1$ to take any value (denoted $0/1$);
regardless of the other variables (irrelevant values are marked $-$), $t_1 =
\false$ satisfies the clause
{} $\neg t_1 \vee \cdots \vee \neg t_n \vee
{}  \neg c_1 \vee \cdots \vee \neg c_m \vee
{}  \neg a \vee b$
without the need to also satisfy its subset $\neg a \vee b$; this subclause can
be false and still $A$ is true. In the second model the values of $x_i$ and
$e_i$ are opposite to each other for every $i$, but the clause $f_2 \in F$ is
false; $c_1$ can take any value, including $\false$; this again allows $A$ to
be true even if $\neg a \vee b$ is false. In the third model, the variables
$x_i$ and $e_i$ are all opposite to each other and all clauses of $F$ true; all
$t_i$ and $c_i$ are forced to be true, making $\neg a \vee b$ the only way to
satisfy the clause
{} $\neg t_1 \vee \cdots \vee \neg t_n \vee
{}  \neg c_1 \vee \cdots \vee \neg c_m \vee
{}  \neg a \vee b$.
When removing the intermediate variables $t_i$ and $c_i$, all that matters is
whether $\neg a \vee b$ was allowed to be false for some values of the removed
variables or not. This is the case for the first two models but not the third,
where $\neg a$ and $b$ are necessary.

\

\noindent {\bf Minimality.} The minimality of $A$ is proved applying
Lemma~\ref{set-value} to remove some clauses so that the remaining ones do not
resolve and Lemma~\ref{no-resolution} applies. Lemma~\ref{minimal} proves $A$
minimal since it only contains superirredundant clauses.

Substituting the variables $a$, $b$ with $\false$ removes
{} $\neg t_1 \vee \cdots \vee \neg t_n \vee
{}  \neg c_1 \vee \cdots \vee \neg c_m \vee
{}  \neg a \vee b$ and
{} $a \vee \neg b$
from $A$. The remaining clauses contain $x_i,e_i$ only negative and $t_i,c_j$
only positive. Therefore, these clauses do not resolve. Since they do not
contain each other, Lemma~\ref{no-resolution} proves them superirredundant.
They are also superirredundant in $A$ by Lemma~\ref{set-value} since $A$ does
not contain any of their supersets.

The superirredundancy of the remaining two clauses is proved by substituting
all $x_i,e_i$ with $\false$. This substitution removes the clauses
{} $\neg x_i \vee \neg e_i$,
{} $\neg x_i \vee t_i$,
{} $\neg e_i \vee t_i$,
{} $\neg x_i \vee c_j$ and
{} $\neg e_i \vee c_j$
because they all contain either $\neg x_i$ or $\neg e_i$. The two remaining
clauses are
{} $\neg t_1 \vee \cdots \vee \neg t_n \vee
{}  \neg c_1 \vee \cdots \vee \neg c_m \vee
{}  \neg a \vee b$ and
{} $a \vee \neg b$.
They have opposite literals, but resolving them results in tautologies. As a
result, $F = \rescn(F)$. Since none of the two entails the other, they are
irredundant in $\rescn(F)$ and are therefore superirredundant. By
Lemma~\ref{set-value}, they are superirredundant in $A$ as well since $A$ does
not contain a superset of them.

\

\noindent {\bf Formula $F$ is unsatisfiable.} Forgetting all variables except
$X_C$ from $A$ is expressed by
{} $B = \{\neg x_i \vee \neg e_i \mid x_i \in X\} \cup \{a \vee \neg b\}$,
a Horn formula of the required variables $X_C = X \cup E \cup \{a, b\}$ and
size $||B|| = 2 \times n + 2 = k$.

Theorem~\ref{consistent-literals-complete} proves that $B$ expresses forgetting
if every set of literals $S$ that contains exactly all variables $X_C = X \cup
E \cup \{a,b\}$ is satisfiable with $A$ if and only if it is satisfiable with
$B$. Two cases are possible.

\begin{description}

\item[$\{x_i,e_i\} \subseteq S$ for some $i$]; the clause $\neg x_i \vee \neg
e_i$ in both $A$ and $B$ is falsified by $S$; both $A \cup S$ and $B \cup S$
are unsatisfiable;

\item[$\{x_i,e_i\} \subseteq S$ for no $i$]; since $S$ contains either $x_i$
or $\neg x_i$ for each $i$ and either $e_i$ or $\neg e_i$ for each $i$,
either $\neg x_i \in S$ or $\neg e_i \in S$; as a result, all clauses $\neg
x_i \vee \neg e_i$ are satisfied in $A \cup S$ and $B \cup S$, and can
therefore be disregarded from this point on; the only remaining clause of $B$
is $a \vee \neg b$;

if $S$ contains $\neg a$ and $b$, then $B$ is not satisfied; but $A$ contains
the same clause $a \vee \neg b$, so it is not satisfied either; if $S$ contains
both $a$ and $b$ or both $\neg a$ and $\neg b$, then $B$ is satisfied, and $A$
is also satisfied by setting all variables $t_i$ and $c_j$ to $\true$;
therefore, the only sets $S$ that may differ when added to $A$ and $B$ are
those containing $a$ and $\neg b$; these sets are consistent with $B$; they
make the clause $a \vee \neg b$ of $A$ redundant, and resolve with the clause
{} $\neg t_1 \vee \cdots \vee \neg t_n \vee
{}  \neg c_1 \vee \cdots \vee \neg c_m \vee
{}  \neg a \vee b$
making it subsumed by
{} $\neg t_1 \vee \cdots \vee \neg t_n \vee
{}  \neg c_1 \vee \cdots \vee \neg c_m$.

Two subcases are considered:

\begin{description}

\item[$\{\neg x_i,\neg e_i\} \subseteq S$ for some $i$] the remaining clauses
of $A$ are satisfied by setting $t_i$ to $\false$, all $t_z$ with $z \not= i$
to $\true$ and all $c_j$ to $\true$; in particular, the clause
{} $\neg t_1 \vee \cdots \vee \neg t_n \vee
{}  \neg c_1 \vee \cdots \vee \neg c_m$
is satisfied because of $t_i = \false$;

\item[$\{\neg x_i,\neg e_i\} \subseteq S$ for no $i$]; at this point, also
$\{x_i,e_i\} \subseteq S$ for no $i$; as a result, $S$ contains either
$\{x_i,\neg e_i\}$ or $\{\neg x_i,e_i\}$, which means that it implies $x_i
\not\equiv e_i$; the clauses $\neg e_i \vee c_j$ are therefore equivalent to
$x_i \vee c_j$; by assumption, at least a clause of $F$ is false for every
possible value of the variables $X$; let $f_j$ be such a clause for the only
truth evaluation on $X$ that satisfies $S$; by setting all variables $t_i$ and
all $c_z$ with $z \not= j$ to $\true$ and $c_j$ to $\false$, this assignment
satisfies all clauses; in particular, the clauses
{} $\neg x_i \vee c_j$ and $x_i \vee c_j$
are satisfied even if $c_j$ is false because $f_j$ is false in $S$, which
implies that all literals $x_i$ and $\neg x_i$ it contains are false; the
clause
{} $\neg t_1 \vee \cdots \vee \neg t_n \vee
{}  \neg c_1 \vee \cdots \vee \neg c_m$
is satisfied because of $c_j = \false$.

\end{description}

\end{description}

All of this proves that $B$ is the result of forgetting all variables except
$X_C$ from $A$.

% The minimality of $B$ follows from the following point: its size is $k$, and
% the following point proves that every formula expressing forgetting has size
% $k$ or more. It also follows from Lemma~\ref{no-resolution} since its clauses
% do not share variables.

\

\noindent {\bf Minimal number of literals.} Every CNF formula $B$ that
expresses forgetting all variables except $X_C = X \cup E \cup \{a,b\}$ from
$A$ contains at least $k = 2 \times n + 2$ literal occurrences.

This is proved by showing that $B$ contains the literals $\neg x_i$, $\neg
e_i$, $a$ and $\neg b$. This is in turn proved by Lemma~\ref{forget-contains}:
for each of them $l$, a set $S$ is shown consistent with $A$ while
{} $S \backslash \{l\} \cup \{\neg l\}$
is not.

For the literals $\neg x_i$ and $a$ the set $S$ contains all $\neg x_i$, all
$e_i$, $a$ and $b$. It is consistent with $A$ because both are satisfied by the
model that sets all $x_i$ to false and all $e_i$, $t_i$, $c_i$, $a$ and $b$ to
true. Replacing $\neg x_i$ with $x_i$ makes $S$ inconsistent with $\neg x_i
\vee \neg e_i$. Replacing $a$ with $\neg a$ makes $S$ inconsistent with $a \vee
\neg b$.

For the literals $\neg e_i$ and $\neg b$, the set $S$ contains all $x_i$, all
$\neg e_i$, $\neg a$ and $\neg b$. It is consistent with $A$ because both are
satisfied by the model that assigns all $x_i$, $t_i$ and $c_i$ to $\true$ and
all $e_i$, $a$ and $b$ to $\false$. Replacing $\neg e_i$ with $e_i$ makes $S$
inconsistent with $\neg x_i \vee \neg e_i$. Replacing $\neg b$ with $b$ makes
it inconsistent with $a \vee \neg b$.

This proves that every formula obtained by forgetting all variables except
$X_C$ from $A$ contains all the $k = 2 \times n + 2$ literals $\neg x_i$, $\neg
e_i$, $a$ and $\neg b$.

\

\noindent {\bf Formula $F$ is satisfiable.} If this is the case, every CNF
formula $B$ that expresses forgetting all variables except $X_C$ from $A$
contains the literals $\neg a$ and $b$. These two literals are in addition to
the $k$ literals of the previous point, raising the minimal number of literals
to $k + 2$.

That $B$ contains $\neg a$ is proved by exhibiting a set of literals $S$ that
is consistent with $A$ while $S \backslash \{l\} \cup \{\neg l\}$ is not where
$l = \neg a$. This implies that $B$ contains $\neg a$ by
Lemma~\ref{forget-contains}. A similar set with $l = b$ shows that $B$ also
contains $b$.

Let $M$ be a model of $F$. The set of literals $S$ contains $x_i$ or $\neg x_i$
depending on whether $M$ satisfies $x_i$; it contains $e_i$ or $\neg e_i$
depending on whether $M$ falsifies $x_i$; it also contains $\neg a$ and $\neg
b$. This set is consistent with $A$ because they are both satisfied by the
model that extends $M$ by setting each $e_i$ opposite to $x_i$, all $t_i$
and $c_i$ to $\true$ and $a$ and $b$ to $\false$. In particular, the clause
{} $\neg t_1 \vee \cdots \vee \neg t_n \vee
{}  \neg c_1 \vee \cdots \vee \neg c_m \vee
{}  \neg a \vee b$
is satisfied because it contains $\neg a$.

Replacing $\neg a$ with $a$ makes $S$ no longer consistent with $A$. Let $S' =
S \backslash \{\neg a\} \cup \{\neg \neg a\}$. This set has the same literals
over $x_i$ and $e_i$ of $S$. Since $M$ satisfies $F$, for each of its clauses
$f_j$ at least a literal in $f_j$ is true in $M$. If this literal is $x_i$,
then $S'$ contains $x_i$; since $x_i$ is in $f_j$, formula $A$ contains the
clause $\neg x_i \vee c_j$; therefore, $S' \cup A \models c_j$. If the literal
of $f_j$ that is true in $M$ is $\neg x_i$, then $S'$ contains $e_i$; since
$\neg x_i$ is in $f_j$, formula $A$ contains $\neg e_i \vee c_j$; therefore,
$S' \cup A \models c_j$. This proves that regardless of whether the literal of
$f_j$ that is true in $M$ is positive or negative, if $F$ is consistent then
$S' \cup A$ implies $c_j$. This is the case for every $j$ since $M$ satisfies
all clauses of $F$. Since $S'$ contains either $x_i$ or $e_i$ for every $i$ and
$A$ contains both $\neg x_j \vee t_j$ and $\neg e_j \vee t_j$ for every $j$,
$S' \cup A$ also implies all variables $t_j$. Since $S' = S \backslash \{\neg
a\} \cup \{a\}$ also contains $a$ and $\neg b$, it is inconsistent with
{} $\neg t_1 \vee \cdots \vee \neg t_n \vee
{}  \neg c_1 \vee \cdots \vee \neg c_m \vee
{}  \neg a \vee b$.
This proves that $\neg a$ is in $B$.

The similar set $S$ that contains $a$ and $b$ leads to the same point where all
variables $c_j$ and $t_i$ are implied, making
{} $\neg t_1 \vee \cdots \vee \neg t_n \vee
{}  \neg c_1 \vee \cdots \vee \neg c_m \vee
{}  \neg a \vee b$
consistent with $\{a,b\}$ but not with $\{a,\neg b\}$. This proves that $b$ is
also in $B$.~\qed

\relemma{horn-np}
{\em
There exists a polynomial algorithm that turns a CNF formula $F$ into a
minimal-size Horn formula $A$, a subset $X_C \subseteq \var(A)$ and a number
$k$ such that forgetting all variables except $X_C$ from $A$ is expressed by a
Horn formula of size $k$ if $F$ is satisfiable and only by Horn formulae of size
greater than $k$ otherwise.
}

\

\proof Let the formula be $F = \{f_1,\ldots,f_m\}$ and $X=\{x_1,\ldots,x_n\}$
its variables. The formula $A$ is built over an extended alphabet comprising
the variables
{} $X = \{x_1,\ldots,x_n\}$
and the additional variables
{} $O = \{o_1,\ldots,o_n\}$,
{} $E = \{e_1,\ldots,e_n\}$,
{} $P = \{p_1,\ldots,p_n\}$,
{} $T = \{t_1,\ldots,t_n\}$,
{} $C = \{c_1,\ldots,c_m\}$,
{} $R = \{r_1,\ldots,r_n\}$,
{} $S = \{s_1,\ldots,s_n\}$ and
{} $q$.

The formula $A$, the set of variables $X_C$ and the integer $k$ are as follows.

\begin{eqnarray*}
A &=& A_F \cup A_T \cup A_C \cup A_B					\\
A_F &=&
\{x_i \vee \neg o_i, o_i \vee \neg q \mid x_i \in X\} \cup
\{e_i \vee \neg p_i, p_i \vee \neg q \mid x_i \in X\}			\\
A_T &=&
\{\neg x_i \vee t_i, \neg e_i \vee t_i \mid x_i \in X\}			\\
A_C &=&
\{\neg x_i \vee c_j \mid      x_i \in f_j ,~ f_j \in F\} \cup
\{\neg e_i \vee c_j \mid \neg x_i \in f_j ,~ f_j \in F\}				\\
A_B &=&
\{\neg t_1 \vee \cdots \vee \neg t_n \vee
  \neg c_1 \vee \cdots \vee \neg c_m \vee
  x_i \vee \neg r_i, r_i \vee \neg q \mid x_i \in X\} \cup		\\
&&
\{\neg t_1 \vee \cdots \vee \neg t_n \vee
  \neg c_1 \vee \cdots \vee \neg c_m \vee
  e_i \vee \neg s_i, s_i \vee \neg q \mid x_i \in X\}			\\
X_C &=& X \cup E \cup T \cup C \cup R \cup S \cup \{q\}			\\
k &=& 2 \times n + ||A_T|| + ||A_C|| + ||A_B||
\end{eqnarray*}

Before formally proving that the reduction works, a short summary of why it
works is given. The variables to forget are $O \cup P$. A way to forget them is
to turn $A_F$ into $A_R = \{x_i \vee \neg q, e_i \vee \neg q \mid x_i \in X\}$.
The other clauses of $A$ are superirredundant; therefore, all minimal
equivalent formulae contain them. The bound $k$ allows only one clause of $A_R$
for each $i$. Combined with the clauses of $A_T$ they entail $t_i \vee \neg q$.
If $F$ is satisfiable, they also combine with the clauses $A_C$ to imply all
clauses $c_j \vee \neg q$. Resolving these clauses with $A_B$ produces all
clauses $x_i \vee \neg q$ and $e_i \vee \neg q$, including the ones not in the
formula. This way, a formula that contains one clause of $A_R$ for each index
$i$ implies all of $A_R$, but only if $F$ is satisfiable.

The following figure shows how $e_1 \vee \neg q$ is derived from $x_1 \vee \neg
q$ and $e_2 \vee \neg q$, when the formula is $F = \{f_1,f_2\}$ where $f_1 =
x_1 \vee x_2$ and $f_2 = \neg x_1 \vee \neg x_2$. These clauses translate into
{} $A_C = \{
{}	\neg x_1 \vee c_1,
{}	\neg x_2 \vee c_1,
{}	\neg e_1 \vee c_2,
{}	\neg e_2 \vee c_2
{} \}$.
The rest of the formula $A$ does not depend on the specific clauses of $F$ but
only on the number of variables and clauses it contains.

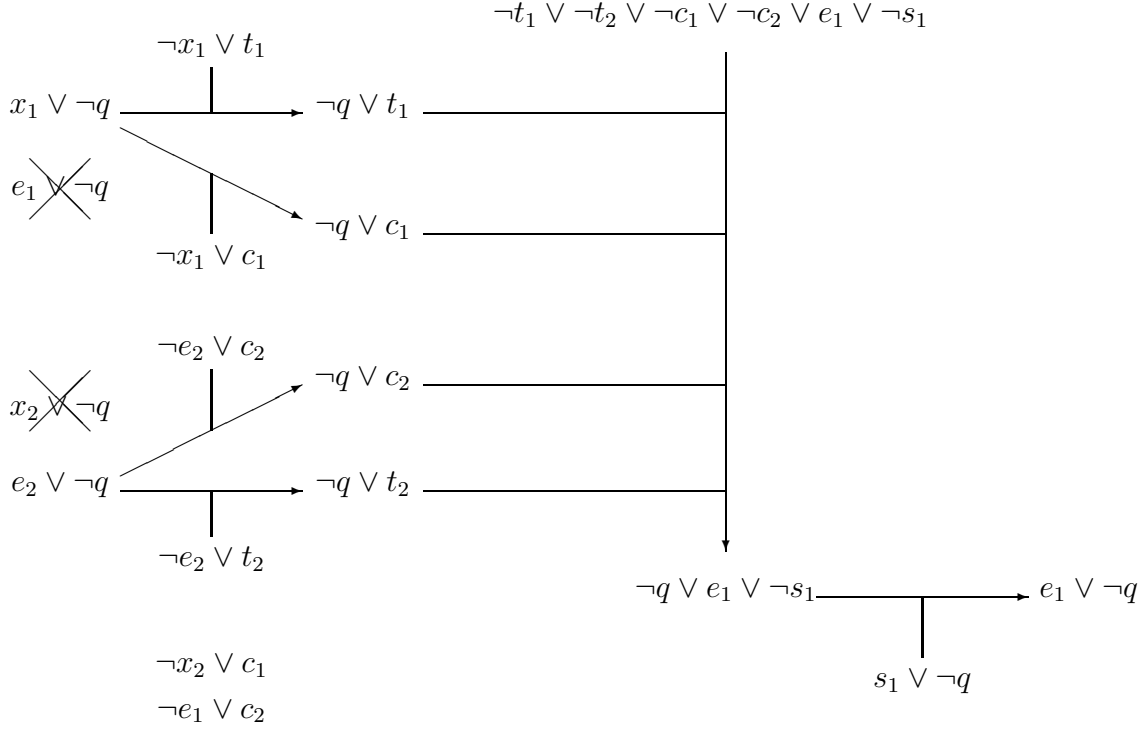
\begin{center}
\setlength{\unitlength}{5000sp}%
\begingroup\makeatletter\ifx\SetFigFont\undefined%
\gdef\SetFigFont#1#2#3#4#5{%
  \reset@font\fontsize{#1}{#2pt}%
  \fontfamily{#3}\fontseries{#4}\fontshape{#5}%
  \selectfont}%
\fi\endgroup%
\begin{picture}(5277,3675)(6589,-7726)
\thinlines
{\color[rgb]{0,0,0}\put(7501,-4486){\line( 0,-1){225}}
}%
{\color[rgb]{0,0,0}\put(7501,-5011){\line( 0,-1){300}}
}%
{\color[rgb]{0,0,0}\put(7501,-5986){\line( 0,-1){300}}
}%
{\color[rgb]{0,0,0}\put(6601,-4936){\line( 1,-1){300}}
}%
{\color[rgb]{0,0,0}\put(6601,-5236){\line( 1, 1){300}}
}%
{\color[rgb]{0,0,0}\put(7501,-6586){\line( 0,-1){225}}
}%
{\color[rgb]{0,0,0}\put(8551,-5311){\line( 1, 0){1500}}
}%
{\color[rgb]{0,0,0}\put(8551,-6061){\line( 1, 0){1500}}
}%
{\color[rgb]{0,0,0}\put(8551,-6586){\line( 1, 0){1500}}
}%
{\color[rgb]{0,0,0}\put(8551,-4711){\line( 1, 0){1500}}
}%
{\color[rgb]{0,0,0}\put(6601,-5986){\line( 1,-1){300}}
}%
{\color[rgb]{0,0,0}\put(6601,-6286){\line( 1, 1){300}}
}%
{\color[rgb]{0,0,0}\put(10501,-7111){\vector( 1, 0){1050}}
}%
{\color[rgb]{0,0,0}\put(10051,-4411){\vector( 0,-1){2475}}
}%
{\color[rgb]{0,0,0}\put(7051,-4711){\vector( 1, 0){900}}
}%
{\color[rgb]{0,0,0}\put(7051,-4786){\vector( 2,-1){900}}
}%
{\color[rgb]{0,0,0}\put(7051,-6511){\vector( 2, 1){900}}
}%
{\color[rgb]{0,0,0}\put(7051,-6586){\vector( 1, 0){900}}
}%
{\color[rgb]{0,0,0}\put(11026,-7411){\line( 0, 1){300}}
}%
\put(7501,-4411){\makebox(0,0)[b]{\smash{{\SetFigFont{12}{24.0}
{\rmdefault}{\mddefault}{\updefault}{\color[rgb]{0,0,0}$\neg x_1 \vee t_1$}%
}}}}
\put(7501,-6961){\makebox(0,0)[b]{\smash{{\SetFigFont{12}{24.0}
{\rmdefault}{\mddefault}{\updefault}{\color[rgb]{0,0,0}$\neg e_2 \vee t_2$}%
}}}}
\put(6751,-4711){\makebox(0,0)[b]{\smash{{\SetFigFont{12}{24.0}
{\rmdefault}{\mddefault}{\updefault}{\color[rgb]{0,0,0}$x_1 \vee \neg q$}%
}}}}
\put(6751,-5116){\makebox(0,0)[b]{\smash{{\SetFigFont{12}{24.0}
{\rmdefault}{\mddefault}{\updefault}{\color[rgb]{0,0,0}$e_1 \vee \neg q$}%
}}}}
\put(6751,-6211){\makebox(0,0)[b]{\smash{{\SetFigFont{12}{24.0}
{\rmdefault}{\mddefault}{\updefault}{\color[rgb]{0,0,0}$x_2 \vee \neg q$}%
}}}}
\put(6751,-6586){\makebox(0,0)[b]{\smash{{\SetFigFont{12}{24.0}
{\rmdefault}{\mddefault}{\updefault}{\color[rgb]{0,0,0}$e_2 \vee \neg q$}%
}}}}
\put(8251,-4711){\makebox(0,0)[b]{\smash{{\SetFigFont{12}{24.0}
{\rmdefault}{\mddefault}{\updefault}{\color[rgb]{0,0,0}$\neg q \vee t_1$}%
}}}}
\put(8251,-6586){\makebox(0,0)[b]{\smash{{\SetFigFont{12}{24.0}
{\rmdefault}{\mddefault}{\updefault}{\color[rgb]{0,0,0}$\neg q \vee t_2$}%
}}}}
\put(7501,-5461){\makebox(0,0)[b]{\smash{{\SetFigFont{12}{24.0}
{\rmdefault}{\mddefault}{\updefault}{\color[rgb]{0,0,0}$\neg x_1 \vee c_1$}%
}}}}
\put(7501,-7711){\makebox(0,0)[b]{\smash{{\SetFigFont{12}{24.0}
{\rmdefault}{\mddefault}{\updefault}{\color[rgb]{0,0,0}$\neg e_1 \vee c_2$}%
}}}}
\put(7501,-7486){\makebox(0,0)[b]{\smash{{\SetFigFont{12}{24.0}
{\rmdefault}{\mddefault}{\updefault}{\color[rgb]{0,0,0}$\neg x_2 \vee c_1$}%
}}}}
\put(8251,-5311){\makebox(0,0)[b]{\smash{{\SetFigFont{12}{24.0}
{\rmdefault}{\mddefault}{\updefault}{\color[rgb]{0,0,0}$\neg q \vee c_1$}%
}}}}
\put(7501,-5911){\makebox(0,0)[b]{\smash{{\SetFigFont{12}{24.0}
{\rmdefault}{\mddefault}{\updefault}{\color[rgb]{0,0,0}$\neg e_2 \vee c_2$}%
}}}}
\put(8251,-6061){\makebox(0,0)[b]{\smash{{\SetFigFont{12}{24.0}
{\rmdefault}{\mddefault}{\updefault}{\color[rgb]{0,0,0}$\neg q \vee c_2$}%
}}}}
\put(9976,-4261){\makebox(0,0)[b]{\smash{{\SetFigFont{12}{24.0}
{\rmdefault}{\mddefault}{\updefault}{\color[rgb]{0,0,0}$\neg t_1 \vee \neg t_2 \vee \neg c_1 \vee \neg c_2 \vee e_1 \vee \neg s_1$}%
}}}}
\put(10051,-7111){\makebox(0,0)[b]{\smash{{\SetFigFont{12}{24.0}
{\rmdefault}{\mddefault}{\updefault}{\color[rgb]{0,0,0}$\neg q \vee e_1 \vee \neg s_1$}%
}}}}
\put(11026,-7561){\makebox(0,0)[b]{\smash{{\SetFigFont{12}{24.0}
{\rmdefault}{\mddefault}{\updefault}{\color[rgb]{0,0,0}$s_1 \vee \neg q$}%
}}}}
\put(11851,-7111){\makebox(0,0)[b]{\smash{{\SetFigFont{12}{24.0}
{\rmdefault}{\mddefault}{\updefault}{\color[rgb]{0,0,0}$e_1 \vee \neg q$}%
}}}}
\end{picture}%
\nop{
      -x1 v t1           -t1 v -t2 v -c1 v -c2 v e1 v -s1
          |                    |
       ---+---> -q v t1 -------+
x1 v -q                        |
       ---+---> -q v c1 -------+
          |                    |
      -x1 v c1                 |
                               |
(e1 v -q)                      |
                               |
                               |
(x2 v -q)                      |
                               |
      -e2 v c2                 |
          |                    |
      ----+---> -q v c2 -------+
e2 v -q                        |
      ----+---> -q v t2 -------+
          |                    |
      -e2 v t2                 V
                         -q v e1 v -s1 ----+----> e1 v -q
                                           |
      -x2 v c1                          s1 v -q
      -e1 v c2
}
\end{center}

For each index $i$, at least one among $x_i \vee \neg q$ and $e_i \vee \neg q$
is necessary for deriving $\neg q \vee t_i$, which is required for these
derivations to work. Alternatively, $\neg q \vee t_i$ may be selected. Either
way, for each index $i$ at least a two-literal clause is necessary.

The claim is formally proved in four steps: first, a non-minimal way to forget
all variables except $X_C$ is shown; second, its superirredundant clauses are
identified; third, an equivalent formula of size $k$ is built if $F$ is
satisfiable; fourth, the necessary clauses in every equivalent formula are
identified; fifth, if $F$ is unsatisfiable every equivalent formula is proved
to have size greater than $k$.

\

{\bf Effect of forgetting.}

Theorem~\ref{resolve-out} proves that forgetting all variables not in $X_C$,
which are $O \cup P$, is expressed by resolving out these variables. Since
$o_i$ occurs only in $x_i \vee \neg o_i$ and $o_i \vee \neg q$, the result is
$x_i \vee \neg q$. The same holds for $p_i$. The resulting clauses are denoted
$A_R$:

\[
A_R = \{x_i \vee \neg q, e_i \vee \neg q \mid x_i \in X\}
\]

\

{\bf Superirredundancy.}

The claim requires $A$ to be minimal, which follows from all its clauses being
superirredundant by Lemma~\ref{minimal}. Most of them survive forgetting; the
reduction is based on these being superirredundant. Instead of proving
superirredundancy in two different but similar formulae, it is proved in their
union.

In particular, the clauses
{} $A_F \cup A_T \cup A_C \cup A_B$
are shown superirredundant in
{} $A_F \cup A_R \cup A_T \cup A_C \cup A_B$.
Lemma~\ref{superset} implies that they are also superirredundant in its subsets
{} $A_F \cup A_T \cup A_C \cup A_B$
and
{} $A_R \cup A_T \cup A_C \cup A_B$,
the formula before and after forgetting.

To be precise, the latter is just one among the formulae expressing forgetting.
Yet, its superirredundant clauses are in all minimal CNF formulae equivalent to
it. Therefore, all minimal CNF formulae expressing forgetting contain them.

Superirredundancy is proved via Lemma~\ref{set-value}: a substitution
simplifies
{} $A_F \cup A_R \cup A_T \cup A_C \cup A_B$
enough to prove superirredundancy easily, for example because its clauses do
not resolve and Lemma~\ref{no-resolution} applies.

\begin{itemize}

\item

Replacing all variables $x_i$, $e_i$, $t_i$ and $c_j$ with $\true$ removes from
{} $A_F \cup A_R \cup A_T \cup A_C \cup A_B$
all clauses in $A_R \cup A_T \cup A_C$, all clauses of $A_F$ but $o_i \vee \neg
q$ and $p_i \vee \neg q$ and all clauses of $A_B$ but $r_i \vee \neg q$ and
$s_i \vee \neg q$. The remaining clauses contain only the literals $o_i$,
$p_i$, $r_i$, $s_i$ and $\neg q$. Therefore, they do not resolve. Since none is
contained in another, they are all superirredundant by
Lemma~\ref{no-resolution}. This proves the superirredundancy of all clauses
$o_i \vee \neg q$, $p_i \vee \neg q$, $r_i \vee \neg q$ and $s_i \vee \neg q$.

\item

Replacing all variables $q$, $o_i$, $p_i$, $r_i$ and $s_i$ with $\false$
removes from
{} $A_F \cup A_R \cup A_T \cup A_C \cup A_B$
all clauses but $A_T \cup A_C$. These clauses contain only the literals $\neg
x_i$, $\neg e_i$, $t_i$ and $c_j$. Therefore, they do not resolve. Since they
are not contained in each other, Lemma~\ref{no-resolution} proves them
superirredundant.

\item

Replacing all variables $q$, $r_i$ and $s_i$ with $\false$ and all variables
$t_i$ and $c_i$ with $\true$ removes from
{} $A_F \cup A_R \cup A_T \cup A_C \cup A_B$
all clauses but $x_i \vee \neg o_i$ and $e_i \vee \neg p_i$. They do not
resolve because they do not share variables. Lemma~\ref{no-resolution} proves
them superirredundant because they do not contain each other.

\item

Replacing all variables with $\false$ except for all variables $t_i$ and $c_j$
and the two variables $x_h$ and $r_h$ removes all clauses from
{} $A_F \cup A_R \cup A_T \cup A_C \cup A_B$
but
{} $\neg x_h \vee t_h$,
{} $\neg t_1 \vee \cdots \vee \neg t_n \vee
{}  \neg c_1 \vee \cdots \vee \neg c_m \vee
{}  x_h \vee \neg r_h$
and all clauses
{} $\neg x_h \vee c_j$ with $x_h \in f_j$.
They only resolve in tautologies. Therefore, their resolution closure only
contains them. Removing
{} $\neg t_1 \vee \cdots \vee \neg t_n \vee
{}  \neg c_1 \vee \cdots \vee \neg c_m \vee
{}  x_h \vee \neg r_h$
from the resolution closure leaves only
{} $\neg x_h \vee t_h$
and all clauses
{} $\neg x_h \vee c_j$ with $x_h \in f_j$.
They do not resolve since they do not contain opposite literals. Since
{} $\neg t_1 \vee \cdots \vee \neg t_n \vee
{}  \neg c_1 \vee \cdots \vee \neg c_m \vee
{}  x_h \vee \neg r_h$
is not contained in them, it is not entailed by them. This proves it
superirredundant. A similar replacement proves the superirredundancy of each
{} $\neg t_1 \vee \cdots \vee \neg t_n \vee
{}  \neg c_1 \vee \cdots \vee \neg c_m \vee
{}  e_h \vee \neg s_h$.

\end{itemize}

These points prove that the clauses
{} $A_F \cup A_T \cup A_C \cup A_B$
are superirredundant in the formula before forgetting and the clauses
{} $A_T \cup A_C \cup A_B$
are superirredundant in the formula after forgetting. The only clauses that may
be superredundant are $A_R$ in the formula after forgetting.

\

{\bf Formula $F$ is satisfiable.}

Let $M$ be a model satisfying $F$. Forgetting all variables except $X_C$ is
expressed by $A_R' \cup A_T \cup A_C \cup A_B$, where $A_R'$ comprises the
clauses $x_i \vee \neg q$ such that $M \models x_i$ and the clauses $e_i \vee
\neg q$ such that $M \models \neg x_i$. This Horn formula has size $k$. It
expresses forgetting because it is equivalent to $A_R \cup A_T \cup A_C \cup
A_B$. This is proved by showing that it entails every clause in $A_R$.

Since $M$ satisfies every clause $f_j \in F$, it satisfies at least a literal
of $f_j$: for some $x_i$, either $x_i \in f_j$ and $M \models x_i$ or $\neg x_i
\in f_j$ and $M \models \neg x_i$. By construction, $x_i \in f_j$ implies $\neg
x_i \vee c_j \in A_C$ and $\neg x_i \in f_j$ implies $\neg e_i \vee c_j \in
A_C$. Again by construction, $M \models x_i$ implies $x_i \vee \neg q \in A_R'$
and $M \models \neg x_i$ implies $e_i \vee \neg q \in A_R'$. As a result,
either
{} $x_i \vee \neg q \in A_R'$ and $\neg x_i \vee c_j \in A_C$
or 
{} $e_i \vee \neg q \in A_R'$ and $\neg e_i \vee c_j \in A_C$.
In both cases, the two clauses resolve in $c_j \vee \neg q$.

Since $M$ satisfies either $x_i$ or $\neg x_i$, either $x_i \vee \neg q \in
A_R'$ or $e_i \vee \neg q \in A_R'$. The first clause resolves with $\neg x_i
\vee t_i$ and the second with $\neg e_i \vee t_i$. The result is $t_i \vee \neg
q$ in both cases.

Resolving all these clauses $t_i \vee \neg q$ and $c_j \vee \neg q$ with
{} $\neg t_1 \vee \cdots \vee \neg t_n \vee
{}  \neg c_1 \vee \cdots \vee \neg c_m \vee
{}  x_i \vee \neg r_i$
and then with $r_i \vee \neg q$, the result is $x_i \vee \neg q$. In the same
way, resolving these clauses with
{} $\neg t_1 \vee \cdots \vee \neg t_n \vee
{}  \neg c_1 \vee \cdots \vee \neg c_m \vee
{}  e_i \vee \neg s_i$
and $s_i \vee \neg q$ produces $e_i \vee \neg q$. This proves that all clauses
of $A_R$ are entailed.

\

{\bf Necessary clauses}

All CNF formulae that are equivalent to $A_R \cup A_T \cup A_C \cup A_B$ and
have minimal size contain
{} $A_T \cup A_C \cup A_B$
because these clauses are superirredundant. Therefore, these formulae are
{} $A_N \cup A_T \cup A_C \cup A_B$
for some set of clauses $A_N$. This set $A_N$ is now proved to contain either
{} $x_h \vee \neg q$,
{} $x_h \vee \neg r_h$,
{} $e_h \vee \neg q$,
{} $e_h \vee \neg s_h$
or
{} $t_h \vee \neg q$
for each index $h$. Let $M$ and $M'$ be the following models.

\begin{eqnarray*}
M &=&
\{x_i = e_i = t_i = \true \mid i \not= h\} \cup
\{x_h = e_h = t_h = \false\} \cup \\
&&
\{c_j = \true\} \cup
\{q = \true\} \cup
\{r_i = \true, s_i = \true\}
\\
M' &=&
\{x_i = e_i = t_i = \true \mid i \not= h\} \cup
\{x_h = e_h = t_h = \true \} \cup \\
&&
\{c_j = \true\} \cup
\{q = \true\} \cup
\{r_i = \true, s_i = \true\}
\end{eqnarray*}

The five clauses are falsified by $M$. Since the two of them $x_h \vee \neg q$
and $e_h \vee \neg q$ are in $A_R$, this set is also falsified by $M$. As a
result, $M$ is not a model of
{} $A_R \cup A_T \cup A_C \cup A_B$.
This formula is equivalent to
{} $A_N \cup A_T \cup A_C \cup A_B$,
which is therefore falsified by $M$. In formulae,
{} $M \not\models A_N \cup A_T \cup A_C \cup A_B$.

The formula $A_N \cup A_T \cup A_C \cup A_B$ contains a clause falsified by
$M$. Since $M \models A_T \cup A_C \cup A_B$, this clause is in $A_N$ but not
in $A_T \cup A_C \cup A_B$. In formulae, $M \not\models c$ for some $c \in A_N$
and $c \not\in A_T \cup A_C \cup A_B$. This clause is entailed by
{} $A_R \cup A_T \cup A_C \cup A_B$
because this formula entails all of
{} $A_N \cup A_T \cup A_C \cup A_B$,
and $c$ is in $A_N$.
In formulae,
{} $A_R \cup A_T \cup A_C \cup A_B \models c$.

This clause $c$ contains either
{} $x_h$, $e_h$ or $t_h$.
This is proved by deriving a contradiction from the assumption that $c$ does
not contain any of these three literals. Since $M \not\models c$, the clause
$c$ contains only literals that are falsified by $M$. Not all of them: it does
not contain
{} $x_h$, $e_h$ and $t_h$
by assumption. It does not contain $\neg x_h$, $\neg e_h$ and $\neg t_h$ either
because it would otherwise be satisfied by $M$. As a result, $c$ is also
falsified by $M'$, which is the same as $M$ but for the values of
{} $x_h$, $e_h$ and $t_h$.
At the same time, $M'$ satisfies
{} $A_R \cup A_T \cup A_C \cup A_B$,
contradicting
{} $A_R \cup A_T \cup A_C \cup A_B \models c$.
This contradiction proves that $c$ contains either
{} $x_h$, $e_h$ or $t_h$.

From the fact that $c$ contains either $x_h$, $e_h$ or $t_h$, that is a
consequence of $A_R \cup A_T \cup A_C \cup A_B$, and that is in a minimal-size
formula, it is now possible to prove that $c$ contains either
{} $x_h \vee \neg q$,
{} $x_h \vee \neg r_h$,
{} $e_h \vee \neg q$,
{} $e_h \vee \neg s_h$
or
{} $t_h \vee \neg q$.

Since $c$ is entailed by
{} $A_R \cup A_T \cup A_C \cup A_B$,
a subset of $c$ follows from resolution from it:
{} $A_R \cup A_T \cup A_C \cup A_B \vdash c'$ with $c' \subseteq c$.
This implies
{} $A_N \cup A_T \cup A_C \cup A_B \models c'$
by equivalence. If $c' \subset c$, then
{} $A_N \cup A_T \cup A_C \cup A_B$
would not be minimal because it contained a non-minimal clause $c \in A_N$.
Therefore,
{} $A_R \cup A_T \cup A_C \cup A_B \vdash c$.

The only two clauses of
{} $A_R \cup A_T \cup A_C \cup A_B$
that contain $x_h$ are $x_h \vee \neg q$ and
{} $\neg t_1 \vee \cdots \vee \neg t_n \vee
{}   \neg c_1 \vee \cdots \vee \neg c_m \vee
{}   x_h \vee \neg r_h$.
They contain either $\neg q$ or $\neg r_h$. These literals are only resolved
out by clauses containing their negations $q$ and $r_h$. No clause contains $q$
and the only clause that contains $r_h$ is $r_h \vee \neg q$, which contains
$\neg q$. If a result of resolution contains $x_h$, it also contains either
$\neg q$ or $\neg r_h$. This applies to $c$ because it is a result of
resolution.

The same applies if $c$ contains $e_h$: it also contains either $\neg q$ or
$\neg s_i$.

The case of $t_h \in c$ is a bit different. The only two clauses of
{} $A_R \cup A_T \cup A_C \cup A_B$
that contain $t_h$ are $\neg x_h \vee t_h$ and $\neg e_h \vee t_h$. Since both
are in $A_T$ and $c \not\in A_T$, they are not $c$. The first clause $\neg x_h
\vee t_h$ only resolves with $x_i \vee \neg q$ or
{} $\neg t_1 \vee \cdots \vee \neg t_n \vee
{}   \neg c_1 \vee \cdots \vee \neg c_m \vee
{}   x_h \vee \neg r_h$,
but resolving with the latter generates a tautology. The result of resolving
$\neg x_h \vee t_h$ with $x_i \vee \neg q$ is $t_h \vee \neg q$; no clause
contains $q$. Therefore, $c$ can only be $t_h \vee \neg q$. The second clause
$\neg e_h \vee t_h$ leads to the same conclusion.

In summary, $c$ contains either
{} $x_h \vee \neg q$,
{} $x_h \vee \neg r_i$,
{} $e_h \vee \neg q$,
{} $e_h \vee \neg s_i$ or
{} $t_h \vee \neg q$.
In all these cases it contains at least two literals. This is the case for
every index $h$; therefore, $A_N$ contains at least $n$ clauses of two
literals. Every minimal CNF formula equivalent to
{} $A_R \cup A_T \cup A_C \cup A_B$
has size at least $2 \times n$ plus the size of $A_T \cup A_C \cup A_B$. This
sum is exactly $k$. This proves that every minimal CNF formula expressing
forgetting contains at least $k$ literal occurrences. Worded differently, every
CNF formula expressing forgetting has size at least $k$.

\

{\bf Formula $F$ is unsatisfiable}

The claim is that no CNF formula of size $k$ expresses forgetting if $F$ is
unsatisfiable. This is proved by deriving a contradiction from the assumption
that such a formula exists.

It has been proved that every CNF formula expressing forgetting is equivalent
to
{} $A_R \cup A_T \cup A_C \cup A_B$
and that the minimal equivalent CNF formulae are
{} $A_N \cup A_T \cup A_C \cup A_B$
for some set $A_N$ that contains clauses that include either
{} $x_h \vee \neg q$,
{} $x_h \vee \neg r_i$,
{} $e_h \vee \neg q$,
{} $e_h \vee \neg s_i$ or
{} $t_h \vee \neg q$
for each index $h$.

If $A_N$ contains other clauses, or more than one clause for each $h$, or these
clauses contain other literals, the size of
{} $A_N \cup A_T \cup A_C \cup A_B$
is larger than
{} $k = 2 \times n + ||A_T|| + ||A_C|| + ||A_B||$,
contradicting the assumption. This proves that every formula of size $k$ that
is equivalent to
{} $A_R \cup A_T \cup A_C \cup A_B$
is equal to
{} $A_N \cup A_T \cup A_C \cup A_B$
where $A_N$ contains exactly one clause among
{} $x_h \vee \neg q$,
{} $x_h \vee \neg r_i$,
{} $e_h \vee \neg q$,
{} $e_h \vee \neg s_i$ or
{} $t_h \vee \neg q$
for each index $h$.

The case
{} $x_h \vee \neg r_h \in A_N$
is excluded. It would imply
{} $A_R \cup A_T \cup A_C \cup A_B \models x_h \vee \neg r_h$,
which implies the redundancy of
{} $\neg t_1 \vee \cdots \vee \neg t_n \vee
{}   \neg c_1 \vee \cdots \vee \neg c_m \vee
{}   x_h \vee \neg r_h \in A_B$
contrary to its previously proved superirredundancy.
A similar argument proves
{} $e_h \vee \neg s_h \not\in A_N$.

The conclusion is that every formula of size $k$ that is equivalent to
{} $A_R \cup A_T \cup A_C \cup A_B$
is equal to
{} $A_N \cup A_T \cup A_C \cup A_B$
where $A_N$ contains exactly one clause among
{} $x_h \vee \neg q$,
{} $e_h \vee \neg q$,
{} $t_h \vee \neg q$
for each index $h$.

If $F$ is unsatisfiable, all such formulae are proved to be satisfied by a
model that falsifies
{} $A_R \cup A_T \cup A_C \cup A_B$,
contrary to the assumed equivalence.

Let $M$ be the model that assigns $q = \true$ and $t_i=\true$, and assigns
$x_i=\true$ and $e_i=\false$ if $x_i \vee \neg q \in A_N$ and $x_i=\false$ and
$e_i=\true$ if $e_i \vee \neg q \in A_N$ or $t_i \vee \neg q \in A_N$. All
clauses of $A_N$ and $A_T$ are satisfied by $M$.

This model $M$ can be extended to satisfy all clauses of $A_C \cup A_B$. Since
$F$ is unsatisfiable, $M$ falsifies at least a clause $f_j \in F$. Let $M'$ be
the model obtained by extending $M$ with the assignments of $c_j$ to false, all
other variables in $C$ to $\true$ and all variables $r_i$ and $s_i$ to $\true$.
This extension satisfies all clauses of $A_B$ either because it sets $c_j$ to
false or because it sets $r_i$ and $s_i$ to true. It also satisfies all clauses
of $A_C$ that do not contain $c_j$ because it sets all variables of $C$ but
$c_j$ to true.

The only clauses that remain to be proved satisfied are the clauses of $A_C$
that contain $c_j$. They are
{} $\neg x_i \vee c_j$ for all $x_i \in f_j$
and
{} $\neg e_i \vee c_j$ for all $\neg x_i \in f_j$.
Since $M'$ falsifies $f_j$, it falsifies every $x_i \in f_j$; therefore, it
satisfies $\neg x_i \vee c_j$. Since $M'$ falsifies $f_j$, it falsifies every
$\neg x_i \in f_j$; since by construction it assigns $e_i$ opposite to $x_i$,
it falsifies $e_i$ and therefore satisfies $\neg e_i \vee c_j$.

This proves that $M'$ satisfies
{} $A_N \cup A_T \cup A_C \cup A_B$.
It does not satisfy
{} $A_R \cup A_T \cup A_C \cup A_B$.
If $x_1 \vee \neg q \in A_N$, then $M'$ sets $x_1$ to $\true$ and $e_1$ to
$\false$; therefore, it does not satisfy $e_1 \vee \neg q \in A_R$. Otherwise,
$M'$ sets $x_1$ to false and $e_1$ to true; therefore, it does not satisfy $x_1
\vee \neg q \in A_N$.

This contradicts the assumption that
{} $A_N \cup A_T \cup A_C \cup A_B$
is equivalent to
{} $A_R \cup A_T \cup A_C \cup A_B$.
The assumption that it has size $k$ is therefore false.~\qed

\relemma{horn-hard}
{\em
Checking whether forgetting some variables from a minimal-size Horn formula is
expressed by a Horn formula bounded by a certain size is \Dp-hard.
}

\

\proof For every CNF formula $F$, Lemma~\ref{horn-conp} ensures the existence
of a minimal-size Horn formula $A$, a set of variables $X_A$ and an integer $k$
such that forgetting all variables except $X_A$ from $A$ is expressed by a Horn
formula of size $k$ if $F$ is unsatisfiable and is only expressed by larger CNF
formulae otherwise.

For every CNF formula $G$, Lemma~\ref{horn-np} ensures the existence of a
minimal-size Horn formula $B$, a set of variables $X_B$ and an integer $l$ such
that forgetting all variables except $X_B$ from $B$ is expressed by a Horn
formula of size $l$ if $G$ is satisfiable and is only expressed by larger CNF
formulae otherwise.

The prototypical \Dp-hard problem is that of establishing whether a formula $F$
is satisfiable and another $G$ is unsatisfiable. If the alphabets of the two
formulae $G$ and $F$ are not disjoint, they can be made so by renaming one of
them to fresh variables because renaming does not affect satisfiability. The
same applies to the formulae $B$ and $A$ respectively build from them according
to Lemma~\ref{horn-conp} and Lemma~\ref{horn-np} because renaming does not
change the minimal size of forgetting either. Lemma~\ref{independent} proves
that $A \cup B$ can be minimally expressed by $C \cup D$ where $C$ minimally
expresses forgetting from $A$ and $D$ from $B$. The size of these two formulae
are $l$ and $k$ if $G$ is unsatisfiable and $F$ satisfiable. If $G$ is
satisfiable, then $D$ is larger than $k$ while $C$ is still large at least $l$;
the minimal expression of forgetting $A \cup B$ is therefore strictly larger
than $k+l$. The same happens if $F$ is unsatisfiable.

This proves that the problem of checking the satisfiability of a formula and
the unsatisfiability of another reduces to the problem of checking the size of
the minimal expression of forgetting from Horn formulae.~\qed

\retheorem{horn-complexity}
{\em
Checking whether forgetting some variables from a Horn formula is expressed by
a Horn formula bounded by a certain size expressed in unary is \Dp-hard
and in \S{2}, and remains hard even if the formula is restricted to be of
minimal size.
}

\

\proof The problem belongs to \S{2} because it can be expressed as the
existence of a formula of the given size or less that expresses forgetting the
given variables from the formula. In turn, expressing forgetting is by
Theorem~\ref{consistent-literals-complete} the same as the equiconsistency with
a set of literals containing all variables not to be forgotten. This condition
can be expressed by the following metaformula where $A$ is the formula,
$Y${\plural} are the variables not to be forgotten and $k$ the size bound.

\[
\exists B ~.~
	||B|| \leq k \mbox{ and }
	\forall S ~.~ \var(S) \subseteq Y \Rightarrow
		(S \cup A \not\models \bot
		\Leftrightarrow
		S \cup B \not\models \bot)
\]

Both $B$ and $S$ are bounded in size: the first by $k$, the second by the
number of variables in $Y$. Since consistency is polynomial for Horn formulae,
this is a $\exists\forall$QBF, which proves membership to \S{2}.

Hardness for \Dp\  is proved by Lemma~\ref{horn-hard}.~\qed

\relemma{general-p2}
{\em
There exists a polynomial algorithm that turns a CNF formula $F$ into a
minimal-size CNF formula $A$, a subset $X_C \subseteq \var(A)$ and a number $k$
such that forgetting all variables from $A$ except $X_C$ is expressed by a CNF
formula of size $k$ if $\forall X \exists Y . F$ is valid and only by CNF
formulae of size $k+2$ or greater otherwise.
}

\

\proof Let $F = \{f_1, \ldots, f_m\}$ and $X = \{x_1,\ldots,x_n\}$. Checking
the validity of $\forall X \exists Y . F$ remains \P{2}-hard even if $F$ is
satisfiable: if $F$ is not satisfiable, $\forall X \exists Y . F$ can be turned
into the equivalent formula $\forall X \cup \{s\} \exists Y . s \vee F$, and $s
\vee F$ is satisfiable.

The reduction is based on an extended alphabet with the additional fresh
variables
{} $E = \{e_1,\ldots,e_n\}$,
{} $C = \{c_1,\ldots,c_m\}$ and
{} $\{a,b,q,r\}$.
The formula $A$, the set of variables $X_C$ and the number $k$ are:

\begin{eqnarray*}
A &=&
	\{ f_j \vee c_j \vee q \mid f_j \in F\} \cup			\\
&&
	\{\neg c_j \vee r \mid f_j \in F\} \cup				\\
&&
	\{\neg r \vee \neg a \vee b \vee q\} \cup			\\
&&
	\{a \vee \neg b \vee q\} \cup					\\
&&
	\{x_i \vee e_i \mid x_i \in X\}					\\
X_C &=& X \cup E \cup \{a,b,q\}						\\
k &=& 2 \times n + 3
\end{eqnarray*}

A short explanation of how the reduction works precedes its formal proof. The
key is how a model over $X \cup \{q\}$ extends to a model of $A$, in particular
its possible values of $a$ and $b$. All models over $X \cup \{q\}$ that satisfy
$q$ can be extended to satisfy $A$: all clauses not containing $q$ are
satisfied by setting $r=\true$ and $e_i$ opposite to $x_i$; satisfaction is not
affected by the values $a$ and $b$. The remaining models set $q = \false$. For
these models, the truth of a clause $f_j$ makes $f_j \vee c_j \vee q$ satisfied
even if $c_j = \false$. In turn, $c_j = \false$ satisfies
{} $\neg c_j \vee r$
even if $r = \false$, which satisfies
{} $\neg r \vee \neg a \vee b \vee q$
regardless of the values of $a$ and $b$; the values of $a$ and $b$ only need to
satisfy $a \vee \neg b \vee q$. Otherwise, the falsity of $f_j$ imposes $c_j =
\true$ to satisfy $f_j \vee c_j \vee q$, which makes
{} $\neg c_j \vee r$ 
require $r=\true$, which turns
{} $\neg r \vee \neg a \vee b \vee q$
into
{} $\neg a \vee b \vee q$,
making the literals $\neg a$ and $b$ necessary in addition to $a$ and $\neg b$.

% The only function of the clauses $x_i \vee e_i$ is to make all literals $x_i$
% necessary in $A$.

The proof comprises four steps: first, $A$ is proved minimal as required by the
claim of the lemma; second, $k$ literals that are in every formula that
expresses forgetting regardless of the validity of the QBF are identified;
third, a formula of size $k$ expressing forgetting when the QBF is valid is
determined; fourth, every formula expressing forgetting contains at least two
further literals if the QBF is invalid.

\

\noindent {\bf Minimality of $A$.}

Follows from Lemma~\ref{minimal} since all clauses of $A$ are superirredundant.
This is in turn proved by showing substitutions that disallow all resolutions,
which proves the superredundancy of the remaining clauses by
Lemma~\ref{set-value} and Lemma~\ref{no-resolution}.

The substitution that replaces with $\true$ the variables $a$, $b$, $r$, all
$e_i$ and all $c_j$ with $j \not= h$ for every given $h$ such that $f_h \in F$
removes all clauses but $f_h \vee c_h \vee q$, which is therefore
superirredundant.

The clauses $\neg c_j \vee r$ are proved superirredundant by substituting $q$
and all variables $e_i$ with $\true$, which removes all other clauses. The
clauses $\neg c_j \vee r$ do not resolve because they do not contain opposite
literals.

Two other clauses are proved superirredundant by the substitution that replaces
all variables $e_i$ with $\true$, all $c_j$ with $\false$, and $X \cup Y$ with
some values that satisfy $F$; such values exist because $F$ is by assumption
satisfiable. This substitution removes all clauses but
{} $\neg r \vee \neg a \vee b \vee q$
and
{} $a \vee \neg b \vee q$,
which only resolve in tautologies.

Finally, the clauses $x_h \vee e_h$ are proved superirredundant by replacing
$q$ and $r$ with $\true$, which removes all other clauses. Since the clauses
$x_h \vee e_h$ only contain positive literals, they do not resolve.

\

\noindent {\bf Necessary literals.}

Regardless of the validity of $\forall X \exists Y . F$, the literals $X \cup E
\cup \{a,\neg b,q\}$ are necessary in every CNF formula that expresses
forgetting all variables except $X_C$ from $A$. This is proved by
Lemma~\ref{forget-contains}, exhibiting a set of literals $S$ such that $S \cup
A$ is consistent, but $S \backslash \{l\} \cup \{\neg l\} \cup A$ is not for
every $l \in X \cup E \cup \{a,\neg b,q\}$.

The first set is
{} $S = \{x_i, \neg e_i, a, b, \neg q\}$,
which is consistent with $A$ because of the model that satisfies $S$ and
assigns $r$ and all variables $c_j$ to $\true$. Changing $x_i$ to $\neg x_i$
violates the clause $x_i \vee e_i$. Changing $a$ to false violates $a \vee \neg
b \vee q$. This proves that $a$ and all variables $x_i$ are necessary by
Lemma~\ref{forget-contains}.

The second set is
{} $S = \{\neg x_i, e_i, \neg a, \neg b, \neg q\}$,
which is consistent with $A$ because of the model that satisfies $S$ and
assigns $r$ and all variables $c_j$ to $\true$. Changing $e_i$ to $\neg e_i$
violates $x_i \vee e_i$, changing $b$ to $\true$ violates $a \vee \neg b \vee
q$. This proves that $e_i$ and $\neg b$ are necessary by
Lemma~\ref{forget-contains}.

The third set is
{} $S = \{x_i, e_i, \neg a, b, q\}$,
which is consistent with $A$ because of the model that satisfies $S$ and
assigns $r$ and all variables $c_j$ to $\true$. Changing $q$ to false violates
the clause $a \vee \neg b \vee q$, proving that $q$ is necessary.

In summary, all literals in $X \cup E \cup \{a, \neg b, q\}$ occur in every CNF
formula expressing forgetting all variables except $X_C$ from $A$. These
literals are $2 \times n + 3$. This is a part of the claim: no CNF formula
expressing forgetting is smaller than $2 \times n + 3$.

\

\noindent {\bf Forgetting when $\forall X \exists Y . F$ is valid}

If $\forall X \exists Y . F$ is valid, forgetting is expressed by
{} $B = \{a \vee \neg b \vee q\} \cup \{x_i \vee e_i \mid x_i \in X\}$,
which has the required size $k = 2 \times n + 3$ and variables $X_C = X \cup E
\cup \{a,b,q\}$. Theorem~\ref{consistent-literals-complete} proves that this
formula expresses forgetting: every set $S$ of literals of $X_C$ that contains
all variables of $X_C$ is consistent with $B$ if and only if it is consistent
with $A$.

Since $B$ only contains clauses of $A$, every set of literals $S$ that is
consistent with $A$ is also consistent with $B$. The claim follows from proving
the converse for every set of literals $S$ over $X_C$ that mentions all
variables of $X_C$.

The assumption is that $S \cup B$ is consistent; the claim is that $S \cup A$
is consistent. Since $S \cup B$ is consistent, it has a model $M$. Let $M_X$ be
its restriction to the variables $X$ and $M_Y'$ to $Y$. By assumption, $\forall
X \exists Y . F$ is valid. Therefore, $M_X \cup M_Y$ satisfies $F$ for some
truth evaluation $M_Y$ over $Y$. Since $S$ is satisfied by $M$ and does not
mention any variable $Y$, it is also satisfied by $M \backslash M_Y' \cup M_Y$.
The truth evaluation
{} $M_C = \{c_j = \false \mid f_j \in F\} \cup \{r = \false\}$
satisfies all clauses $\neg c_j \vee r$ and $\neg r \vee \neg a \vee b \vee q$.
Since $M_X \cup M_Y$ satisfies all clauses $f_j \in F$, the union $M \backslash
M_Y' \cup M_Y \cup M_C$ satisfies all clauses $f_j \vee c_j \vee q$ of $A$.
This proves that $M \backslash M_Y' \cup M_Y \cup M_C$ satisfies all clauses of
$A$ that $B$ does not contain.

\

\noindent {\bf Forgetting when $\forall X \exists Y . F$ is invalid}

All CNF formulae that express forgetting have been proved to mention $X \cup E
\cup \{a, \neg b, q\}$. If $\forall X \exists Y . F$ is invalid, they all
mention $\neg a$ and $b$ as well.

This is proved by Lemma~\ref{forget-contains}: a set of literals $S$ over $X_C$
is shown to be consistent with $A$ while $S \backslash \{\neg a\} \cup \{a\}$
is not. A similar set is shown for $b$.

Since $\forall X \exists Y . F$ is invalid, for some interpretation $M_X$ over
$X$ the interpretation $M_X \cup M_Y$ falsifies $F$ for every interpretation
$M_Y$ over $Y$. The required set $S$ is built from $M_X$: it contains the
literals over $x_i$ that are satisfied by $M_X$ and $\neg a$, $\neg b$ and
$\neg q$.

\[
S =
\{x_i \mid M_X \models x_i\} \cup \{\neg x_i \mid M_X \models \neg x_i\}
\cup
\{\neg a, \neg b, \neg q\}
\]

By construction, $M_X$ satisfies the first part of $S$.
The model
{} $M_O = \{a = \false, b = \false, q = \false\}$
satisfies the second. Therefore, $M_X \cup M_O$ satisfies $S$.

The consistency of $S \cup A$ is shown by proving that $M_X \cup M_O$ can be
extended to the other variables to satisfy $A$. This extension is $M_X \cup M_Y
\cup M_O \cup M_N \cup M_C$, where $M_Y$ is an arbitrary model over $Y$, $M_N$
assigns every $e_i$ opposite to $x_i$ in $M_X$ and
{} $M_C$ is $\{c_j = \true \mid f_j \in F\} \cup \{r = \true\}$.
The clauses $f_j \vee c_j \vee q$ are satisfied because $c_j$ is true, the
clauses $\neg c_j \vee r$ because $r$ is $\true$, the clause $\neg r \vee \neg
a \vee b \vee q$ because $a$ is $\false$, $a \vee \neg b \vee q$ because $b$ is
$\false$, the clauses $x_i \vee e_i$ because $M_N \models e_i$ if $M_X
\not\models x_i$.

This proves that $M_X \cup M_O \cup M_N \cup M_C$ satisfies $S \cup A$, which
is therefore satisfiable.

The claim is a consequence of $S' = S \backslash \{\neg a\} \cup \{a\}$ being
inconsistent with $A$.

\[
S' =
\{x_i \mid M_X \models x_i\} \cup \{\neg x_i \mid M_X \models \neg x_i\}
\cup
\{a, \neg b, \neg q\}
\]

This is proved by contradiction: a model $M'$ is assumed to satisfy $S' \cup
A$. Since $M'$ satisfies $S'$, it assigns the variables $x_i$ the same as
$M_X$. Let $M_Y$ be the restriction of $M'$ to the variables $Y$. By
assumption, $M_X$ is a model over $X$ that cannot be extended to $Y$ to satisfy
$F$. As a result, $M_X \cup M_Y \not\models F$. Therefore, $M'$ falsifies at
least a clause $f_j \in F$. Since $M'$ satisfies $f_j \vee c_j \vee q$ but
falsifies both $f_j$ and $q$, it satisfies $c_j$. It also satisfies $r$ because
it satisfies $\neg c_j \vee r$ and falsifies $c_j$. Since $M'$ satisfies $S'$
it satisfies $a$ and falsifies $b$ and $q$. The conclusion is that all literals
of $\neg r \vee \neg a \vee b \vee q \in A$ are false, contrary to the
assumption that $M'$ satisfies $A$.

A similar set $S$ with $a$ and $b$ in place of $\neg a$ and $\neg b$ proves
that expressing forgetting also requires $b$.~\qed

\relemma{general-s2}
{\em
There exists a polynomial algorithm that turns a DNF formula $F = f_1 \vee
\cdots \vee f_m$ over variables $X \cup Y$ into a minimal-size CNF formula $A$,
a subset $X_C \subseteq \var(A)$ and a number $k$ such that forgetting all
variables except $X_C$ from $A$ is expressed by a CNF formula of size $k$ if
$\exists X \forall Y . F$ is valid, and only by larger CNF formulae otherwise.
}

\

\proof Let $F = f_1 \vee \cdots \vee f_m$ be the DNF formula over variables $X
\cup Y$. The reduction employs additional variables:
{} $O = \{o_i \mid x_i \in X\}$,
{} $E = \{e_i \mid x_i \in X\}$,
{} $P = \{p_i \mid x_i \in X\}$,
{} $T = \{t_i \mid x_i \in X\}$,
{} $D = \{d_j \mid f_j \in F\}$,
{} $R = \{r_i \mid x_i \in X\}$,
{} $S = \{s_i \mid x_i \in X\}$
and $q$. The formula $A$, the alphabet $X_C$ and the number $k$ are as follows.

\begin{eqnarray*}
A &=& A_F \cup A_T \cup A_D \cup A_B					\\
A_F &=&
\{ x_i \vee \neg o_i, o_i \vee q \mid x_i \in X \} \cup
\{ e_i \vee \neg p_i, p_i \vee q \mid x_i \in X \}			\\
A_T &=& \{ \neg x_i \vee t_i ,~ \neg e_i \vee t_i \mid x_i \in X \}	\\
A_D &=& \{ \neg (f_j[e_i/\neg x_i]) \vee d_j \mid f_j \in F \}		\\
A_B &=&
\{ \neg t_1 \vee \cdots \vee \neg t_n \vee \neg d_j \vee x_i \vee \neg r_i,
   r_i \vee q \mid x_i \in X ~, f_j \in F\} \cup			\\
&&
\{ \neg t_1 \vee \cdots \vee \neg t_n \vee \neg d_j \vee e_i \vee \neg s_i,
   s_i \vee q \mid x_i \in X ,~ f_j \in F\}				\\
X_C &=& X \cup E \cup Y \cup T \cup D \cup R \cup S \cup \{q\}		\\
k &=& 2 \times n + ||A_T \cup A_D \cup A_B||
\end{eqnarray*}

% $\neg f_j[e_i/\neg x_i]$ is $f_l$ with every $\neg x_i$ replaced by $e_i$,
% the result negated and turned into a disjunction; therefore, it contains
% $x_i$ and $e_i$ only negated; it is not the converse: first negation and
% conversion into a disjunction, then substitution; that would still allow a
% model over $X$ to be extended to $E$ by setting each $e_i$ opposite to $x_i$,
% but would not allow the resulting disjunction to resolve with $x_i \vee q$ or
% $e_i \vee q$

% forgetting is proved to be expressed by the following formula, which may not
% be minimal
% \[ A_R = \{x_i \vee q, e_i \vee q \mid x_i \in X\} \]

The reduction works because every minimal CNF formula that expresses forgetting
contains at least one among $x_h \vee q$, $e_h \vee q$ and $t_h \vee q$ for
each $h$, and all of $A_T \cup A_D \cup A_B$. This proves the lower bound $k$.
If the QBF is valid, for some evaluation over $X$ the formula $F$ is true
regardless of $Y$. Choosing the clauses $x_h \vee q$, $e_h \vee q$ or $t_h \vee
q$ that correspond to this model, some clause of $A_D$ implies $q \vee d_j$,
which allows $A_B$ to entail all remaining clauses. If the QBF is not valid, no
clause $q \vee d_j$ is entailed.

The formal proof requires five steps: first, every formula expressing
forgetting is equivalent to a certain formula $A_R \cup A_T \cup A_D \cup A_B$;
second, $A$ is a minimal CNF formula and the clauses of $A_T \cup A_D \cup A_B$
are in all minimal CNF formulae equivalent to $A_R \cup A_T \cup A_D \cup A_B$;
third, forgetting is expressed by a formula of size $k$ if the QBF is valid;
fourth, every minimal CNF formula expressing forgetting contains either $x_h
\vee q$, $e_h \vee q$ or $t_h \vee q$ for each $h$; fifth, if the QBF is
invalid then forgetting is only expressed by formulae larger than $k$.

\

{\bf Effect of forgetting.}

The variables to forget are $O \cup P$. Each is contained only in two clauses
of $A$, with opposite signs. Resolving them produces the clauses in the
following set $A_R$.

\[
A_R = \{x_i \vee q, e_i \vee q \mid x_i \in X\}
\]

By Theorem~\ref{resolve-out}, forgetting is expressed by $A_R \cup A_T \cup A_D
\cup A_B$. Therefore, all formulae that express forgetting are equivalent to
this formula.

\

{\bf Superirredundancy.}

All clauses of
{} $A_F \cup A_T \cup A_D \cup A_B$
are proved superirredundant in
{} $A_F \cup A_R \cup A_T \cup A_D \cup A_B$.
Both $A$ and
{} $A_R \cup A_T \cup A_D \cup A_B$
are subsets of this formula; therefore, the superirredundant clauses are
superirredundant in both formulae by Lemma~\ref{superset}. Since $A$ comprises
exactly them, it is minimal thanks to Lemma~\ref{minimal}. Since all formulae
expressing forgetting are equivalent to
{} $A_R \cup A_T \cup A_D \cup A_B$,
where
{} $A_T \cup A_D \cup A_B$
are superirredundant, these clauses are in all formulae expressing forgetting.

Superirredundancy is proved applying a substitution to the formula so that the
resulting clauses do not resolve and are not contained in one another. This
condition proves them superirredundant by Lemma~\ref{no-resolution}.
Lemma~\ref{set-value} implies their superirredundancy in the original formula.

Replacing all variables $X$, $E$, $T$ and $D$ with $\true$ removes from the
formula
{} $A_F \cup A_R \cup A_T \cup A_D \cup A_B$
all clauses but
{} $o_i \vee q$, $p_i \vee q$, $r_i \vee q$ and $s_i \vee q$.
These clauses do not resolve because they only contain positive literals. None
is contained in another.

Replacing all variables $R$ and $S$ with $\false$ and all variables $T$, $D$
and $q$ with $\true$ removes from the formula
{} $A_F \cup A_R \cup A_T \cup A_D \cup A_B$
all clauses but the clauses $x_i \vee \neg o_i$ and $e_i \vee \neg p_i$. They
are not contained in one another; they do not resolve because they do not
contain opposite literals.

Replacing all variables $O$, $P$, $R$ and $S$ with $\false$ and $D$ and $q$
with $\true$ removes all clauses but $\neg x_i \vee t_i$ and $\neg e_i \vee
t_i$. These clauses do not resolve because they do not contain opposite
literals; they are not contained in one another.

Replacing all variables $O$, $P$, $R$ and $S$ with $\false$ and $T$, $D
\backslash \{d_h\}$ and $q$ with $\true$ removes all clauses but
{} $(\neg f_h[e_i/\neg x_i]) \vee d_h$,
which is therefore superirredundant.

The last substitution replaces all variables $X \backslash \{x_h\}$, $E$, $O$,
$P$, $R \backslash \{r_h\}$ and $S$ with $\false$, all variables $D \backslash
\{d_l\}$ and $q$ with $\true$, all variables $y_i$ such that $y_i \in \neg
f_l[e_i/\neg x_i]$ to $\true$ and all such that $\neg y_i \in \neg f_l[e_i/\neg
x_i]$ to $\false$. This substitution removes all clauses but
{} $\neg x_h \vee t_h$,
{} $\neg t_1 \vee \cdots \vee \neg t_n \vee \neg d_l \vee x_h \vee \neg r_i$
and possibly
{} $\neg (f_l[e_i/\neg x_i]) \vee d_l$.
The latter clause is removed if it contains some variable $y_i$. It is removed
if it contains some literal $\neg x_i$ with $i \not= h$. It is removed if it
contains some literal $\neg e_i$. The only other literals it may contain are
$\neg x_h$ and $d_l$; it contains both: $d_l$ by construction, $\neg x_h$
because otherwise $f_l$ would be empty. The remaining clauses are therefore
{} $\neg x_h \vee t_h$,
{} $\neg t_1 \vee \cdots \vee \neg t_n \vee \neg d_l \vee x_i \vee \neg r_i$
and possibly
{} $\neg x_h \vee d_h$.
These clauses only resolve in tautologies, which proves the second
superirredundant. A similar argument holds for
{} $\neg t_1 \vee \cdots \vee \neg t_n \vee \neg d_l \vee e_i \vee \neg s_i$.

\

{\bf Validity of $\exists X \forall Y . F$.}

Let $M$ be a model over variables $X$ that makes $F$ true regardless of the
values of $Y$. Let $A_R' \subseteq A_R$ be the set of clauses $x_i \vee q$ such
that $M \models x_i$ and $e_i \vee q$ such that $M \models \neg x_i$. This set
has size $2 \times n$. Therefore, $A_R' \cup A_T \cup A_D \cup A_B$ has size $k
= 2 \times n + ||A_T \cup A_D \cup A_B||$. This formula expresses forgetting if
it is equivalent to $A_R \cup A_T \cup A_D \cup A_B$, which is the case if
$A_R' \cup A_T \cup A_D \cup A_B \models A_R$. The claim is proved by showing
that $A_R' \cup A_T \cup A_D \cup A_B$ entails $A_R$.

Either $x_i \vee q$ or $e_i \vee q$ is in $A_R'$ for every $i$ and these
clauses respectively resolve with $\neg x_i \vee t_i$ and $\neg e_i \vee t_i$,
producing $t_i \vee q$ in both cases. Each clause
{} $\neg t_1 \vee \cdots \vee \neg t_n \vee \neg d_j \vee x_i \vee \neg r_i$
resolve with them and with $r_i \vee q$ to $\neg d_j \vee x_i \vee q$. This
clause further resolves with
{} $\neg (f_j[e_i/\neg x_i]) \vee d_j$
to produce
{} $\neg (f_j[e_i/\neg x_i]) \vee x_i \vee q$.
This proves that 
{} $A_R' \cup A_T \cup A_D \cup A_B$ implies every clause
{} $\neg (f_j[e_i/\neg x_i]) \vee x_i \vee q$ with $f_j \in F$.
The following equivalence holds.

\ttytex{
\begin{eqnarray*}
\{ \neg (f_j[e_i/\neg x_i]) \vee x_i \vee q | f_j \in F \}
& \equiv &
	\left(\bigwedge \{\neg (f_j[e_i/\neg x_i]) \mid f_j \in F\} \right)
	\vee x_i \vee q							\\
& \equiv &
	\neg \left(\bigvee \{f_j[e_i/\neg x_i] \mid f_j \in F\} \right)
	\vee x_i \vee q							\\
& \equiv & \neg F[e_i/\neg x_i] \vee x_i \vee q
\end{eqnarray*}
}{
{ -fj[ni/-xi] v xi v q | fj in F } =
                           = &{ -fj[ni/-xi] | fj in F } v xi v q
                           = -(v { fj[ni/-xi] | fj in F }) v xi v q
                           = -F[ni/-xi] v x_i v q
}

Since $A_R' \cup A_T \cup A_D \cup A_B$ implies the first set, it implies the
last formula: $A_R' \cup A_T \cup A_D \cup A_B \models \neg F[e_i/\neg x_i]
\vee x_i \vee q$.

Since $M$ satisfies $F$ regardless of $Y$, it follows that
{} $\{x_i \mid M \models x_i\} \cup \{\neg x_i \mid M \models \neg x_i\}
{}  \models F$.
Replacing each $\neg x_i$ with $e_i$ in both sides of this entailment turns
into
{} $\{x_i \mid M \models x_i\} \cup \{e_i \mid M \models \neg x_i\}
{}  \models F[e_i/\neg x_i]$.
Disjoining both terms with $q$ results into
{} $A_R' \models F[e_i/\neg x_i] \vee q$.

This entailment and the previously proved
{} $A_R' \cup A_T \cup A_D \cup A_B \models
{}  \neg F[e_i/\neg x_i] \vee x_i \vee q$
imply $A_R' \cup A_T \cup A_D \cup A_B \models x_i \vee q$.

The same holds for $e_i \vee q$ by symmetry. Therefore, $A_R' \cup A_T \cup A_D
\cup A_B$ implies every clause of $A_R$.

\

{\bf Necessary clauses.}

All formulae that express forgetting are equivalent to $A_R \cup A_T \cup A_D
\cup A_B$ and therefore contain all its superirredundant clauses $A_T \cup A_D
\cup A_B$. As a result, they have the form $A_N \cup A_T \cup A_D \cup A_B$ for
some set of clauses $A_N$. It is now shown that all equivalent CNF formulae of
minimal size contain a clause that include either
{} $x_h \vee q$,
{} $x_h \vee \neg r_h$,
{} $e_h \vee q$,
{} $e_h \vee \neg s_h$,
or
{} $t_h \vee q$
for each $h$.

Since
{} $A_N \cup A_T \cup A_D \cup A_B$
is equivalent to
{} $A_R \cup A_T \cup A_D \cup A_B$,
it entails
{} $x_h \vee q \in A_R$.
This clause is not satisfied by the following model.

\begin{eqnarray*}
M &=&
\{x_i = e_i = t_i = \true \mid i \not= h\}		\cup
\{x_h = e_h = t_h = \false\}				\cup		\\
&&
\{d_j = \true \mid f_j \in F\}				\cup
\{r_i = s_i = \true\}					\cup
\{q = \false\}
\end{eqnarray*}

This model satisfies all clauses of $A_T \cup A_D \cup A_B$. If $A_N$ also
satisfied it,
{} $A_N \cup A_T \cup A_D \cup A_B$
would have a model that falsifies $x_h \vee q$, which it instead entails. As a
result, $A_N$ contains a clause $c$ that $M$ falsifies. Since
{} $A_N \cup A_T \cup A_D \cup A_B$
is a formula of minimal size, it entails no proper subset of $c$. By
equivalence, the same applies to
{} $A_R \cup A_T \cup A_D \cup A_B$.

\begin{eqnarray*}
&& M \not\models c				\\
&& A_R \cup A_T \cup A_D \cup A_B \models c	\\
&& A_R \cup A_T \cup A_D \cup A_B \models c'
   \mbox{ implies } c' \not\subset c
\end{eqnarray*}

If $c$ contains neither $x_h$, $e_h$ nor $t_h$, it would still be falsified by
the model that is the same as $M$ except that it assigns $x_h$, $e_h$ and $t_h$
to $\true$. This model satisfies
{} $A_R \cup A_T \cup A_D \cup A_B$.
As a result,
{} $A_R \cup A_T \cup A_D \cup A_B \cup \neg(c)$
is consistent, contradicting $A_R \cup A_T \cup A_D \cup A_B \models c$. This
proves that $c$ contains either $x_h$, $e_h$ or $t_h$.

Since these three variables are negative in $M$ and $M \not\models c$, they are
positive in $c$. In other words, $c$ contains either $x_h$, $e_h$ or $t_h$
unnegated.

Since $c$ is entailed by $A_R \cup A_T \cup A_D \cup A_B$, but none of its
proper subsets does, it follows from resolution:
{} $A_R \cup A_T \cup A_D \cup A_B \vdash c$.

If $c$ contains $x_h$, it also contains either $q$ or $\neg r_h$. This is
proved as follows. Since $c$ is the root of a resolution tree and contains
$x_h$, this literal is also in one of the leaves of resolution. The only
clauses of
{} $A_R \cup A_T \cup A_D \cup A_B$
containing $x_h$ are $x_h \vee q$ and all clauses
{} $\neg t_1 \vee \cdots \vee \neg t_n \vee \neg d_j \vee x_h \vee \neg r_h$.
The first does not resolve over $q$ because the formula does not contain $\neg
q$. The other clauses only resolve over $r_h$ with $r_h \vee q$, which
introduces $q$, which again cannot be removed by resolution. Since $c$ is
obtained by resolution, if it contains $x_h$ it also contains either $\neg r_h$
or $q$.

By symmetry, if $c$ contains $e_h$ it also contains either $\neg s_h$ or $q$.

The other case is that $c$ contains $t_h$. The only clauses of
{} $A_R \cup A_T \cup A_D \cup A_B$
that contain $t_h$ are $\neg x_h \vee t_h$ and $\neg e_h \vee t_h$. These
clauses are satisfied by $M$ while $c$ is not, therefore $c$ is not one of
them. The first clause $\neg x_h \vee t_h$ only resolves over $x_h$ with $x_h
\vee q$ and all clauses
{} $\neg t_1 \vee \cdots \vee \neg t_n \vee \neg d_j \vee x_h \vee \neg r_h$,
but resolving with the latter only generates tautologies. Therefore, the first
step of resolution is necessarily $\neg x_h \vee t_h, x_h \vee q \vdash t_h
\vee q$. Since none of the involved clauses contains $\neg q$, every clause
obtained from resolution that contains $t_h$ also contains $q$. This also
includes $c$. The same holds by symmetry for $\neg e_h \vee t_h$.

This proves that every minimal-size CNF formula expressing forgetting contains
a clause that includes either
{} $x_h \vee q$,
{} $x_h \vee \neg r_h$,
{} $e_h \vee q$,
{} $e_h \vee \neg s_h$,
or
{} $t_h \vee q$
for each $h$.

\

{\bf Falsity of $\exists X \forall Y . F$.}

The falsity of $\exists X \forall Y . F$ contradicts the existence of a
minimal-size CNF formula of size $k$ expressing forgetting. The relevant
results proved so far are: every CNF formula expressing forgetting has size $k$
or more and is equivalent to $A_R \cup A_T \cup A_D \cup A_B$; the minimal-size
such formulae are $A_N \cup A_T \cup A_D \cup A_B$ where $A_N$ contains, for
each $h$, a clause that includes either
{} $x_h \vee q$,
{} $x_h \vee \neg r_h$,
{} $e_h \vee q$,
{} $e_h \vee \neg s_h$,
or
{} $t_h \vee q$.

A formula $A_N \cup A_T \cup A_D \cup A_B$ of size $k$ expressing forgetting,
if any, is minimal since no smaller formula expresses forgetting. Therefore,
$A_N$ includes, for each $h$, a clause containing one of the five disjunctions.
Since these are not in $A_T \cup A_D \cup A_B$, the size of such formulae is $k
= 2 \times n + ||A_T \cup A_D \cup A_B||$ if every clause of $A_N$ is exactly
one of the above disjunction for each $h$. If $A_N$ contains more than one
clause for some $h$ or the clause for some $h$ contains more than two literals
or $A_N$ contains other clauses, the formula is not minimal.

The case
{} $x_h \vee \neg r_h \in A_N$
can be excluded: it makes
{} $\neg t_1 \vee \cdots \vee \neg t_n \vee \neg d_j \vee x_h \vee \neg r_h \in
{}   A_N$
redundant in
{} $A_N \cup A_T \cup A_D \cup A_B$,
contradicting the minimality of this formula. The case $e_h \vee \neg s_h \in
A_N$ is excluded in the same way.

These exclusions leave $A_N$ to contain exactly one among
{} $x_h \vee q$,
{} $e_h \vee q$,
and
{} $t_h \vee q$
for each $h$ and nothing else.

The final step of the proof is that no such $A_N$ makes
{} $A_N \cup A_T \cup A_D \cup A_B$
equivalent to
{} $A_R \cup A_T \cup A_D \cup A_B$
if $\exists X \forall Y . F$ is invalid. Nonequivalence is proved by exhibiting
a model of the first formula that does not satisfy the second.

Let $M_X$ be the model over $X$ that contains $x_i = \true$ if $x_i \vee q \in
A_N$ and $x_i = \false$ otherwise. Let $M_N$ be the model that assigns every
$e_i$ opposite to $x_i$ and $M_T=\{t_i = \true \mid t_i \in T\}$. By
construction, $M_X \cup M_N \cup M_T \cup \{q=\false\}$ satisfies all clauses
of $A_N$. It also falsifies either $x_i \vee q$ or $e_i \vee q$ for each $i$
because it assigns $\false$ to $q$ and to either $x_i$ or $e_i$. It therefore
falsifies $A_R$.

Since $\exists X \forall Y . F$ is invalid, every interpretation over $X$
falsifies $F$ with an interpretation over $Y$. Let $M_Y$ be the interpretation
over $Y$ such that $M_X \cup M_Y \models \neg F$. Since $F = f_1 \vee \cdots
\vee f_m$, it holds $M_X \cup M_Y \models \neg f_j$ for every $f_j \in F$. It
follows $M_X \cup M_Y \cup M_N \models \neg f_j[e_i/\neg x_i]$ since $M_N$
assigns every $e_i$ opposite to $x_i$.

Merging the results proved in the preceding two paragraphs, $M_X \cup M_T \cup
\{q=\false\} \cup M_N \cup M_Y$ satisfies both $A_N$ and $\neg f_j[e_i/\neg
x_i]$ for every $f_j \in F$.

This model can be extended to a model of $A_N \cup A_T \cup A_D \cup A_B$ by
adding $M_O = \{d_j = \false\} \cup \{r_i = \true\} \cup \{s_i = \true\}$. The
clauses of $A_N$ are already proved satisfied. The clauses $\neg x_i \vee t_i
\in A_T$ are satisfied because $M_T$ contains $t_i = \true$. The clauses $(\neg
f_j[e_i/\neg x_i]) \vee d_j$ are satisfied because $\neg f_j[e_i/\neg x_i]$ is.
The clauses of $A_B$ are satisfied because each contains either $\neg d_j$,
$r_i$ or $s_i$, and these literals are true in $M_O$.

This proves that $M_X \cup M_T \cup \{q=\false\} \cup M_N \cup M_Y \cup M_O$
satisfies $A_N \cup A_T \cup A_D \cup A_B$. It does not satisfy $A_R$, which
means that it falsifies $A_R \cup A_T \cup A_D \cup A_B$. This proves that
{} $A_N \cup A_T \cup A_D \cup A_B$
is not equivalent to
{} $A_R \cup A_T \cup A_D \cup A_B$.

In summary, assuming that the QBF is not valid and that a CNF formula of size
$k$ expresses forgetting, it is proved that the formula does not express
forgetting. This contradiction shows that no formula of size $k$ expresses
forgetting if the QBF is not valid.~\qed

\relemma{general-hard}
{\em
Checking whether forgetting a given set of variables from a minimal-size CNF
formula is expressed by a CNF formula bounded by a certain size is \Dptwo-hard.
}

\

\proof For every $\forall$QBF Lemma~\ref{general-p2} ensures the existence of a
minimal-size CNF formula $A$, a set of variables $X_A$ and an integer $k$ such
that forgetting all variables except $X_A$ from $A$ is expressed by a CNF
formula of size $k$ if the QBF is valid and is only expressed by larger CNF
formulae otherwise.

For every $\exists$QBF Lemma~\ref{general-s2} ensures the existence of a
minimal-size CNF formula $B$, a set of variables $X_B$ and an integer $l$ such
that forgetting all variables except $X_B$ from $B$ is expressed by a CNF
formula of size $l$ if the QBF is valid and is only expressed by larger CNF
formulae otherwise.

% the class \Dptwo\  is defined as the set of problems that are the
% intersection of a \S{2} problem and a \P{2} problem; since $\exists$QBF is
% hard for \S{2} and $\forall$QBF is hard for \P{2}, their combination is hard
% for \Dptwo

A \Dptwo-hard problem is that of establishing whether an $\exists$QBF and a
$\forall$QBF are both valid. If their alphabets are not disjoint, they can be
made so by renaming one of them to fresh variables since renaming does not
affect validity. The same applies to the formulae $B$ and $A$ respectively
build from them according to Lemma~\ref{general-p2} and Lemma~\ref{general-s2}
because renaming does not change the minimal size of forgetting either.
Lemma~\ref{independent} proves that forgetting from $A \cup B$ is expressed by
$C \cup D$ where $C$ expresses forgetting from $A$ and $D$ from $B$. The
minimal size of two such CNF formulae are respectively $k$ and $l$. If the QBFs
are both valid, they are exactly $k$ and $l$ large. Otherwise, they are
strictly larger than either $k$ or $l$. The sum is $k+l$ if both QBFs are valid
and is larger than $k+l$ otherwise.~\qed

\retheorem{general-complexity}
{\em 
Checking whether forgetting some variables from a CNF formula is expressed by a
CNF formula of a certain size expressed in unary is \Dptwo-hard and in \S{3},
and remains hard even if the CNF formula is restricted to be of minimal size.
}

\

\proof Membership to \S{3} is proved first. The problem is the existence of a
CNF formula of the given size or less that expresses forgetting the given
variables from the formula. Theorem~\ref{consistent-literals-complete}
reformulates forgetting in terms of equiconsistency with a set of literals
containing all variables not to be forgotten. Forgetting withing a certain size
is formalized by the following metaformula where $A$ is the formula, $Y$ the
variables not to be forgotten and $k$ the size bound.

\[
\exists B ~.~
	\var(B) \subseteq Y ,~
	||B|| \leq k \mbox{ and }
	\forall S ~.~ \var(S) \subseteq Y \Rightarrow
		(\exists M ~.~ M \models S \cup A
		\Leftrightarrow
		\exists M' ~.~ M' \models S \cup B)
\]

All four quantified entities are bounded in size: $B$ by $k$, $S$ and $M'$ by
the number of variables in $Y$ and $M$ by the number of variables in $A$. This
is therefore a $\exists\forall\exists$QBF, which proves membership to \S{3}.

Hardness to \Dptwo\  is proved by Lemma~\ref{general-hard} in the restriction
where $A$ is minimal.~\qed

%input{acks.tex}

\let\c=\cedilla
\bibliographystyle{plain}

\end{document}